\documentclass[fleqn,usenatbib]{mnras}
\usepackage[T1]{fontenc}
\usepackage{ae,aecompl}
\usepackage{graphicx}	
\usepackage{amsmath}	
\usepackage{amssymb}	
\usepackage{tikz}
\usepackage{times}
\usepackage[normalem]{ulem}
\usepackage{multirow}
\usepackage{tabularx}
\usepackage{bm}            
\usepackage{url}
\usepackage{color,colortbl}
\usepackage{xcolor}
\usepackage{placeins}

\usepackage{hyperref}

\pdfoutput=1

\newcommand{\el}[2]{$^{#1}$#2}
\newcommand{\Ni}{\el{56}{Ni}}
\newcommand{\Msun}{$M_{\odot}$}
\newcommand{\kms}{\ensuremath{~\mathrm{km~s^{-1}}}}

\newcommand{\NIIsinglet}{[\ion{N}{II}] $\lambda$5754}
\newcommand{\NIIdoublet}{[\ion{N}{II}] $\lambda\lambda$6548, 6583}

\newcommand{\OIdoublet}{[\ion{O}{I}] $\lambda\lambda$6300, 6364}

\newcommand{\NIIdiag}{$f_{[\ion{N}{II}]}$}
\newcommand{\Mi}{$M_{\text{He,i}}$}
\newcommand{\Mf}{$M_{\text{preSN}}$}

\definecolor{Gray}{gray}{0.9}

\usepackage{color}
\definecolor{mycolor}{rgb}{.0,.3,1.}

\begin{document}



\title[A New Diagnostic for SN Progenitor Mass]{Nebular Nitrogen Line Emission in Stripped-Envelope Supernovae - a New Progenitor Mass Diagnostic}

\author[Barmentloo et al.]
{\parbox{\textwidth}{
    Stan Barmentloo$^{1}$ \thanks{E-mail: stan.barmentloo@astro.su.se},
    Anders Jerkstrand$^{1}$  \thanks{E-mail: anders.jerkstrand@astro.su.se}, 
    Koichi Iwamoto$^{2}$,
    Izumi Hachisu$^{3}$,
    Ken'ichi Nomoto$^{4}$,
    Jesper Sollerman$^{1}$, 
    Stan Woosley$^{5}$
    }
\vspace{.2cm}\\
$^{1}$ The Oskar Klein Centre, Department of Astronomy, Stockholm University, AlbaNova, SE-10691 Stockholm, Sweden \\
$^{2}$ Department of Physics, College of Science and Technology, Nihon University, Tokyo 101-8308, Japan \\
$^{3}$ Department of Earth Science and Astronomy, The University of Tokyo, Meguro-ku, Tokyo, Japan \\
$^{4}$ Kavli Institute for the Physics and Mathematics of the Universe (WPI),
    The University of Tokyo, Kashiwa, Chiba 277-8583, Japan \\
$^{5}$ Department of Astronomy and Astrophysics, University of California at Santa Cruz, Santa Cruz, CA 95060, USA
}


\pubyear{2023}

\label{firstpage}
\pagerange{\pageref{firstpage}--\pageref{lastpage}}
\maketitle

\begin{abstract}
{Nitrogen is produced by CNO-cycling in massive stars, and can be ejected in significant amounts in supernova explosions. While in H-rich SNe, its [\ion{N}{II}] 6548, 6583 emission becomes obscured by strong H$\alpha$, in explosions of He stars, this nitrogen emission becomes more visible. We here explore the formation of this line, using the \texttt{SUMO} code to compute spectra for a grid of 1D models with parameterized mixing informed from new 2D simulations. Because the mass fraction of nitrogen in the ejecta decreases with larger He core masses, 
as more of the He/N zone gets processed by shell helium burning and is lost to winds, the [\ion{N}{II}] luminosity relative to the overall optical flux probes the He core mass. By comparing to large samples of data, we find that low-mass He cores ($M_{\rm preSN}\lesssim\ 3\ M_\odot$) are exclusively associated with Type IIb SNe, with the exception of Type Ib SN 2007Y. Seeing no strong nitrogen emission in other Type Ib SNe, the implication is either an origin from low-mass stars with the He/N layer (but not the He/C) layer peeled away, or from higher-mass He cores. We also see no clear nitrogen emission in Type Ic SNe. We discuss the diagnostic potential of this new line metric, and also dependencies on mass-loss-rate and metallicity.}

\end{abstract}


\section{Introduction} 
\label{sec:introduction}

Once a massive star ($M_{ZAMS} \gtrsim$ 8 -- 11 \Msun{}; \citealt{Janka_2007_CCSNE, Smartt_2009_Progenitors}) has fused most of its inner core into Fe, it is no longer able to gain energy from nuclear fusion. With fusion shut down, the star loses its main source of outward pressure, initiating the collapse of the star. Eventually, interaction with $\sim 10^{53}$ erg of emitted neutrinos \citep{Colgate_1966_SNcanon,Bethe_1985_Neutrinoheating, Janka_2007_CCSNE, Janka_2012_CCSNE_Neutrinos, Burrows_2013_Neutrino_Explosions} leads to most of the original star being ejected outward with a typical total kinetic energy of $\sim 10^{51}$ erg, at velocities of up to $\sim$ 10$^{4}$ km s$^{-1}$, leaving behind only a neutron star (NS) of about $\sim$ 1.5 \Msun{}. While this violent event (known as a Core Collapse SuperNova, CCSN) is powered by roughly the same mechanism for most massive stars, the resulting light curves and spectra vary starkly \citep[see e.g.,][]{Wheeler_1990_LCvariety, Maguire_2010_LCvariety, Taddia_2018_Nimasses_2009K_a}, requiring an extensive taxonomy necessary to describe all of them \citep{Filippenko_1997_Taxonomy}. 

A particular subcategory of CCSNe are the stripped envelope supernovae (SESNe). As the name suggests, these SNe are thought to arise from progenitors for which the outer hydrogen envelope (Type Ib SNe, identified in \citealt{Wheeler_1985_Ib, Uomoto_1985_Ib, Elias_1985_Ib}) and potentially helium envelope (Type Ic SNe, identified in \citealt{Wheeler_1987_Ic}) was lost before the core collapse. Even Type IIb SNe (identified in \citealt{Woosley_1987_IIb, Woosley_1988_IIb, Filippenko_1988_IIb}) may be considered a part of this class, as they only retain a small part of their hydrogen envelope ($\sim$ 10$^{-3}$ -- 1 \Msun{}; \citealt{Woosley_1994_1993J, 2022crv}), resulting in their spectra relatively soon after explosion no longer being distinguishable from SN Ib spectra \citep{Nomoto_1993_IIb, Podsiadlowski_1993_IIb, Filippenko_1993_IIb}. There exist two widely accepted pathways in the literature for envelope stripping: mass loss by stellar winds \citep[e.g.][]{Woosley_1993_Massloss, Heger_2003_Massloss, Smith_2014_MassLoss} and stripping by a binary companion \citep[e.g.][]{Chevalier_1976__Binaries, Shigeyama_1990_Ib_Mixing, Podsiadlowski_1992_Binaries, Woosley_1995_Binaries, Sana_2012_Binaries}. Following the first work pointing out that the single star channel tends to give too massive ejecta to fit Ibc SN light curves \citep{Ensman1988}, several further lines of arguments and evidence have increasingly implicated the binary mass loss channel as being important \citep[e.g.][]{
Nomoto_1995_Mixing, 
Maund_2004_1993J, Smith_2011_BinarymoreMasstransfer, Dessart2011,Bersten_2012_2011dh, Eldridge_2013_BinarymoreMass,Jerkstrand_2015_NII_discovery,  Yoon_2017_BinaryRadii}.

To better understand the progenitor histories of SESNe, detailed computational modeling of their light curves and spectra is an essential tool. Nebular phase ($\gtrsim$ 150 days post-explosion for SESNe) models and observation are in particular useful, as during this phase the ejecta are becoming mostly optically thin, so that the spectra directly reveal the full nucleosynthetic content of the expanding nebula. Comparing strong nebular emission lines such as \OIdoublet{}, [\ion{Ca}{II}] $\lambda\lambda$7291,7323 and \ion{Mg}{I}] $\lambda$4571 to models makes it possible to constrain the ejected amounts of these elements, as well as the composition in the different layers in the ejecta \citep[see][for a review]{Jerkstrand_2017_book}. The late phases requires NLTE modelling. Among state-of-the- art, non-local thermodynamic equilibrium (NLTE) modeling codes, one can mention \texttt{SUMO} \citep{Jerkstrand_2011_SUMOa, Jerkstrand_2012_SUMOb, Jerkstrand_2014_SUMOc}, \texttt{CMFGEN} \citep{Hillier_1998_CMFGENa, Dessart_2010_CMFGENb, Hillier_2012_CMFGENc} and \texttt{JEKYLL} \citep{Ergon_2018_JEKYLLa, Ergon_2022_JEKYLLb}. NLTE capabilities have also been developed for the initially LTE-codes \texttt{SEDONA} \citep{Kasen2006} and \texttt{ARTIS} \citep{Kromer2009} by \citet{Botyanszki2018} and \citet{Shingles2020}, respectively.

One of the opportunities arising from these modeling efforts is to constrain the SN progenitor mass from its inferred nucleosynthesis. An early devised method to this end is to measure the ratio of \OIdoublet{} / [\ion{Ca}{II}] $\lambda\lambda$7291,7323 \citep{Fransson_1989_CaOIratio}. Because oxygen yields increase very strongly with He-core mass \citep{Woosley1995, Thielemann_1996_oxygen, Limongi2000}, whereas calcium yields are more weakly dependent, this ratio has a general tendency to increase with progenitor mass \citep{Nomoto_2013_O-amount}. However, as discussed in \citet{Jerkstrand_2017_book}, some issues with this methodology make it difficult to use with confidence for high accuracy.  

As an alternative, with improved understanding of the morphology of CCSNe ejecta from 3D hydrodynamic simulations, enabling a realistic mixing ansatz for 1D-modelling, and observed spectra with better flux calibrations and background subtractions, methods developed over the last decade have largely focused on direct, individual line luminosity modelling. Today, grids of [\ion{O}{I}] line luminosities versus progenitor mass are the main workhorses to determine SN progenitor masses for both H-rich \citep{Jerkstrand_2012_SUMOb,Jerkstrand_2014_SUMOc,Dessart2021} and H-poor \citep{Jerkstrand_2015_NII_discovery,Dessart_2021_Hestarexpl} SNe. Application of these grids to observations have given good matches with the low and intermediate mass parts of a regular IMF, starting at $M_{ZAMS}\sim\ 8-10\ M_\odot$, but a dearth of massive ($M_{ZAMS}\gtrsim\ 25\ M_\odot$) progenitors exploding as either H-rich or H-poor SNe, broadly in line with results from progenitor imaging \citep{Smartt2015}.

Nevertheless, a robust and complete understanding of CCSNe requires modelling and understanding of all their major emission lines. There are always caveats with any single method, for example the quite uncertain molecule and dust cooling in the oxygen-rich zones in the case of the [O I] lines. For the large future datasets expected from the next generation surveys such as the LSST, distances will also in general become larger for observed SNe, and extinction and background contamination could be expected to become issues, keeping line ratios and line luminosities as fraction of total optical brightness as important diagnostics.

\vspace{0.2 cm}

The most recent new line identified in CCSN nebular spectra is that of \NIIdoublet{}. In the Type IIb models of \citet{Jerkstrand_2015_NII_discovery}, nitrogen was for the first time included in a CCSN nebular model, and it turned out to produce relatively strong \NIIdoublet{} emission - providing a natural explanation for the emission line sometimes seen at these wavelengths in SESNe. Furthermore, it resolved the issue of previous models failing to produce strong enough H$\alpha$ to understand the observations assuming the emission came from hydrogen \citep[see e.g. section 6.3 discussion in][]{2008ax_b}.

Additional analysis of 7 Type IIb and 2 Type Ib SNe by \citet{Fang_2018_NIIdiscovery} found a similar result. Further strengthening this finding, \citet{Dessart_2021_Hestarexpl} and \citet{Dessart_2023_TimeEvolution} performed spectroscopic modeling of the Type Ibc SN models by  \citealt{Woosley_2019_models,Ertl_2020_models} and also found significant emission by \NIIdoublet{} in many of their models. More specifically, it was found in \citet[][their fig. 5]{Dessart_2021_Hestarexpl} that the relative strength of this emission almost monotonically decreases with an increase of the SN progenitor mass.

The nitrogen resides in the outermost part of the He-shell, being a residual of the CNO burning phase. While the CNO cycle mainly turns \el{1}{H} into \el{4}{He}, about 1\% (by mass) of the end products is \el{14}{N}: as the reaction that consumes it (\el{14}{N}($p, \gamma$)\el{15}{O}) is the slowest in the cycle, it becomes a bottleneck, leading to a build up of \el{14}{N}. When eventually the star enters a phase of helium shell-burning, this \el{14}{N} is very effectively consumed via \el{14}{N} ($\alpha, \gamma$)\el{18}{F}($\beta^{+}\nu$)\el{18}{O}, so that the end products of this phase only contain 10$^{-3}$ -- 10$^{-1}$ \% \el{14}{N}. Exactly what part of the He/N envelope is left at the SN-explosion is dependent on the star's temperature in this region through its evolution, and as this temperature is higher for more massive stars, this explains why the relative amount of unburned \el{14}{N} (and thus its emission strength) inversely scales with the progenitor mass.

In this work, we aim to combine spectral modelling with observational analysis to investigate \NIIdoublet{} emission in SESNe progenitors. On the modelling side, we explore a selection of the helium star progenitor models with "standard" mass loss from \citet{Woosley_2019_models} and \citet{Ertl_2020_models} using the \texttt{SUMO} code. We carry out new 2D hydrodynamic simulations of the Rayleigh-Taylor instability (Iwamoto et al. 2024, in preparation) in order to accurately parameterize the mixing of the ejecta for the 1D modelling. By considering the \NIIdoublet{} emission for each model at multiple epochs, so-called nitrogen-curves are obtained, distinct for each model mass. These tracks are then interpolated to fill up the parameter space. Then, we measure nitrogen-tracks for a sample of SESNe and compare these to the modelled space. From this comparison, we will provide progenitor mass estimates (for our mass-loss assumption, see Section \ref{sec:uncertainties}) for all SNe considered in this work, about half of which had no previous estimates in the literature. The other half will be compared to the available literature estimates to determine the success of our diagnostic. Finally, we will discuss the mass distributions that arise for the three different SN types.

The paper is structured as follows. In Section \ref{sec:observations}, the data selection and data processing procedures are discussed. Section \ref{sec:methodology} lays out how the models were computed, including both the hydrodynamic and radiative transfer modelling. A detailed description of how the amount of \NIIdoublet{} emission is determined is also provided. The main results of this work are presented in Section \ref{sec:results}, where both the nitrogen-curves and derived progenitor masses for our SN sample are given. These results are interpreted and discussed in Section \ref{sec:discussion}, with a large focus on the differences between our three considered SESN types. Finally, Section \ref{sec:conclusion} summarises the main results of this work.

\section{Observations}
\label{sec:observations}

In this section, the observations used in this work are described. In Section \ref{sec:data_selection} we describe how the archival observational data was obtained and describe our selection of SNe. Then, we describe the data processing (e.g. host galaxy emission removal) applied to this data set to come to our final collection of spectra in Section \ref{sec:data_reduction}. An overview of this final collection is given in Table \ref{tab:sample}.

\subsection{Data Selection}
\label{sec:data_selection}

In this work, the main source used to acquire observed nebular spectra is the Weizmann Interactive Supernova Data Repository (WISeREP) \citep{Yaron_2012_WISEREP} \footnote{The repository can be accessed at \url{https://www.wiserep.org/}}. WISeREP is an open archive of SN observations that allows users to upload spectra from their own dedicated observational campaigns. With a total of around 50,000 available spectra, it is an ideal tool to get a quick and clear overview of all available observations for a given SN type.

Our selection starts by retrieving all available spectra marked as SNe Type Ib, Ic or IIb, with the requirement that the particular SN needs to have at least 5 spectra available (not necessarily in the nebular-phase) in WISEREP. This requirement ensures that we only select those SNe that have reasonable time coverage, which greatly decreases the uncertainty in the inferences made when later comparing to models.

After this initial cut, each individual spectrum that fulfilled two further requirements was visually examined. This inspection is a necessity, as some spectra posted on WISEREP are simply not useful for our work, due to a variety of reasons (e.g. bad S/N, too strong contamination from the host galaxy). To come to visual examination, the spectra had to 1) at least cover the wavelength range 5000 -- 8000 Å and 2) be taken at least 120 days after explosion. The first requirement ties in directly to our definition of the [\ion{N}{II}] diagnostic (see Section \ref{a:diagnostic}), which contains the total integrated flux in this specific range. The second requirement is to ensure that the spectrum is likely taken during the nebular phase, so that the flux in the region around \NIIdoublet{} can be said to originate mostly from \NIIdoublet{} and not from other ions (see Section \ref{a:diagnostic} for more discussion). The explosion dates were adopted from literature (see Table \ref{tab:sample}), or taken to be 20 days before the time of peak-light when no explosion date was given. This value of 20 days is rather typical for most Type Ib, Ic and IIb SNe \citep{Valenti_2011_risetime_2009jf, Taddia_2015_Risetime}. 

Besides this main group of spectra, some additional spectra were obtained from the Open Supernova Catalog (OSC)\footnote{This catalogue can be found at \url{https://github.com/astrocatalogs/supernovae}}, while those for iPTF13bvn were directly taken from the source paper by \citet{PTF12os_iPTF13bvn_b}. Furthermore, the spectra for SN 2019odp were kindly provided by the authors of \citet{2019odp}, and the same was true for the spectra of SN 2022crv by the authors of \citet{2022crv}. The source papers for the remaining spectra are given in Table \ref{tab:sample}.

\subsection{Data Processing}
\label{sec:data_reduction}

With the sample selected, the data had to be standardised. This process included correcting for host galaxy and Milky Way extinctions (assuming a \citet{Fitzpatrick_1999_Reddening} reddening law with $R_{V} = 3.1$), correcting for redshift and removing any strong host galaxy emission lines. As quite often no further information was given in the metadata files on whether any of these corrections had already been performed, it was necessary to check each individual spectrum. Exactly which corrections were performed for each individual spectrum was recorded and can be retrieved from the final data products in the online repository\footnote{All scripts necessary to reproduce the figures in this paper can be found at \url{https://github.com/StanBarmentloo/NII_nebular_phase}}.

\begin{figure}
        \centering
        \includegraphics[width=.98\linewidth,angle=0]{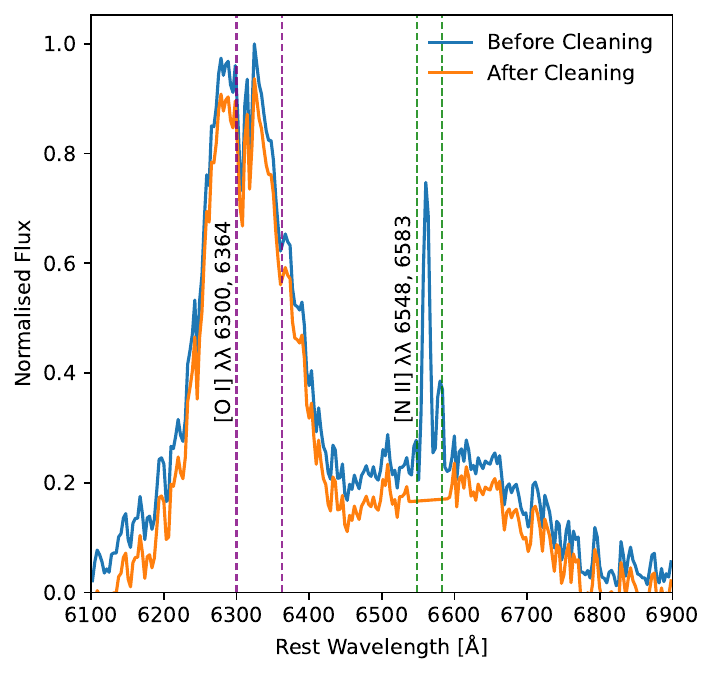}
        \vskip -.2cm
        \caption{An example of the data processing performed (as described in Section \ref{sec:data_reduction}) on the observational data. For this particular observation of SN 2011ei (presented in \citet{2011ei}), narrow H$\alpha$ and [\ion{N}{II}] $\lambda$ 6583  between 6540 Å and 6590 Å were removed. Furthermore, a small amount of continuum flux was removed (see final paragraph in Section \ref{sec:fitting_procedure}). The rest wavelengths of  \OIdoublet{} and \NIIdoublet{} have been indicated with vertical dashed lines.  
        }
          \label{fig:halpha_removal}
\end{figure}

The most important correction that had to be performed was the removal of any narrow line emission in the 6500 -- 6600 Å region (most often H$\alpha$ and \textbf{narrow} [\ion{N}{II}] $\lambda\lambda$ 6548, 6583) from the SN-host). If left unchanged, these features would interfere with our fitting algorithm and thus the inference of the amount of \NIIdoublet{} emission from the ejecta. The features have distinct narrow profiles, not wider than $\sim$ 25 Å (compared to typical widths of 200 -- 250 Å for \NIIdoublet{}). Therefore, if any clear excess narrow emission was present, the region 6540 -- 6590 Å was 'cleaned' by replacing the narrow lines with a straight line between 6540 Å and 6590 Å. However, if these emission features were completely dominating the spectrum or were wider than the 6540 -- 6590 Å region (and thus likely caused by CSM interaction, see e.g. \citealt{Matheson_2000_1993Jboxy} for SN 1993J and \citealt{Maeda_2015_2013df} for SN 2013df), we discarded the spectrum. For clarity, Figure~\ref{fig:halpha_removal} visualises the applied procedure.

At the end of the data selection and reduction process, we were left with a final sample of 14 Ib, 12 Ic and 17 IIb SNe, with a total of 34 Ib, 26 Ic and 71 IIb spectra. The sample is summarised in Table \ref{tab:sample}.

\section{Methodology}
\label{sec:methodology}

In this section the methodology is described that was used to establish a consistent nitrogen-line diagnostic for SESNe. Section \ref{sec:modelspectra} describes the input 1D hydrodynamic models that were used, as well as the 2D hydrodynamic simulations that were performed to parameterize the mixing for the 1D spectral modelling. Section \ref{sec:RTmethod} describes the radiative transfer modelling done with \texttt{SUMO} to produce synthetic spectra of these input models. Finally, section \ref{sec:fitting_procedure} describes the methodology used to estimate the nitrogen emission line luminosity from observed spectra, and how our final diagnostic quantity is defined.

\subsection{Hydrodynamic models}
\label{sec:modelspectra}

\subsubsection{1D stellar evolution and explosions}
\label{sec:inputmodels}

As input models for the \texttt{SUMO} spectral modelling, we adopted 5 models (\Mi{} = 3.3, 4.0, 5.0, 6.0 and 8.0 \Msun{}\footnote{For reference, the $M_{\rm ZAMS}$ of these models were 16.1, 18.1, 20.8, 23.3, and 27.9 \Msun{}. In the remainder of this work, $M_{\rm ZAMS}$ will be sparsely mentioned. The main reason for this is explained in Section \ref{sec:mass_determination}.}, from here on referred to as he3p3, he4p0 etc.) from the helium star grid evolved and exploded by \cite{Woosley_2019_models} and \cite{Ertl_2020_models}, respectively. These models were evolved with solar metallicity and the "standard" mass-loss prescription (see \citet{Woosley_2019_models}; the effects of metallicity and mass-loss are further discussed in Section \ref{sec:uncertainties}). In this work, \Mi{} refers to the mass of the star at the onset of core helium-burning. The ejecta models were obtained from the Garching CCSN archive\footnote{\url{https://wwwmpa.mpa-garching.mpg.de/ccsnarchive/}}.

The stellar evolution of the input models was done using the \texttt{KEPLER} code, which is described in detail in e.g. \cite{Weaver_1978_KEPLERa} and \citet{Woosley_2002_SNcanon}. As initial composition of each model star, the products of hydrogen burning from the 13 \Msun{}, solar metallicity model from \citealt{Woosley_2015_Initcomp} are taken. To consider the star as one evolving in a binary, the authors then instantaneously remove the hydrogen envelope, so that the star left at the beginning of the simulation is a bare helium star. This is an approximate treatment of early Case B mass transfer. Due to the absence of a hydrogen burning shell during the He core burning, the He core loses mass over time, due to winds, rather than gain mass, as happens by hydrogen shell burning when the hydrogen envelope is retained. The final pre-collapse He core masses are in the range $2.7-5.6\ M_\odot$ (see Table \ref{tab:core_params}).

\begin{figure*}
        \centering
        \includegraphics[width=.8\linewidth, height = .5\linewidth, angle=0]{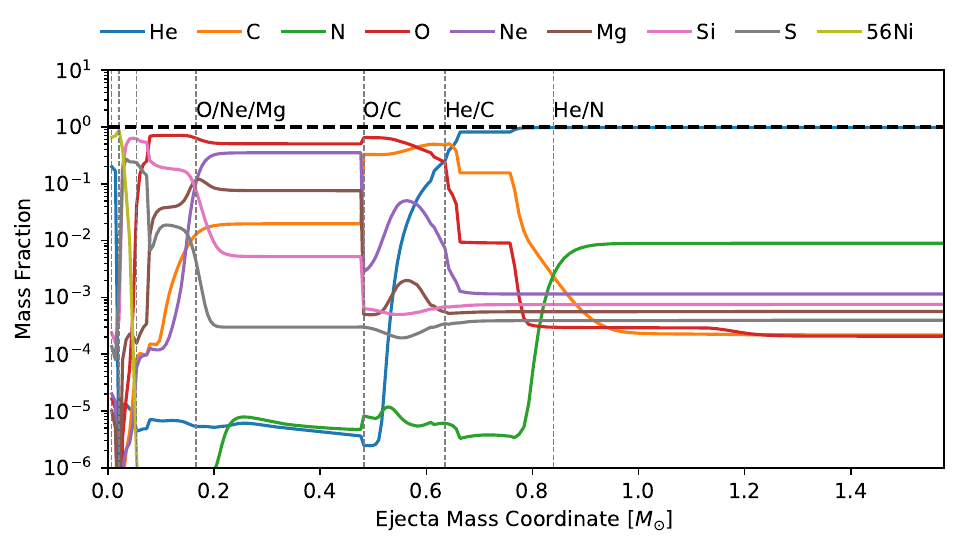}
        \vskip -.2cm
        \caption{Composition of the ejecta for the \Mi{} = 4.0 \Msun{} helium star SN model of \citet{Ertl_2020_models}. The elements shown are He, C, N, O, Ne, Mg, Si, S and \el{56}{Ni}, which cover the most abundant elements in the ejecta. The x-axis shows the mass coordinate in the ejecta (i.e. $M = 0$ is the surface of the neutron star). The boundaries between the different zones as defined in Section \ref{sec:inputmodels} are indicated by vertical dashed lines, while the horizontal dashed line indicates a mass fraction of 1. The outer zones have been indicated with their names, and the inner three zones are from left to right Fe/He, Si/S and O/Si/S. Note that these ejecta are still unmixed, for the mixing procedure, see Section \ref{sec:submix}.
        }
          \label{fig:Woosleymodel}
\end{figure*}

The explosions of the input models were performed using the \texttt{P-HOTB} code \citep{Janka_1996_PHOTBa, Kifonidis_2003_PHOTBb}, and are extensively described in \cite{Ertl_2020_models}. In short, \texttt{P-HOTB} simulates explosions using the neutrino-driven explosion mechanism rather than a simple piston. While most of the parameters in the code are constrained, three of them are left free so that explosions with realistic energies and \Ni{} production can be simulated. To calibrate these parameters, the resulting energies and \Ni{} productions are compared to observed values for SN 1987A and SN 1054 (The Crab), which span energies and \Ni{} productions of 10$^{50}$ -- $\gtrsim$10$^{51}$ erg and $\gtrsim$ 10$^{-3}$ -- 0.07 \Msun{}, respectively. To get the proto-neutron star parameters for a simulated precollapse star, the parameter $M_{3000}$ is determined (the mass enclosed within the central 3000 km). Based on this parameter, interpolated values between the Crab- and 87A-engines are determined (for further details, see \citet{Sukhbold_2016_explosions}). With the engine parameters set, the hydrodynamic evolution is followed from the beginning of core collapse up to the late stages of ejecta expansion. The nucleosynthesis is obtained by postprocessing the thermodynamic trajectories with the \texttt{KEPLER} nuclear network. As an example of the final SN structure, the composition profiles for selected elements (in addition to these also Na, Ca, and Fe are included in the \texttt{SUMO} modelling) in the \Mi{} = 4.0 \Msun{} model is shown in Figure \ref{fig:Woosleymodel}.

The resulting ejecta models specify the mass, velocity, and composition ($\sim$ 1000 isotopes up to \el{128}{Sb} for models he3p3 and he4p0, and $\sim$ 1900 isotopes up to \el{211}{At} for the more massive ones) of several hundred radial shells\footnote{These models are thus the same as used in \citet{Dessart_2021_Hestarexpl}, as modified from the original explosion models from \citet{Ertl_2020_models}}. Because the composition profiles naturally divide into seven regions with relatively uniform abundances in each, corresponding to the seven main burning stages (H$\rightarrow$He/N, He-partial$\rightarrow$He/C, He$\rightarrow$O/C, C$\rightarrow$O/Ne/Mg, Ne$\rightarrow$O/Si/S, O$\rightarrow$Si/S, Si$\rightarrow$Fe), we follow standard methodology \citep{Kozma1998,Jerkstrand_2011_SUMOa,Dessart_2021_Hestarexpl, Dessart_2023_TimeEvolution} to discretise the grid into these zones. The purpose of this is partially to reduce the number of shells to solve the conditions in, but also to enable a manageable artificial mixing using the virtual grid method \citep[][more details below]{Jerkstrand_2011_SUMOa}. An additional gain becomes that, working with a manageable set of discrete zones, we can achieve a graspable analysis of the behaviour of the models.

To choose the exact boundaries between the zones, we follow the rules listed in Table~\ref{tab:rules}. A rule was also added that did not allow a zone to have more than 20\% difference between lowest and highest velocity. If a zone broke this rule, it was divided into subzones of equal composition, but simply with a velocity difference of at most 20\%. Only the He/N zone was affected by this, becoming divided into 5-6 shells. The final composition of each layer (after the mixing procedure described in Section \ref{sec:submix} was performed) for our models can be found in Tables \ref{tab:comp33}-\ref{tab:comp80} of Appendix \ref{appendix:model_compositions}.

As discussed by \cite{Ertl_2020_models}, the \Ni\ production is quite uncertain in the explosion modelling. This is because the 1D models are too simple to reliably determine the neutron richness and degree of $\alpha$-rich freezeout in the innermost layers, which gives uncertainty in the relative abundances of $^{56}$Ni, neutron-rich isotopes (labelled "Tr"), and $\alpha$-particles. We here follow the approach to use observationally inferred values to guide the choice of \Ni\ values, but staying within the range allowed for in the \cite{Ertl_2020_models} simulations. Values estimated for observed Type Ib, Ic and IIb SNe by \cite{Taddia_2018_Nimasses_2009K_a} have a median of 0.1 -- 0.15 \Msun{} for each class, with only 2 out of 33 objects having a \Ni{} mass below 0.07 \Msun{}. In \citet{Prentice_2016_biggersample} and \citet{Prentice_2019_Bigsample_2015ah} the found median is slightly lower, namely 0.07 -- 0.09 \Msun{}. The upper-end values in \citet{Ertl_2020_models}, $0.75 \times \left(\mbox{\Ni+“Tr”+$\alpha$}\right)$, are closest to these values, giving \Ni{} masses of $0.055, 0.061, 0.11, 0.084$, and $0.061\ M_\odot$ for the he3p3, he4p0, he5p0, he6p0, and he8p0 models, respectively, although even these are lower than observationally inferred. We note that our \Ni{} masses by this choice are about 25\% higher than those in the models of \cite{Dessart_2021_Hestarexpl,Dessart_2023_TimeEvolution}, using the same progenitors.

\begin{figure*}
        \centering
        \includegraphics[width=.98\linewidth ,angle=0]{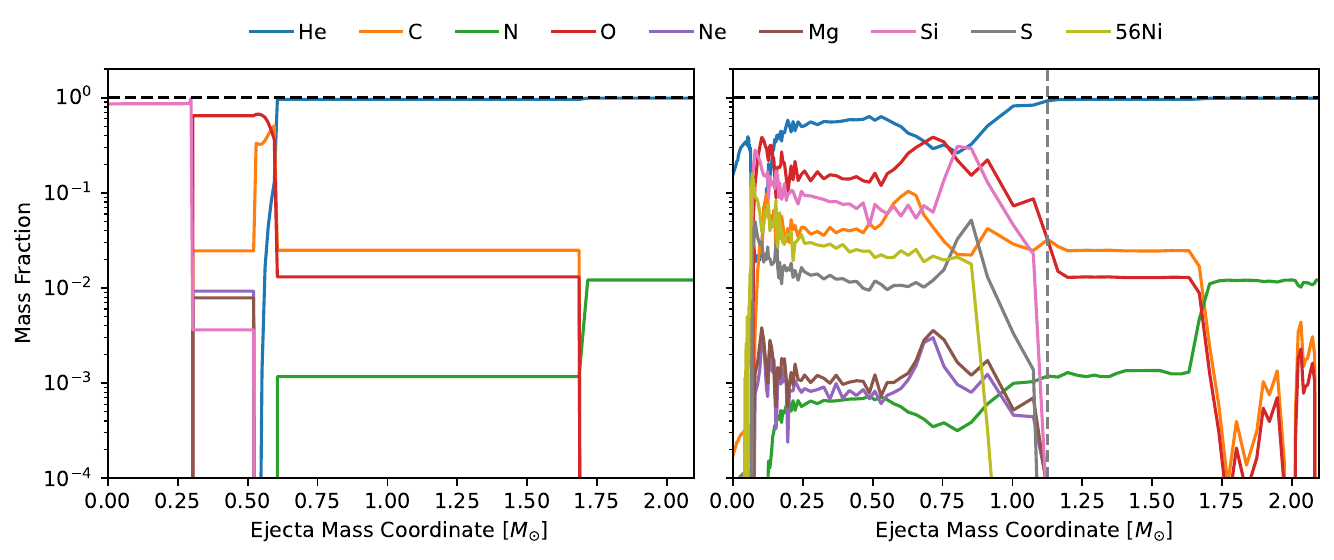}
        \vskip -.2cm
        \caption{Composition of the ejecta for the 2D hydrodynamical models (Iwamoto et al. 2024, in preparation) before (left plot) and after (angle-averaged values, right plot) explosion for a 3.3 \Msun{} progenitor \citep{Nomoto_1988_Hestars}. The elements shown are the same as in Figure \ref{fig:Woosleymodel}. On the x-axis, the ejecta mass coordinate is given, with $M = 0$ being the surface of the neutron star. The boundary between what is considered 'core' and 'envelope' in our mixing prescription (see Section \ref{sec:inputmodels}) is indicated by a vertical dashed line in the rhs plot. One can see that while the He/N layer remains mostly unmixed, some nitrogen, residing in the He/C layer, gets mixed inward, bringing it closer to the \el{56}{Ni}. 
        }
          \label{fig:Mixing_33}
\end{figure*}

\begin{figure*}
        \centering
        \includegraphics[width=.98\linewidth, angle=0]{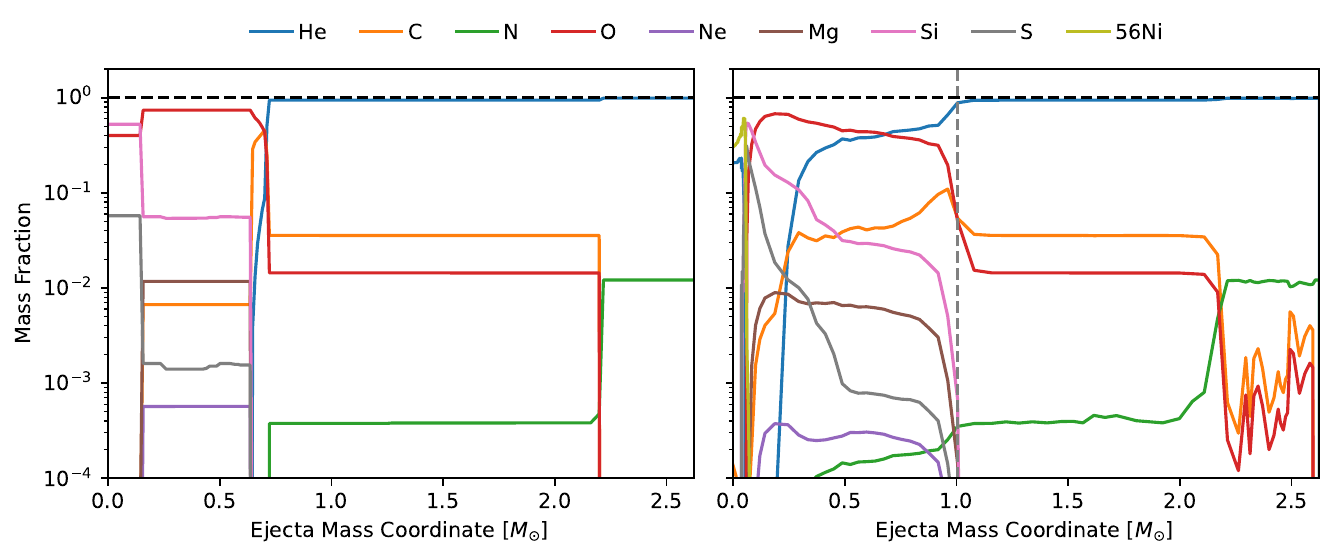}
        \vskip -.2cm
        \caption{Same as Figure \ref{fig:Mixing_33}, but now for a \Mi{} = 4.0 \Msun{} progenitor \citep{Nomoto_1988_Hestars}. Comparing to Figure \ref{fig:Mixing_33}, one can see that the extent of the mixing here is a lot less. Compare e.g. the extent of nitrogen inmixing, which now occurs over both a smaller mass range ($\sim$ 0.7 \Msun{} vs $\sim$ 1.0 \Msun{}) and at a lower abundance ($\sim 2\times 10^{-4}$ vs $\sim 6\times 10^{-4}$).
        }
          \label{fig:Mixing_40}
\end{figure*}

\subsubsection{2D hydrodynamic models and mixing treatment}
\label{sec:submix}

The SN models of \cite{Ertl_2020_models} are computed in spherical symmetry. However, mixing induced by symmetry breaking is a well-established property of SN ejecta \citep{Falk_1973_RT, Chevalier_1978_RT, Ebisuzaki_1989_asymmetry, Hachisu_1992_asymmetry, Kifonidis_2006_mixing, Janka_2007_CCSNE}. Modelling in spherical symmetry therefore requires, in most cases, some kind of artificial, parameterized mixing in order to achieve realistic light curves and spectra.

For SESNe, the 2D hydrodynamical modelling by \cite{Shigeyama_1990_Ib_Mixing}, \cite{Hachisu_1991_Ib_Mixing_Models}, \cite{Nomoto_1995_Mixing} and \cite{Iwamoto_1997_RTinstab} were pioneering in showing that also without a massive H envelope (which seeds strong Rayleigh-Taylor instabilities in Type II SNe), SNe can still achieve significant mixing by action of other instabilities and composition interfaces. They found that the strength of the mixing was inversely related to the He core mass, with the smallest-mass cores experiencing the most extensive mixing. The reason for this is twofold: lower-mass cores have a) a steeper density gradient between the CO-core and b) the ratio of envelope to core mass is larger, so that the core is more decelerated. These two effects lead to stronger Rayleigh-Taylor instabilities in lower-mass cores, so that mixing is more extensive.

Of particular importance for the work here is to obtain a realistic treatment of the degree of mixing of the He/N layer. If this layer is partially mixed into the core region and obtains a close proximity to the $^{56}$Ni power source, the nitrogen (and helium) emission can become significantly stronger than if it remains unmixed. As one of the key goals here is to investigate whether nitrogen lines can be a diagnostic of the He-core mass, it becomes of particular importance to capture the mass-dependent behaviour of the mixing indicated in the studies mentioned above.

Currently available multi-D hydrosimulations of SESNe are either of too limited scope and detail \citep{Shigeyama_1990_Ib_Mixing,Hachisu_1991_Ib_Mixing_Models,Nomoto_1995_Mixing, Iwamoto_1997_RTinstab} or do not include nitrogen (\citealt{Wongwathanarat_2013_multidhydro_a, Wongwathanarat_2015_multidhydrob}, van Baal et al. (2024, in preparation)), so that they can not provide sufficient information for the purposes of this paper. To obtain a clearer picture of the mixing of the He/N layer in Type Ib SNe, we therefore carried out new 2D simulations (Iwamoto et al. 2024, in preparation), following a method similar to the one used in \citet{Hachisu_1991_Ib_Mixing_Models}, with the modification of employing an Harten-Lax-van Leer-Contact (HLLC) solver \citep{Toro_1994_HLLC} utilizing the fifth-order Weighted Essentially Non-Oscillatory (WENO) reconstruction \citep{Jiang_1996_WENO} as approximate Riemann solver. The simulations were done for the 3.3 and 4.0 M\(_\odot\) helium stars described in \citet{Nomoto_1988_Hestars}. Early shock propagation was computed using a 1D Lagrangian code \citep{Shigeyama_1992_Lagrangian}. Nucleosynthesis was dealt with by an $\alpha$-network first and recalculated by a larger-size reaction network in post-processing. Subsequently, the 1D models were mapped onto a cylindrical computational domain with 
\(1025 \times 1025\) grid points, and a small sinusoidal perturbation in the form of \( 1+\epsilon \cos (m \theta)\) with \(m=20,\epsilon=0.05\) was applied to the velocities in shocked regions. To track the elements during mixing, approximately \(4 \times 10^4\) marker particles, characterized by fixed chemical compositions according to the 1D models, were distributed on the 2D grids. Their movement in the ambient flows was traced using a third-order Runge-Kutta method. At a later time, when the instability had ceased to grow, the abundances averaged over angles were computed as a function of enclosed mass.

The structures of the pre- and post-mixing models are shown in Figures \ref{fig:Mixing_33} and \ref{fig:Mixing_40}. The figures show that, in fact, there is no significant inward mixing of the He/N zone in either of the two models. Instead, the simulations clarify that the in-mixing of helium-rich material seen in the 1990s simulations is limited to the He/C zone. Because this zone also has some nitrogen (mass fraction $10^{-4}-10^{-3}$, Tables \ref{tab:comp33}-\ref{tab:comp80}, compare $\sim 10^{-2}$ in the He/N zone) a certain fraction of the total nitrogen emission could in principle come from this component, as it mixes much more with the \Ni. About 40\% of the initial nitrogen in the He/C zone pre-explosion is mixed inwards in the 3.3 $M_\odot$ model, and 18\% in the 4.0 $M_\odot$ model. Based on this, we divide the He/C zone in the input model into two parts, letting the 40(18)\% part become macroscopically (but not microscopically) mixed into the core, and the 60(82)\% part remain in the envelope. Details of the method for macroscopic mixing of material in the core can be found in \citet{Jerkstrand_2011_SUMOa}.

The reader is reminded that the models shown in Figures \ref{fig:Mixing_33} and \ref{fig:Mixing_40} \citep{Nomoto_1988_Hestars} (Iwamoto et al. 2024, in preparation) are only used in our work to determine the inmixing parameter (e.g. 40\% for the 3.3 \Msun{} model), not for their composition. This is relevant, as the compositions used in our ejecta models (Figure \ref{fig:Woosleymodel}) are different than those used to determine the inmixing parameter (Figures \ref{fig:Mixing_33} and \ref{fig:Mixing_40}); the relative sizes of the He/C to He/N zones vary, as well as the total ejecta masses.The models in Figures \ref{fig:Mixing_33} and \ref{fig:Mixing_40} experienced no mass-loss by winds, explaining why they have larger ejecta masses. 

As no models were run for more massive He cores, we had to make an assumption for the inmixing parameter for the He/C zone in these models. A value of 10\% was chosen for all models, with the added note that due to the relatively small envelope layers in these more massive models, we expect the difference in output spectra between 10\% inmixing and say 5\% or 0\% to be small or negligible. Additionally, the observed reduction in mixing extent with mass in e.g. \citet{Nomoto_1995_Mixing} indicates that mixing in these higher mass models should be low, at least lower than for the 4.0 \Msun{} model.

For the metal-rich regions (\Ni, Si/S, O/Si/S, O/Ne/Mg, O/C), Figures \ref{fig:Mixing_33} and \ref{fig:Mixing_40} show that the 3.3 $M_\odot$ model has very strong mixing between all these. The mixing in the 4.0 $M_\odot$ model is still significant, although not as complete. We use the standard "full" macroscopic mixing approach of \texttt{SUMO} of the metal core for all models. This metal core is defined as all the material with $V \leq V_{\text{core}}$. For each model, this $V_{\text{core}}$ is the velocity that includes exactly 40(18,10) \% of the He/C zone. This methodology was chosen to stay consistent with the models in \citet{Woosley_2019_models}, as they have well determined velocities for each radial shell. This choice of $V_{\text{core}}$ is different than in previous works with \texttt{SUMO} (e.g. \citet{Jerkstrand_2015_NII_discovery}), where the authors instead set $V_{\text{core}}$ for each model to 3500 \kms{}, based on a good reproduction of observed linewidths for the SNe they considered. The $V_{\text{core}}$ values here are significantly larger, $4000-7000$ \kms{}. An overview of these and other key properties of the finalised input models is presented in Table \ref{tab:core_params} and the detailed compositional structure per model layer can be found in Tables \ref{tab:comp33}-\ref{tab:comp80}.

\begin{table}
    \centering
    \begin{tabular}{l|lllll}
    \hline
    Model & \Mf{}  & $E_{\text{exp}}$  & $M_{^{56}\text{Ni}}$  & $V_{\text{core}}$  & He/C- \\
     & [\Msun{}] & [10$^{51}$ erg] & [\Msun{}] & [\kms{}] & inmixing \\ \hline \hline
    he3p3 & 2.67 & 0.55 & 0.055 & 4300 & 40\% \\
    he4p0 & 3.15 & 0.64 & 0.061 & 4500 & 18\% \\
    he5p0 & 3.82 & 1.49 & 0.11 & 7000 & 10\% \\
    he6p0 & 4.45 & 1.07 & 0.084 & 5700 & 10\% \\
    he8p0 & 5.64 & 0.70 & 0.061 & 4000 & 10\% \\
    \hline \hline
    
    \end{tabular}
    \caption{An overview of some of the key parameters for our five progenitor models.}
    \label{tab:core_params}
\end{table}

\begin{figure*}
        \centering
        \includegraphics[width=.98\linewidth, angle=0]{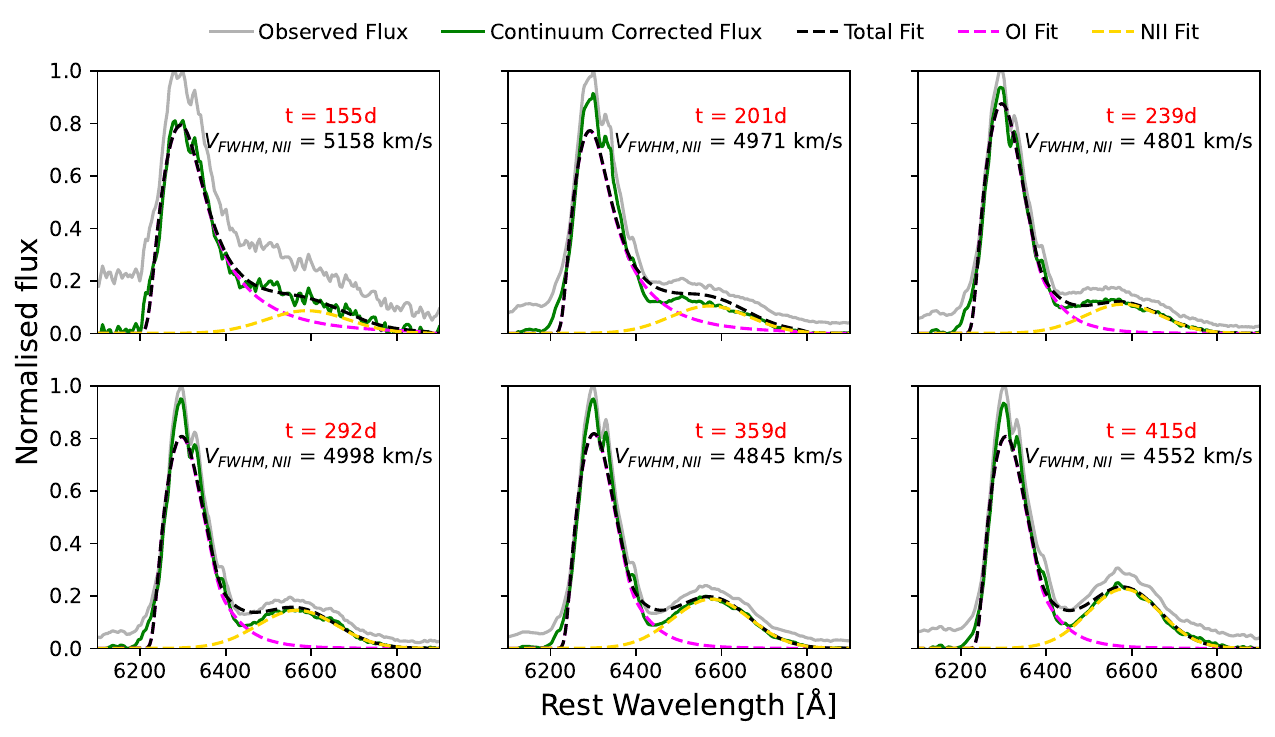}
        \vskip -.2cm
        \caption{Example of a best fit sequence for SN 2011dh. The epoch since explosion and the best fit FWHM velocity for the \NIIdoublet{} contribution is indicated in each subplot. Fits are indicated in dashed lines. The observed flux is shown in grey, while our continuum corrected flux is shown in green. The flux in each spectrum is normalised to the peak of the observed \OIdoublet{} flux. The sequence exemplifies the use of our velocity range constraint: using the clearer shape of the later spectra, the fitting algorithm knows which velocities are reasonable estimates for the earlier epochs, something which is not really possible to know from the first epochs alone.}
        \label{fig:Multi_epoch_fit}
\end{figure*}

\begin{figure}
        \centering
        \includegraphics[width=.98\linewidth,angle=0]{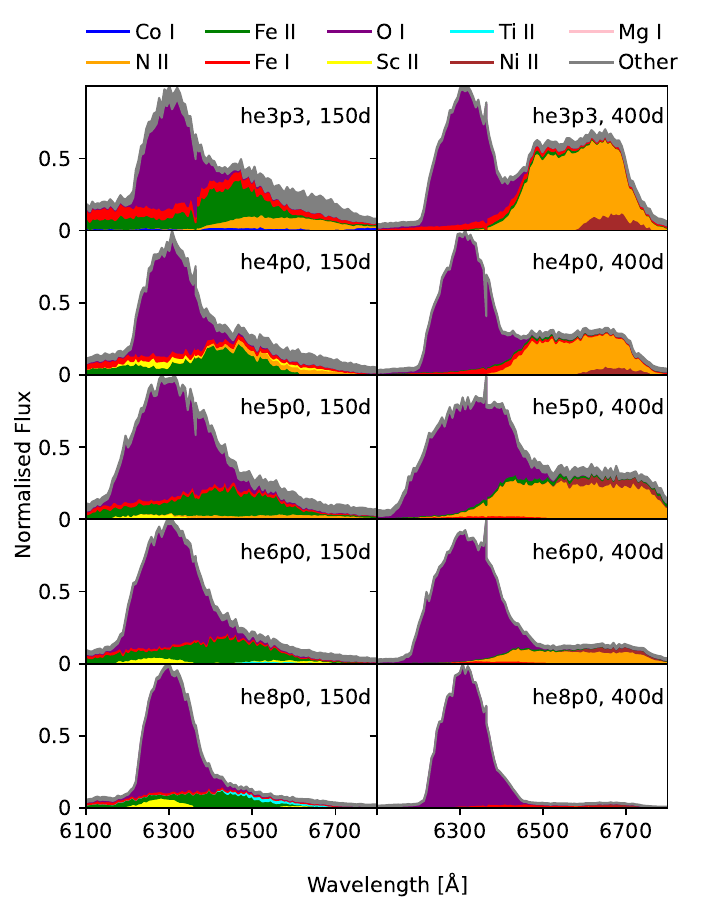}
        \vskip -.2cm
        \caption{Origin of the emission (photon last touch) between 6100 and 6800 Å in our models. The left hand side showcases emission at the earliest epoch (150d), while the right hand side shows emission at the last modelled epoch (400d). The progenitor mass increases downwards. The y-axis runs from 0 to 1, being normalised to the peak flux in the 6100 -- 6800 Å region in each spectrum. Comparing the left and right hand side plots, we see that at earlier epochs the emission in this spectral region originates from a combination of different ions, while at late epochs it is dominated by [\ion{N}{II}] and [\ion{O}{I}] emission, with some minor Fe I and Ni II emission as well. Note how at 150d (the same holds at 200d (not shown)), \NIIdoublet{} emission only constitutes a minor part of the total emission in the 6500 -- 6800 Å region, whereas at 400d is is always dominant.
        }
          \label{fig:emitting_ions}
\end{figure}

\subsection{Spectral modelling}
\label{sec:RTmethod}

The radiative transfer for the input models described in section \ref{sec:modelspectra} was computed using \texttt{SUMO} to arrive at our model spectra. Each model was computed at six different epochs (150, 200, 250, 300, 350 and 400 days post explosion), providing full coverage of that part of the nebular phase that is typically observed in SESNe. Spectra were calculated for a wavelength range between 1000 to 10000 Å, with the standard resolution of $R=1000$ (comparable to the resolution of e.g. X-shooter). We computed the models with modules for charge transfer, molecular formation, and dust switched off. Analysis of the effects of these on the resulting spectra by \citet{Jerkstrand_2015_NII_discovery} found that they had a minor to negligible effect on the strength of the \NIIdoublet{} radiation. We therefore decided that increasing the model complexity by considering these three effects was not beneficial.  

As this work is focused on the emission of \NIIdoublet{}, we mention here also the origin of the atomic data for [\ion{N}{II}] used in \texttt{SUMO}. The energy levels for [\ion{N}{II}] are taken from \citealt{Moore_1993_NIIlevels}, while the transition probabilities (i.e. A-values) for \NIIsinglet{} and \NIIdoublet{} are obtained from \citealt{Tachiev_2001_tp}. The uncertainty on these transitions is between 3 to 10 \%, and is therefore not negligible. However, as this same uncertainty will be propagated in all \texttt{SUMO} models, this means that the relative intensity between models of \NIIdoublet{} is unaffected.

\subsection{Fitting Procedure}
\label{sec:fitting_procedure}

One of the major difficulties of interpreting spectra of SNe are the high ejecta velocities leading to broad emission lines and significant blending. These velocities are of order 5$\times$10$^{3}$\kms{} for the metal-rich cores of SESNe, and above 10$^{4}$\kms{} ($\sim$ 170 Å at 5000 Å) for their He-envelopes. This means that \NIIdoublet{} will partially blend with \OIdoublet{} (the velocity distance between 6364 and 6548 \AA\ is 8400 \kms{}), which typically is one of the three strongest emission lines present in nebular CCSN spectra. Consequently, obtaining an estimate of the  \NIIdoublet{} emission component requires to simultaneously fit for the \OIdoublet{} and \NIIdoublet{} features, as well as making some assumptions on the possible shapes of both of these features. 

The details of this procedure can be found in Appendix \ref{appendix:fitting_procedure}. In Section \ref{A:allowed_shapes}, we first define what shapes we consider when fitting line profiles for the \OIdoublet{} + \NIIdoublet{} complex. Next, Section \ref{A:constraints} lists five constraints that we used to aid our fitting algorithm in deciding how much of the overlapping emission is due to \OIdoublet{}, and how much is due to \NIIdoublet{}. The most important of these is that best fit line widths should show consistency in their time evolution, i.e. that line widths for the same SN should not vary too much between different epochs. Furthermore, the section describes how we define what the 'best' fit for the observed line profiles is. Lastly, Section \ref{a:diagnostic} defines the diagnostic \NIIdiag{} that traces the relative amount of \NIIdoublet{} compared to other emission in the 5000-8000 Å range,  and allows for easy comparison between different SNe. The equation defining this diagnostic is Eq. \ref{eq:definition}, and we refer the reader to Section \ref{a:diagnostic} for more information on it.

To give an example on what the resulting best fits can look like for an individual SN, Figure \ref{fig:Multi_epoch_fit} exemplifies most of the constraints used and the resulting best fits for a sequence of spectra of SN 2011dh. Another important figure guiding many of the constraints used in the fitting algorithm is Figure \ref{fig:emitting_ions}; it shows ion by ion contributions to the emergent spectrum for all He core masses in our model grid, at the first (150d) and final (400d) considered epochs. The figure will be further analysed in section \ref{sec:results}.

\section{Results}
\label{sec:results}

\begin{figure*}
        \centering
        \includegraphics[width=.98\linewidth,angle=0]{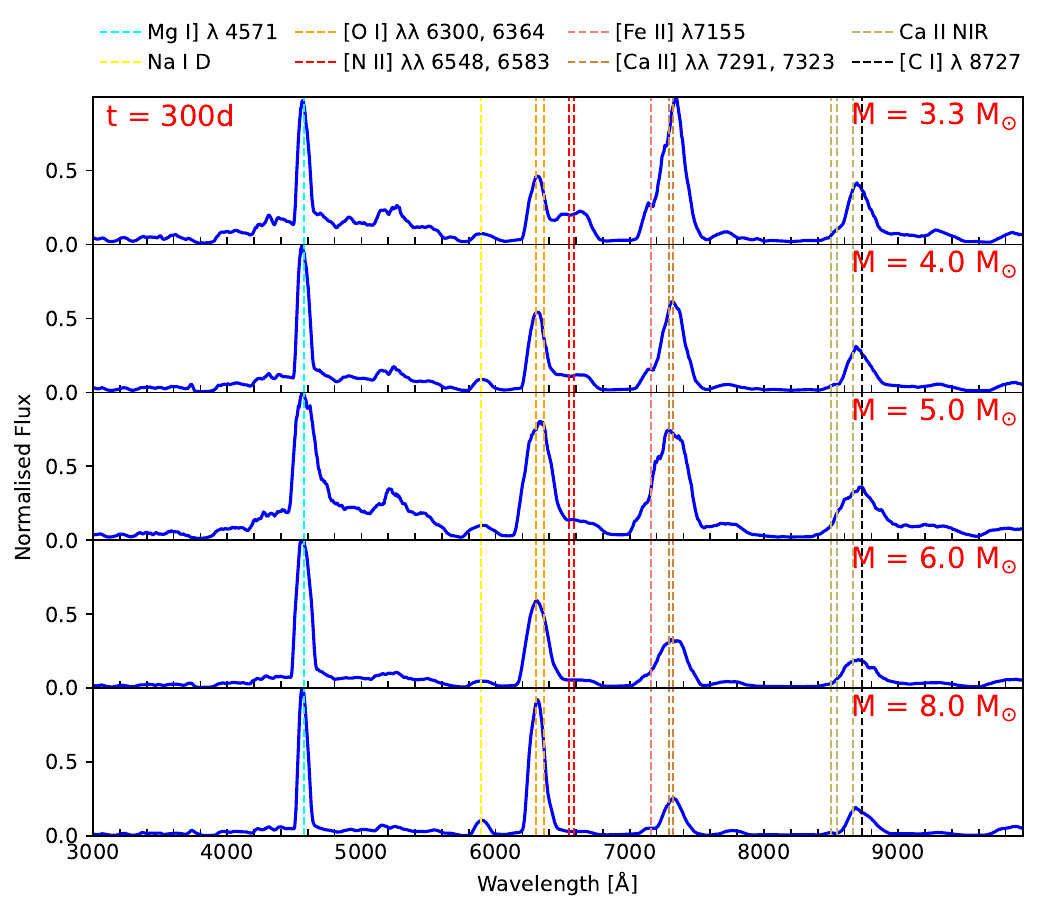}
        \vskip -.2cm
        \caption{The \texttt{SUMO} model spectra at 300 days post-explosion, with each individual panel showcasing the spectrum for a specific \Mi{} progenitor mass. The wavelength range is 3000 Å -- 10000 Å and is indicated on the x-axis, while the y-axis gives the flux normalised for each spectrum's maximum flux value. For visual purposes, the Monte Carlo spectra have been smoothed by taking the moving average per 5 data points. Important/strong emission lines are indicated with vertical dashed lines (doublets and triplets have two and three coloured lines, respectively).  }
          \label{fig:models300d}
\end{figure*}

The spectra for our five different progenitors are shown in Figures \ref{fig:models300d} and \ref{fig:models150d} -- \ref{fig:models400d}, covering 150 -- 400d. Each figure compares the spectra of the five progenitors at a single epoch, resulting in a total of six figures. The strongest time evolution is seen in the blue parts; whereas spectra at early times (e.g. Figures \ref{fig:models150d} and \ref{fig:models200d}) show a significant amount of emission in the Fe-forest around 4000 -- 5000 Å, this gets weaker in the later spectra (e.g. Figures \ref{fig:models350d} and \ref{fig:models400d}). The \ion{Mg}{I}] $\lambda$4571, \OIdoublet{}, [\ion{Ca}{II}] $\lambda\lambda$7291,7323 and \ion{Ca}{II} NIR triplet + [C I] 8727 features are strong throughout. We will now discuss the \NIIdoublet{} emission strength and the resulting \NIIdiag{} diagnostic (defined in Eq. \ref{eq:definition}) in both the model spectra and the observed spectra, as well as the progenitor mass estimates we derive. We end the section with a short analysis of the \NIIsinglet{} line.

\subsection{Comparing \NIIdiag{} trends between models and observations }
\label{sec:NII_per_type_results}

\begin{figure*}
        \centering
        \includegraphics[width=.98\linewidth,angle=0]{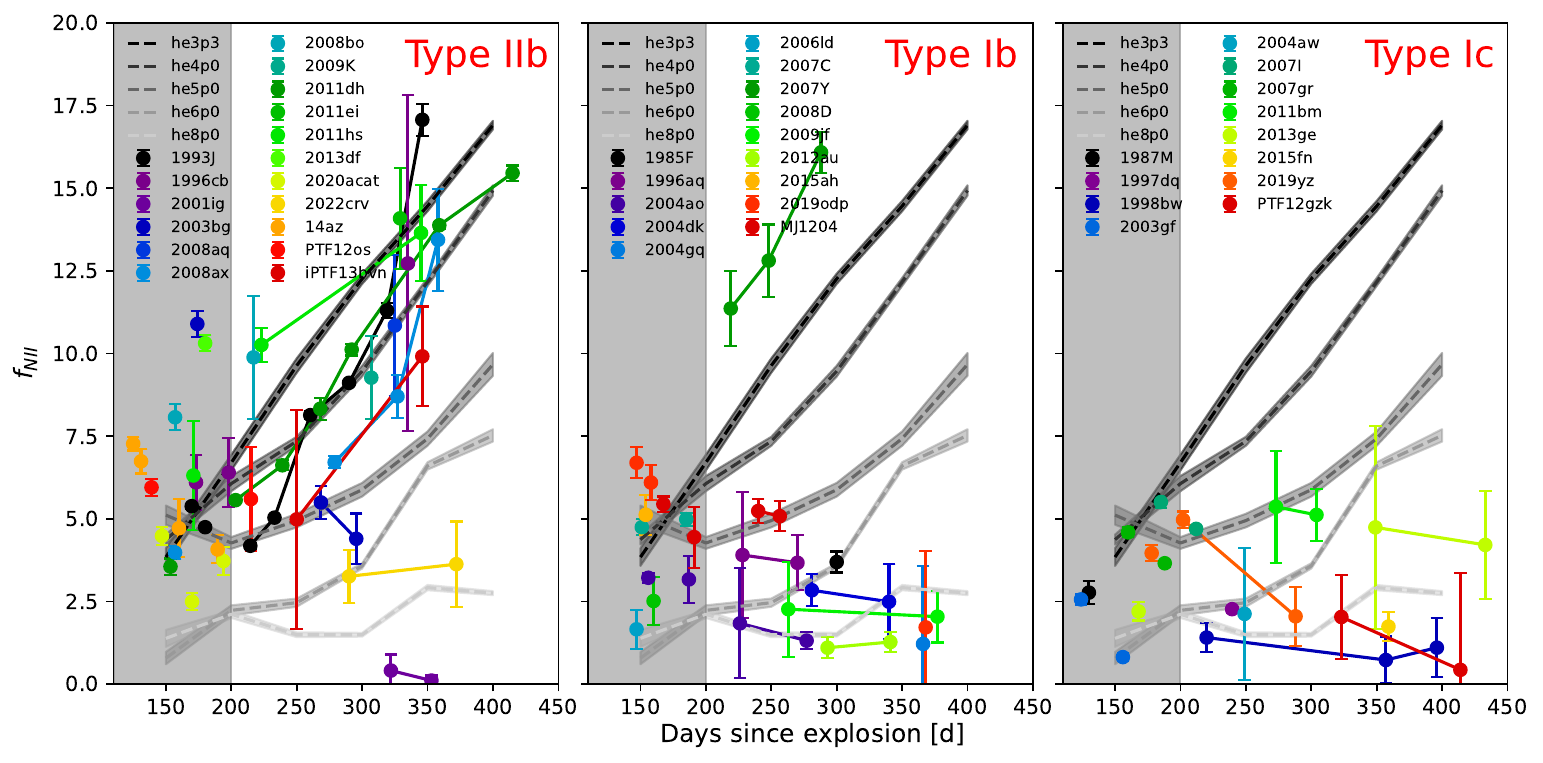}
        \vskip -.2cm
        \caption{A comparison between modeled nitrogen tracks to observations. The figure is divided into three subplots, based on the observed SN classification (IIb on the left, Ib in the middle, Ic on the right). The x-axis records the time since explosion, while the y-axis gives the calculated value for \NIIdiag{}, i.e. the relative strength of \NIIdoublet{} emission compared to the rest of the spectrum (see section \ref{a:diagnostic} for definition). In each subplot, our model tracks are plotted, with the most likely values in dashed lines, and their uncertainty estimates (which are only based on SNR, not model uncertainties, see Section \ref{a:diagnostic}) indicated by the shaded areas. The observed tracks per SN are plotted, again with  1$\sigma$-confidence intervals. To avoid cluttering, observations were binned in 25-day intervals (to clarify: this binning \textit{only} occurs in the plot, not in the analysis). The region before 200 days is shaded gray, as this region is not considered when determining our progenitor mass estimates (see Section \ref{sec:mass_determination}). To amplify this fact, observations in this region are not connected by lines to later observations of the same SN. As discussed in Section \ref{sec:NII_per_type_results}, the only objects showing strong \NIIdoublet{} emission are typed as IIb (except for SN 2007Y), while at the same time there also exist IIb SNe that show as little \NIIdoublet{} emission as Ib and Ic SNe (e.g. SN 2001ig, SN 2003bg and SN 2022crv). We remind the read that, as mentioned and discussed in Section \ref{sec:NII_per_type_results}, a \NIIdiag{} > 0 should not be interpreted as an unambiguous confirmation of \NIIdoublet{} emission in the spectrum.
        }
          \label{fig:NII_per_type}
\end{figure*}

Applying the method described in Section \ref{sec:fitting_procedure}, we create so-called \textit{nitrogen-curves} for each model and for each observed SN. Figure \ref{fig:NII_per_type} plots these curves, along with observational data as divided into the Type IIb, Ib, and Ic classes.

Turning first to the modeling results, there is a clear distinction visible between the low-mass (3.3 and 4.0 \Msun{}) models and the high-mass (8.0 \Msun{} and, to some extent, 6.0 \Msun{}) ones. While all models start off relatively close together in their \NIIdiag{} values at 150 days, the low-mass models show a clear increase with time, reaching as high as \NIIdiag{} = 17 for the he3p3 model. On the other hand, the high-mass models increase little or even stay flat in the case of the he8p0 model. This shows us the value of late-time nebular spectra; at 150 days, the predictive power of our diagnostic is poor as all models have weak \NIIdoublet{} emission and there is in addition no one-to-one relation to the progenitor mass, while at later times the \NIIdoublet{} feature becomes stronger and a clear dependency on progenitor mass is seen. Once more it should be emphasised that a non-zero diagnostic value does not mean that the presence of \NIIdoublet{} emission is confirmed - when the \NIIdiag{} is below around 3, visual inspection of the spectra do not unambiguously confirm the presence of an emission line, and quasi-continuum contributions may dominate the signal. Furthermore, at early ($\lesssim$ 200d) epochs, especially small signals may be dominated by other ions such as \ion{Fe}{II} and \ion{Fe}{I} (see Figure \ref{fig:emitting_ions}).  

Considering now the observational data, we see that there is a clear distinction for the \NIIdiag{} distributions between Types Ib and Ic SNe on the one hand, and Type IIb SNe on the other. In the SN Ic panel, no SNe follow a low-mass track, with most data (considering also error bars) being consistent with non-detection of \NIIdoublet{} - possibly with the exception of SN 2011bm. For the SN Ib panel the picture is similar, except for SN 2007Y which shows the highest \NIIdiag{} values of all SNe at its three observed epochs. All other SNe Ib show tracks more closely resembling higher mass models, with a \NIIdoublet{} signal not clearly distinguishable from zero. Considering the SN IIb panel, there is much more variety in the evolution of the observed nitrogen-curves, with multiple objects following tracks similar to the he3p3 and he4p0 models, as well as some following the he8p0 model.

It is of importance that the observed \NIIdiag{} values appear to span the range of predicted values and, more importantly, that for the cases of clear detections (\NIIdiag{}$\gtrsim 3-4$), the slopes match the model track slopes. These facts strengthen the case for that we are indeed measuring \NIIdoublet{} emission in the observations, and not some other emission (e.g. H$\alpha$) that happens to occur in the 6400 -- 6800 Å region. For H$\alpha$, e.g., the relative strength would be expected to decline with time (see e.g. figure 9 in \citealt{Jerkstrand_2015_NII_discovery}), so the slopes would go the other way. While at first sight there seem to be a few exceptions to the upward sloping curves, closer inspection reveals unusual properties of these data. For e.g. SN 2003bg and SN 2019odp, the early spectra appear quasi-photospheric still, potentially making them unsuitable for our method. These two examples illustrate that the earliest spectra (150 -- 200 days post explosion) are less useful when trying to determine the progenitor mass. Lastly, SN 1993J seems to suddenly experience a rapid slope increase for \NIIdiag{} at around 325d. This particular supernova became powered by CSM interaction around these epochs \citep{Filippenko_1994_1993J, Matheson_2000_1993Jspectra}, which is revealed by a strong, broad top-hat emission around the 6400 -- 6700 Å region, due to H$\alpha$. This fact is also the reason why we do not consider any of the later spectra (high S/N spectra are available well past 450d) that are available for SN 1993J, as the CSM clearly has come to dominate the \NIIdoublet{} emission. At the same time, this example shows that distinguishing between \NIIdoublet{} and H$\alpha$ emission from CSM interaction can indeed be guided using these model nitrogen curves.

\subsection{Mass Determination}
\label{sec:mass_determination}

Comparing the evolution of \NIIdiag{} in our different models with the evolution for observed SNe, we can get an estimate for the progenitor mass using the knowledge of the masses of our model progenitors. With progenitor mass we here mean the mass of the He-core at the moment of explosion, which we will call \Mf{}. This is different from the mass that names our models (as in he3p3 etc.), which is based on the mass at He core ignition, \Mi{}. To translate between \Mi{} and \Mf{}, \citet[][their table 5]{Ertl_2020_models} was used. The reason for us to choose \Mf{} is based on the findings of \citet{Dessart_2023_TimeEvolution}: the emergent SN spectra for two model SNe with the same \Mf{} but different \Mi{} are roughly the same (see their figure 1 and the top panel of their figure 5). This also means that we can not simply extrapolate our \Mf{} back to \Mi{} (or to $M_{\text{ZAMS}}$), as this would require knowledge of the full mass-loss histories of the objects in our sample. 

To obtain the \Mf{} estimates, we make use of a bootstrap algorithm. For a total of 2500 times per object, a new 'perturbed' instance of its nitrogen-curve is created, with the perturbation per epoch a random draw from a Gaussian with $\mu$ = 0 and $\sigma$ equal to the 1$\sigma$ uncertainty in \NIIdiag{} at that epoch. As we saw in Section \ref{sec:NII_per_type_results}, data points before 200 days post-explosion have a poor predictive value. Therefore, the created instances of the nitrogen-curve will not include any \NIIdiag{} values observed earlier than 200 days post-explosion. The resulting perturbed track is then compared to a linearly interpolated grid (with a model track generated for each 0.1 \Msun{} between \Mi{} = 3.3 and 8.0 \Msun{}), in between our five model nitrogen-curves (where the model nitrogen-curves were perturbed in the same manner as mentioned above). Each comparison yields a score $\phi_{M}$ via Equation \ref{eq:score}:

\begin{equation}
    \phi_{M} = \sum_{i} (f_{NII \ model, i} - f_{NII \ obs, i})^{2}
    \label{eq:score}
\end{equation}

Here, $f_{NII \ obs}$ is the value at a given epoch \textit{i} for the observed SN, and $f_{NII \ model}$ is the interpolated value at epoch \textit{i} for the model that it is being compared to. The sum is over all epochs \textit{i} for which the SN has available spectra. The resulting curve in \Mf{} -- $\phi$ space is stored for each of the 2500 iterations, and these are in the end all added up to give a final score curve. The \Mf{} with the lowest score $\phi_{M, opt}$ is taken as the best-fit value, and the uncertainty estimate is taken to be those \Mf{} where  $\phi_{M}$ = 2 $\phi_{M, opt}$.

\vspace{0.2 cm}

With the method described above, we obtained \Mf{} for the SNe in our sample. The resulting masses are compared to mass estimates (by multiple methods, e.g. hydrodynamical light-curve modeling, nebular \OIdoublet{} modelling etc.)  found in the literature in Figure  \ref{fig:mass_estimates}\footnote{The sources of these mass estimates are given in the last column of Table \ref{tab:mass_predictions}. Some objects in our sample were studied independently by multiple groups, so that there exist multiple mass estimates. We give the additional references in Table \ref{tab:multi_estimates}}. For 14 of the SNe in our sample, no such estimates for \Mf{} exist (according to our best knowledge).

\begin{figure}
        \centering
        \includegraphics[width=.98\linewidth,angle=0]{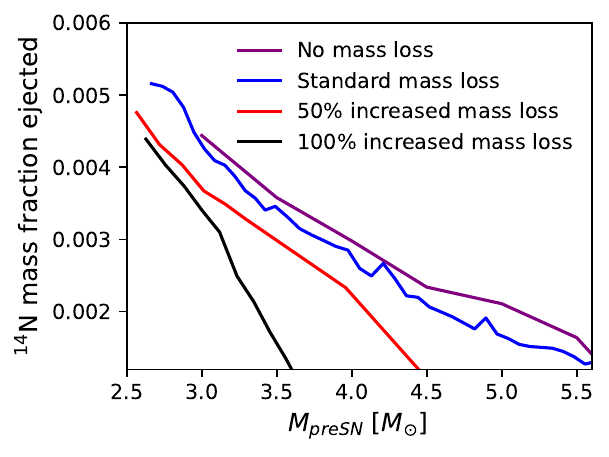}
        \vskip -.2cm
        \caption{ Relation between the mass fraction of nitrogen in the ejecta to \Mf{} of the models in \citet{Woosley_2019_models} for four different mass loss rates. The comparison shows that for decreasing mass fraction of ejected \el{14}{N} (i.e. for decreasing values of \NIIdiag{}), the uncertainty on our \Mf{} estimate increases.
        }
          \label{fig:mass_loss_effect}
\end{figure}

Moving through Figure \ref{fig:mass_estimates} from bottom left to top right, the first thing to note is that both the scatter in and the uncertainties on the datapoints increase. This has two relatively simple explanations. First, Figure \ref{fig:NII_per_type} shows that the differences in \NIIdoublet{} emission relative to the increase in \Mf{} of the models decrease when moving to higher mass models, automatically leading to higher uncertainty in mass estimation. Secondly, when SNe approach our highest mass model of 8.0 \Msun{}, our resulting mass estimates can only be given as lower limits, as any mass above 8.0 \Msun{} is not modeled by our methodology, and is simply returned as 8.0 \Msun{}. This is not simply a limit of our grid-selection, but rather of our method itself, as the amount of \NIIdoublet{} emission becomes too small at even higher masses to give an accurate mass estimate. As the literature methods do not have this limiting mass, a horizontal spread arises in the figure. 

Despite the above observation, most of our SNe still touch the diagonal\footnote{Excluding objects for which we only can give lower/upper-limits, this still remains true.}, meaning our estimates are compatible with previous literature estimates using other methods. Out of the 18 plotted SNe, 5 (SN 1993J, SN 2004aw, SN 2011bm, SN 2013ge, SN 1997dq) do not touch the diagonal, all of them being Type Ic SNe except for SN 1993J (which is Type IIb). For SNe 2013ge, 1997dq and 1993J, the difference in mass estimates can be said to be relatively minor, disappearing if stronger mass loss would have occurred than in our models (e.g. \Mf{} = 5 \Msun{} for the standard mass-loss rate would equal \Mf{} = 4.2 \Msun{} for the 50\% increased mass-loss case for SNe 2013ge and 1997dq). However, for SN 2004aw and SN 2011bm, our estimates are definitely inconsistent with those from the literature. 

For SN 2004aw the argument can be made that our way of translating nitrogen-curve to mass estimate distorts the view given by the nitrogen track, as seen in Figure \ref{fig:NII_per_type}: the uncertainty on \NIIdiag{} for SN 2004aw is so considerable that the lower limit is almost at \NIIdiag{} = 0, but due to our mass cut off at \Mi{} = 8.0 \Msun{}, the uncertainty estimate also stops there. If we could have measured up to higher masses, the uncertainty estimate may indeed have made it so that our estimate and the literature estimate would still be consistent. 

SN 2011bm was found by \citet{2011bm} 
to be one of the slowest evolving SESNe, and one could argue that the spectra used in this work show signs of not being fully nebular. As exemplified in Figure \ref{fig:emitting_ions}, this may mean that there still are a lot of other ions emitting in the spectral region surrounding \NIIdoublet{}. This makes the 'low' mass estimate very uncertain.

All in all, the general compatibility of our results with previous estimates (made by multiple different methods e.g. light curve modelling, [\ion{Ca}{II}]/[\ion{O}{I}] ratio) strengthens the confidence in the method, as well as in the mass-loss rate used.

\begin{figure}
        \centering
        \includegraphics[width=.98\linewidth,angle=0]{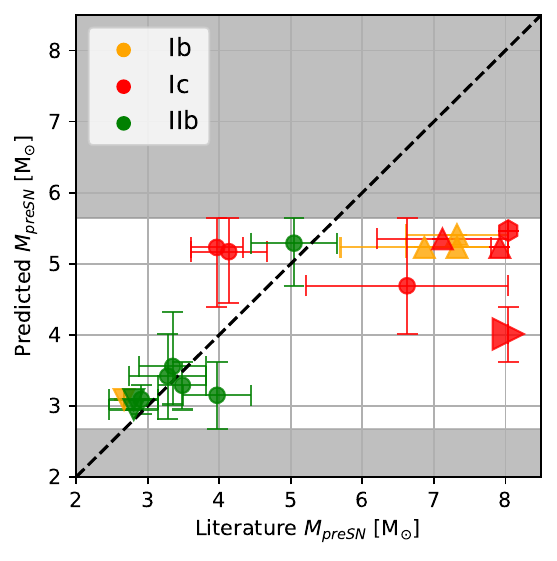}
        \vskip -.2cm
        \caption{ Comparison between \Mf{} masses predicted from our \NIIdoublet{} analysis (x-axis) and literature estimates using other methods (y-axis, numeric values given in Table \ref{tab:mass_predictions}). When a literature estimate for \Mf{} was missing, but an estimate for \Mi{} or $M_{\text{ej}}$  was present, this value was converted to an \Mf{} value using the mappings of \citet[][their table 5]{Ertl_2020_models}. Triangles indicate upper- and lower limits on a mass estimate, which for our estimates arise when the most likely \Mf{} value was at one of the bounds of our model-grid (which in \Mf{} space are 2.67 \Msun{} and 5.64 \Msun{}). The region outside of these limits has been marked grey. If an errorbar touches either of these boundaries, it should be interpreted as being larger in that direction in reality. For the literature values, a cut off of accepted values was made at \Mf{} = 8 \Msun{} for visual purposes. Any errorbars/values ending there should therefore be interpreted to continue to higher values. 
        Hexagons indicate that our estimate is a lower limit, and that simultaneously the literature value is a lower limit. Objects are colour-coded by SN Type. A dashed 1:1 diagonal line is plotted as a visual aid.   
        }
          \label{fig:mass_estimates}
\end{figure}

\begin{table}
\centering
\begin{tabular}{lllll}
\hline
Name & Type &  $M_{pred}$ & $M_{lit}$ & Ref \\
 & & [\Msun{}] & [\Msun{}] & \\ 
\hline \hline  \\[-.25cm]
SN 2019odp & Ib & $ \geq $ 5.2 & 6.9$^{+1.2}_{-1.2}$ & \citet{2019odp} \\  
SN 2012au & Ib & $ \geq $ 5.4 & 7.3$^{+0.7}_{-0.7}$ & \citet{Takaki_2013_2012au} \\  
SN 2009jf & Ib & $ \geq $ 5.2 & 7.3$^{+0.7}_{-0.7}$ & \citet{Valenti_2011_risetime_2009jf} \\  
SN 2007Y & Ib & $ \leq $ 3.1 & 2.7$^{+0.0}_{-0.0}$ & \citet{2007Y} \\    
SN 2015fn & Ic & $ \geq $ 5.5 & $\geq$ 8.0 & \citet{2015fn_a} \\  
SN 2013ge & Ic & 5.2$^{+0.5}_{-0.7}$ & 4.1$^{+0.5}_{-0.5}$ & \citet{2013ge} \\  
SNPTF12gzk & Ic & $ \geq $ 5.2 & 7.9$^{+0.1}_{-0.1}$ & \citet{PTF12gzk_a} \\  
SN 2011bm & Ic & 4.0$^{+0.4}_{-0.4}$ & $\geq$ 8.0 & \citet{2011bm} \\  
SN 2004aw & Ic & 4.7$^{+1.0}_{-0.7}$ & 6.6$^{+1.4}_{-1.4}$ & \citet{2004aw} \\  
SN 1998bw & Ic & $ \geq $ 5.4 & 7.1$^{+0.9}_{-0.9}$ & \citet{Woosley_1999_1998bw} \\
SN 1997dq & Ic & 5.2$^{+0.4}_{-0.8}$ & 4.0$^{+0.4}_{-0.4}$ & \citet{1997dq_a} \\  
SN 2022crv & IIb & 5.3$^{+0.3}_{-0.6}$ & 5.0$^{+0.6}_{-0.6}$ & \citet{2022crv} \\  
iPTF13bvn & IIb & 3.6$^{+0.8}_{-0.5}$ & 3.3$^{+0.5}_{-0.5}$ & \citet{PTF12os_iPTF13bvn_b} \\  
SNPTF12os & IIb & 3.4$^{+0.6}_{-0.6}$ & 3.3$^{+0.5}_{-0.5}$ & \citet{PTF12os_iPTF13bvn_b} \\  
SN 2011hs & IIb & $ \leq $ 3.1 & 2.8$^{+0.3}_{-0.3}$ & \citet{2011hs} \\  
SN 2011ei & IIb & $ \leq $ 3.0 & 2.8$^{+0.3}_{-0.3}$ & \citet{2011ei} \\  
SN 2011dh & IIb & 3.1$^{+0.2}_{-0.2}$ & 3.1$^{+0.7}_{-0.4}$ & \citet{2011dh_C} \\  
SN 2008ax & IIb & 3.3$^{+0.3}_{-0.3}$ & 3.5$^{+0.3}_{-0.3}$ & \citet{Folatelli_2015_2008ax} \\  
SN 1993J & IIb & 3.1$^{+0.5}_{-0.5}$ & 4.0$^{+0.5}_{-0.5}$ & \citet{Woosley_1994_1993J} \\  
SN 1985F & Ib & 4.5$^{+0.2}_{-0.2}$ & -- & -- \\  
SN 1996aq & Ib & 4.1$^{+0.9}_{-0.5}$ & --& -- \\  
SN 2004gq & Ib & $ \geq $ 5.2 & -- & -- \\  
SN 2004dk & Ib & 5.5$^{+0.1}_{-0.6}$ & -- & -- \\  
SN 2004ao & Ib & $ \geq $ 4.6 & -- & -- \\  
SN J1204 & Ib & 3.8$^{+0.3}_{-0.3}$ & -- & -- \\  
SN 2019yz & Ic & 5.2$^{+0.5}_{-1.5}$ & -- & -- \\  
SN 2007I & Ic & 3.8$^{+0.1}_{-0.1}$ & -- & -- \\  
SN 2009K & IIb & 3.2$^{+0.3}_{-0.2}$ & -- & -- \\  
SN 2008bo & IIb & $ \leq $ 3.3 & -- & -- \\ 
SN 2008aq & IIb & 3.1$^{+0.4}_{-0.5}$ & -- & -- \\  
SN 2003bg & IIb & 4.0$^{+0.4}_{-0.4}$ & -- & -- \\  
SN 2001ig & IIb & $ \geq $ 5.3 & -- & -- \\  
SN 1996cb & IIb & 3.0$^{+0.8}_{-0.3}$ & -- & -- \\

\hline \hline

\end{tabular}
\caption{Comparison between \Mf{} as estimated from our models using the \NIIdoublet{} nebular diagnostic ($M_{pred}$) with literature estimates using other methods ($M_{lit}$). Objects are sorted by type, and objects without literature estimates are placed at the bottom. The confidence intervals for $M_{pred}$ indicate the region where the fitting score was twice as high as the best-fit score (see Section \ref{sec:mass_determination}), while the literature values are typically the average of the lower and upper boundaries given in the respective papers. The final column indicates the sources for the literature estimates. Some objects in our sample were studied independently by multiple groups, so that there exist multiple mass estimates besides the one adopted here. We give the additional references in Table \ref{tab:multi_estimates}}
\label{tab:mass_predictions}

\end{table}

\subsection{Prevalence of \NIIsinglet{}}
\label{sec:NIIsinglet}

\begin{figure*}
    \centering
    \includegraphics[width = .85\linewidth]{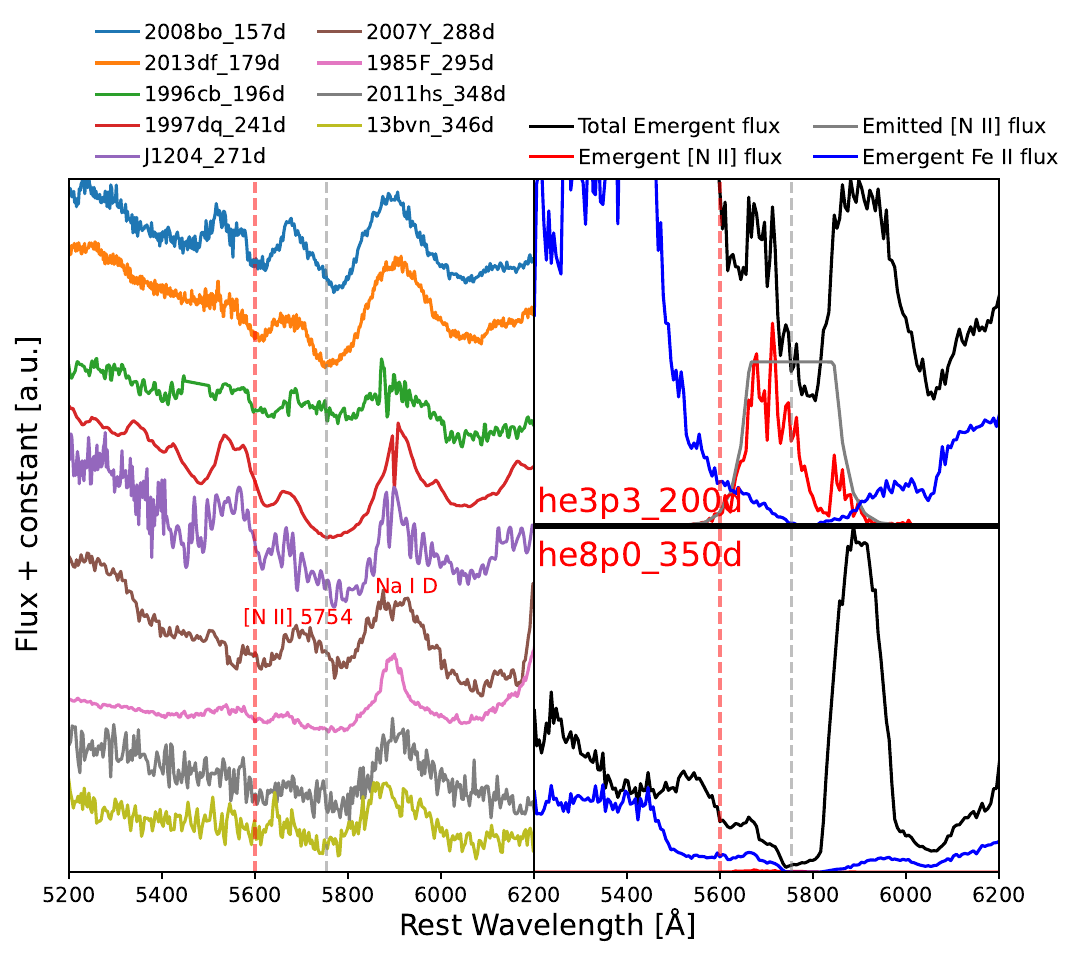}
    \vskip -.2cm
    \caption{\textit{left}: collection of spectra from all SNe with a potential \NIIsinglet{} identification (9 out of the 43 SNe studied here). For each SN, the epoch with the clearest emission peak around $\sim$ 5670 Å was chosen. The grey dashed line is at 5754 Å, while the red dashed line is at 5600 Å, the typical wavelength for the blueshifted edge of observed \NIIsinglet{} emission. As discussed in Section \ref{sec:NIIsinglet}, the expected centre of the \NIIsinglet{} feature is blueshifted from its central wavelength due to scattering by Na I-D lines.  \textit{top- and bottom right}: similar to the \textit{left} subplot, but now for the he3p3 model at 400 days and the he8p0 model at 350 days (different epochs to present the best possible examples of contamination). The flux in both subplots has been normalised to the peak of the Na I-D complex. The figure shows that both \ion{Fe}{II} and [\ion{N}{II}] emission can, combined with the partial absorption into the Na I D lines, contribute to forming a bump between 5600 and 5700 \AA. }
    \label{fig:NII_5754}
\end{figure*}

Of particular interest for our work is the presence or absence of the \NIIsinglet{} feature. In \citealt{Jerkstrand_2015_NII_discovery}, the authors identify the blue side of this line in SNe 1993J, 2008ax and 2011dh, strengthening the identification of \NIIdoublet{} in these SNe. Furthermore, they are able to reproduce it in their lowest mass model spectra. The red side of the line is scattered away by Na I-D lines, so that in the observed spectrum the line peak becomes blueshifted to $\sim$ 5670 Å. In the r.h.s subplot of Figure \ref{fig:NII_5754} we show this region for our model spectra of the he3p3 model (at 200 days) and he8p0 model (at 350 days), making it analogous to figure 14 in \citealt{Jerkstrand_2015_NII_discovery}. 

In our lightest model, the feature is primarily due to \NIIsinglet{} emission, as was found in \citealt{Jerkstrand_2015_NII_discovery}. At the same time, we find that the contribution of \NIIsinglet{} decreases with time, so that it only contributes $\sim$10 \% at 400 days (not shown), compared to $\sim$75 \% for the same epoch in \citealt{Jerkstrand_2015_NII_discovery}. This difference between the models seems to arise due to the different ejecta velocity structures used; our he3p3 model has on average higher velocities in the He/N layer than the 12C model (4700 -- 17000 \kms{} compared to 3500 -- 11000 \kms{}). This leads to decreased density and thereby less efficient trapping of gamma-rays at equal epochs, resulting in less heating in our models. As the strength of \NIIsinglet{} is highly sensitive to temperature (significantly more so than e.g. \NIIdoublet{}, see the discussion in section 5.4.2 of \citealt{Jerkstrand_2015_NII_discovery}), we should expect to see weaker \NIIsinglet{} emission in our spectra in these later epochs.

To allow for comparison with observations, the l.h.s of Figure \ref{fig:NII_5754} shows all SNe in our sample that show a reasonably clear feature around this wavelength. Reassuringly, this subgroup lines up well with our results in Figure \ref{fig:NII_per_type}, in that most SNe that show high \NIIdiag{} values, also show a feature around 5600 -- 5700 Å. Also SNe 1985F, 1997dq and J1204 show signs of an emission feature at this wavelength, while they show no strong \NIIdoublet{} emission. This does not necessarily mean that \NIIsinglet{} is emitting here, however, as in some models (e.g the he8p00 model, see r.h.s. of Figure \ref{fig:NII_5754}) also \ion{Fe}{II} can create significant emission\footnote{We note that this line was not significantly emitting in the models of \citealt{Jerkstrand_2015_NII_discovery}, see their figure A.4}. This showcases the difficulty of unambigously using \NIIsinglet{} across a sample of SNe. 

In summary, identifying an emission feature around 5600 -- 5700 Å may strengthen identification of \NIIdoublet{} as the source of any emission found around 6400 -- 6700 Å (especially if the 5600 -- 5700 Å feature is identified before $\sim$ 300 days), but can not by itself be used as a secure identification of [\ion{N}{II}]-emission due to its relative weakness and the possibility of being confused with \ion{Fe}{II}. 

\section{Discussion}
\label{sec:discussion}

One of the main results of this work is the first systematic analysis of the temporal evolution of the \NIIdoublet{} line strength for a sample of SESNe, as presented in Figure \ref{fig:NII_per_type}. In this section, we dicuss the consequences of the results on current theories concerning the different evolutionary paths for Type Ib, Type Ic and Type IIb SNe. Another main result is the development of a new diagnostic using the \NIIdoublet{} line  for determining the progenitor mass of observed SNe. Its usefulness will be discussed in the remainder of the section. 

\subsection{Type Ic SNe show no signs of strong \NIIdoublet{} emission}
\label{sec:nohe_Ic}

The first result we discuss is that there is no sign in Figure \ref{fig:NII_per_type} of strong \NIIdoublet{} emission among Type Ic SNe. Given the long-standing debate on whether Type Ic SNe may contain "hidden" helium zones, it is of interest to search for signs of such helium layers through other emission than the hard-to-excite He lines. We clearly see here that low-mass stellar cores with a retained He/N zone can be ruled out for Type Ic SNe. Most Ic SNe show nitrogen-curves residing around those of the he6p0 and he8p0 models, which in turn have so weak \NIIdoublet{} emission that it for practical purposes cannot be distinguished from zero. We note that the he6p0 and he8p0 models evolved in \citealt{Dessart_2023_TimeEvolution} show similar patterns in the fractional luminosities for \NIIdoublet{}, residing at such low values that they are hard to distinguish from zero. The values for SN 2011bm and SN 2007I lie up towards the he5p0 model (for which the \NIIdoublet{} line is distinct) , but both have caveats that shed doubt on any nitrogen identification. As mentioned in Section \ref{sec:mass_determination}, the spectra of SN 2011bm may not have been nebular yet. A similar argument could be made for SN 2007I when inspecting the spectra, also because the epochs of the spectra are quite early, at roughly 200 days post explosion. It is only starting around this epoch that the \NIIdiag{} diagnostic starts to become really useful. 

Let us attempt to answer the question what this seeming non-detection of \NIIdoublet{} emission in Type Ic SNe means. In the standard paradigm, Type Ic SNe are explosions of stars that have lost both their hydrogen- and helium-rich envelopes. However, in recent years there has been an ongoing debate (\citealt{Sauer_2006_HiddenHelium, Dessart_2012_Heliumhidden, Hachinger_2012_HiddenHe, Dessart_2020_Hiddenhe, Williamson_2021_HiddenHe}) on whether or not Type Ic SNe could have "hidden" helium-envelopes which do not show any clear helium lines due to the difficulty of exciting them. For nitrogen, there is no such difficulty as it's a simple low-lying cooling (thermal) line, thus by using nitrogen as probe instead of helium we get around this issue. As \NIIdoublet{} emission in our models overwhelmingly originates from the He/N-layer, which is the outer part of the helium-rich envelope (the inner part being the He/C-layer, see Figures \ref{fig:Woosleymodel}-\ref{fig:Mixing_40} ), finding no significant \NIIdoublet{} emission in Type Ic SNe would be in agreement with the scenario that there are no helium-rich envelopes in Type Ic SNe. However, because the N emission from the He/N zone (mass $\gtrsim 0.4\ M_\odot$) in the more massive He cores is predicted to be indistinct, we cannot rule out such progenitors.

The question that remains is whether our results can be used to \textit{confirm} the presence of a He/N-layer in any of the Type Ic SNe in our sample. The answer to this question should once more be negative: there are no tracks in our sample that are inconsistent with originating from the same progenitor as our he8p0 model. From detailed inspection of Figure \ref{fig:emitting_ions}, it is found that only about half of the already small amount of emission in the $\sim$ 6400 -- 6800 Å region is truly \NIIdoublet{} emission. If we would divide all our SN Ic tracks by two, then there are no points left that can be definitely said to have both a) fully nebular spectra and b) \NIIdoublet{} emission significantly different from zero. 

To conclude this subsection, the main conclusion we can draw is that from nitrogen line analysis, low-mass He-core progenitors appear to be ruled out for Type Ic SNe. Thus, if Ic SNe are He core explosions in which the He gets "hidden", those He cores must be quite massive. Even for this statement there are some things to consider. Additional mass loss compared to the one used in the stellar evolution modelling here might be able to peel off the outer lying He/N layer while leaving the inner He/C one. Such models would give no \NIIdoublet{} emission either, and could be consistent with the Type Ic SN data. One might possibly try to exclude this scenario using [C I] 8727 from the He/C zone, analogously to using \NIIdoublet{} for the He/N zone. A complication for that would be that C emission comes also from the deeper lying O/C zone, containing more C than the He/C zone. The [C I] 8727 line can also blend quite severely with the Ca II NIR triplet in SESNe. 

\subsection{Type IIbs are the only low mass He-cores?}

\subsubsection{Type Ib vs IIb classification} 
When interpreting the results from the Type IIb and Type Ib panels in Figure \ref{fig:NII_per_type}, some caution is necessary. As Type IIb SNe will eventually (from $\sim$ 50 days post explosion) be indistinguishable from Type Ib spectra, reassuring ourselves that each individual classification for an object in either of these classes is correct, is important. This is especially true as supernova classification often still is a human job, based on visual identification of certain lines in the earliest available spectra of the SN.

A comprehensive effort to make an objective distinction between SNe Ib and IIb was made by \citet{Liu_2016_IbIIb}. The authors defined a pseudo-Equivalent Width (pEW) diagnostic, which they then determined for the H$\alpha$ line for a sample of $\sim$ 700 photospheric spectra of Type Ib, Ic and IIb SNe, for in total 52 objects. Comparing the pEW evolution of all SNe Ib and IIb in their sample, they find a clear boundary in pEW$_{H\alpha}$-time space separating the two classes. Similar comparisons between the pEW evolution of the major \ion{He}{I} lines (i.e. at 5876, 6678 and 7065 Å) show that these do not give any clear-cut boundary between Type Ib and Type IIb SNe (see also \citet{Fremling_2018_pEW}), although the velocity-distributions for these He I lines do seem to show slightly differing trends between the two classes.

While initially we used the classifications as provided on WISEREP and in the Open Supernova Catalog, we opted instead to use the metric from \citet{Liu_2016_IbIIb} when creating the final version of Figure \ref{fig:NII_per_type}. In the initial figure, two Type Ib SNe were found to show nitrogen-curves following the lower mass models, namely SN 2007Y and iPTF13bvn. Both of these objects had some doubt shed on their classifications though, with \citet{2007Y} (for 2007Y) and \citet{PTF12os_iPTF13bvn_b} (for iPTF13bvn) both suggesting that some H$\alpha$ absorption may be visible in early photospheric spectra.

Following the pEW$_{H\alpha}$ diagnostic for iPTF13bvn, \citet{Liu_2016_IbIIb} find that it is right at the boundary between Ib and IIb SNe in the pEW$_{H\alpha}$-time plane, showcasing a track that is very similar to Type IIb SN 2008ax. This finding is even more interesting when considering that these two objects also have very similar nitrogen-curves (see Figure \ref{fig:NII_per_type}). These results, combined with the fact that the NIR spectra of both SNe are also very similar (see sections 7.1 and 7.2 in \citet{PTF12os_iPTF13bvn_b}), make a strong case to grant both objects the same, Type IIb classification\footnote{Note that this would not discredit the pEW$_{H\alpha}$ diagnostic; it would simply move the cut-off boundary 1 object 'down' in pEW$_{H\alpha}$-time space.}. 

When considering SN 2007Y, it is firmly classified as a Type Ib SN in the work by \citet{Liu_2016_IbIIb}. Interestingly enough, when looking at the He I lines, SN 2007Y shows low absorption velocities compared to other Ib's, having a track in a region mostly populated by IIb's. Moreover, it is a clear outlier when considering the pEW track of He I $\lambda$7065, but not when considering the other He I lines. While these findings confirm that SN 2007Y has some unusual properties, we consider they do not form a strong enough argument for reclassifying SN 2007Y as a Type IIb.

All other Type Ib and Type IIb SNe considered in our work are consistent with the work of \citet{Liu_2016_IbIIb}. The classifications as shown in Figure \ref{fig:NII_per_type} will in the remainder of our analysis assumed to be 'true' classifications, with the urgent note that iPTF13bvn could be classified as either Type Ib and Type IIb.

\subsubsection{Analysis} From the Type IIb and Type Ib panels in Figure \ref{fig:NII_per_type}, a clear pattern emerges: almost all SNe with nitrogen-curves similar to lower-mass models are of Type IIb, with SN 2007Y being the only exception. Additionally, only 3 out of the 17 considered Type IIbs show tracks that follow the higher-mass models: SN 2001ig, SN 2003bg, and SN 2022crv. For the Type Ib and Ic SNe, on the other hand, all objects except SN 2007Y follow high-mass tracks.

\paragraph*{Comparison to literature results.} 

\citet{Dessart_2021_Hestarexpl} and \citet{Dessart_2023_TimeEvolution} studied the same stellar evolution models from \citet{Woosley_2019_models} and \citet{Ertl_2020_models}. However, as they used the radiative transfer code \texttt{CMFGEN} instead of \texttt{SUMO}, as well as a different mixing prescription than used in this work, it is of great interest to compare their results to our findings. \citet{Dessart_2021_Hestarexpl} compare their entire model set to spectra of six SESNe (IIb 1993J and 2011dh, Ib 2007C and Ic 2007gr, 2004aw and 2013ge), to find which are the likely progenitors of observed SNe. In this comparison, they find that Type Ic SNe are all quite well reproduced by the he8p0 model, in agreement with our findings from Figures \ref{fig:NII_per_type} and \ref{fig:mass_estimates} that Type Ic SNe mostly have high-mass He-core progenitors. For SNe 2011dh and 1993J, they find the he5p0 and he7p0 to be best fits respectively, whereas our work prefers he4p0 for both. This difference could be due to our works making different comparisons: whereas we only compare to the values of \NIIdiag{}, \citet{Dessart_2021_Hestarexpl} compare the entire spectra to determine their preferred models. In this particular case however, the difference seems to arise due to actual differences between our spectral modelling results: we obtain consistently higher \OIdoublet{} to [\ion{Ca}{II}] $\lambda\lambda$7291,7323 ratios at the same heXpX model than in \citet{Dessart_2021_Hestarexpl} (compare Figure \ref{fig:models200d} with their figure 4). As these are the two strongest lines in SNe 2011dh and 1993J, \citet{Dessart_2021_Hestarexpl} will prefer higher mass cores than found in our work. The above example shows that calibrating progenitor models to observed SNe is no trivial task and has a non negligible dependence on the employed spectral synthesis code.
 
With the same model set, \citet{Woosley_2021_LC} instead modeled the light curves for these progenitors. Comparing these to a sample of observed Type IIb, Ib and Ic SNe, they find based on matching to rise times that models with \Mf{} $\lesssim$ 3.5 \Msun{} can reproduce most objects (see their figure 4). When matching to decline times instead, the fitting model range is rather 3.0 \Msun $<$ \Mf{} $<$ 5.0 \Msun{}. In either case, models with \Mf{} $>$ 5 \Msun{} seem to give to broad light curves to match the observed Type Ibc population. However, for many of the Type Ib and Ic SNe studied in this work, the \NIIdiag{} curves come closest to the he8p0 model (see Figure \ref{fig:NII_per_type}), which has \Mf{} = 5.6 \Msun{}. It is important to note here that the masses we infer for our sample make the assumption that our employed grid of models is representative for the entire Type Ibc population, i.e. that the only cores without significant \NIIdoublet{} emission are high-mass ones. If this is not true in nature (e.g. there exist low-mass cores that lost their entire He/N zones to winds/stripping), finding no significant \NIIdiag{} emission can no longer directly be linked to a high-mass core. The mismatch with the findings from \citet{Woosley_2021_LC} may be an indication that this is indeed the case. This tension could be resolved with for example a 50\% higher mass-loss rate (see also Section \ref{sec:uncertainties}), which would bring \Mf{} for he8p0 down to 4.4 \Msun{} (see Figure \ref{fig:mass_loss_effect}), within the suggested range from \citet{Woosley_2021_LC}. 

Two fully observational studies of the properties of the different SESNe classes do not find Type IIb SNe to be clearly different or have lower mass cores than Type Ibc SNe. Using $M_{\text{ej}}$ estimates from light curve fitting, \citet{Prentice_2019_Bigsample_2015ah} instead found Type Ib SNe to have a slightly lower average progenitor mass than Type IIb SNe. Additionally, \citet{Fang_2019_NIIusage} found luminosity ratios of \NIIdoublet{}/\OIdoublet{} for Type Ib and IIb SNe to be distinctly different from those for Type Ic and Ic-BL SNe, rather than Type IIb SNe being different. In this work however, the authors considered mostly spectra at 200d post peak, which as we have shown in this work may still contain significant amounts of emission from other elements than nitrogen in the 6400 -- 6700 Å region (e.g. Figure \ref{fig:emitting_ions}), so that a difference between Type IIb and Ib SNe may easily be missed. 

Coming back to our own results now and under the assumption that they are true, the dichotomy in progenitor mass between Type IIb SNe on the one hand, and Type Ib and Type Ic SNe on the other warrants the question whether these two groups could be formed via physically different formation channels.

\paragraph*{A possible correlation with progenitor radii.} An important lead to explain this dichotomy is found when the different progenitor radii are considered. When one does this for the Type IIb SNe in our sample, one finds that objects with strong \NIIdoublet{} emission are objects with large radii (R $\gtrsim$ 200 R$_{\odot}$), while those with little emission are typically compact (R $\sim$ 1 R$_{\odot}$). These two groups are bridged by medium-sized progenitors (50 R$_{\odot}$ $\gtrsim$ R $\gtrsim$ 10  R$_{\odot}$), which show intermediate \NIIdoublet{} curves. The derived radii for our Type IIb sample (when present in literature) are displayed in Table \ref{tab:radii}. We note that such a seemingly continuous trend is not present in the analysis from \citet{Liu_2016_IbIIb}, where the pEW$_{H\alpha}$ of the intermediate-mass SN 2008ax is for example a lot more similar to that of Type Ibs than the heavier cores of SN 2022crv \footnote{pEW$_{H\alpha}$ value from \citealt{2022crv}, not from \citealt{Liu_2016_IbIIb}} and SN 2003bg. 

A subset of models from the same model suite that we use from \citet{Woosley_2019_models} also have significantly larger radii for very low-mass progenitors: below \Mi{} $= 3.3$ \Msun{}, the stars experience a strong increase in radius after He core burning. However, as discussed in \citet{Ertl_2020_models}, the SNe resulting from these progenitors are dissimilar from typical observed SESNe and are more similar to the class of "fast blue transients". Furthermore, the fact that even for these (single star) progenitors the modeled radii only expand up to 100 $R_{\odot}$ (and not $\sim$ 500 $R_{\odot}$) suggests that the mechanism leading to the expansion is more likely due to a binary origin.

Another explanation one could think of to couple radius with He-core mass, is different binary separations. \citealt{Yoon_2017_BinaryRadii} find that SN-progenitors in binaries that start off with a larger initial separation have larger radii (for equal progenitor mass). Binaries at solar metallicity with an inital orbital period of $\sim$ 1000 days are found to exclusively result in red supergiant progenitors (see their figure 2). On the other hand, binaries with separations less than $\sim$ 300 days and the same primary mass only turn into (compact) Ibs. These results could explain the trend between \Mf{} and $R_{\text{preSN}}$ that we see, if the lower-mass end of stars are not formed at close ($\lesssim$ 300 days) separations. This would however not explain the existence of the heavier Type IIb He-cores (i.e. SN 2001ig, 2022crv).

Finally, an important question that should be asked when suggesting different formation channels for Type IIbs and Type Ibs, is how SN 2007Y fits into the picture, as both our work and the findings from \citet{Woosley_2021_LC} indicate this SN to clearly originate from a very low-mass (\Mf{} $\lesssim$ 2.7 \Msun{}) progenitor. Answering this question is outside the scope of this work, but the easiest answer to this question may be that SN 2007Y may simply have been a one-off event: its extremely low explosion energy ($\sim$ a few 10$^{50}$ erg) and brightness is unlike what has been observed for any other SN \citep{2007Y, Woosley_2021_LC}.

\begin{table}
\begin{tabular}{llll}
\hline
Name & $R_{lit}$ [$R_{\odot}$] & $M_{pred}$ [\Msun{}] & Ref \\ \hline \hline
SN 1993J & $\sim$ 600 & 3.1$^{+0.5}_{-0.5}$ & \citealt{Maund_2004_1993J}  \\
SN 2011hs & 500 -- 600 & $\leq$ 3.1 & \citealt{2011hs} \\
SN 2013df & 500 -- 600 & $<$ 3.0 & \citealt{vanDyk_2014_2013df} \\
SN 2011dh & $\sim$200 & 3.1$^{+0.2}_{-0.2}$ & \citealt{Bersten_2012_2011dh} \\
SN 2008ax & 30 -- 50 & 3.3$^{+0.3}_{-0.3}$ & \citealt{Folatelli_2015_2008ax} \\
iPTF13bvn & 10 & 3.6$^{+0.8}_{-0.5}$ & \citealt{PTF12os_iPTF13bvn_b} \\
SN 2022crv & $\sim$ 3 & 5.3$^{+0.3}_{-0.6}$ & \citealt{funky} \\
SN 2003bg & Compact ($\sim$ 1) & 4.0$^{+0.4}_{-0.4}$ & \citealt{Söderberg_2006_2003bg} \\
SN 2001ig & Compact ($\sim$ 1)& $\geq$ 5.3 & \citealt{2001ig_a} \\
\hline \hline

\end{tabular}

\caption{A comparison between estimated progenitor radii and masses for Type IIb SNe in our sample. The mass is taken to be our own derived mass estimate, as given in Table \ref{tab:mass_predictions}. Note that while no estimate was made for SN 2013df due to the early epoch, we include the object here as its $\sim$ 180d spectrum appears fully nebular, and it has a \NIIdoublet{} emission + [\ion{Ca}{II}]/[\ion{O}{I}] ratio indicating a very low-mass object. This low mass is also found in the literature \citep{2013df}.}
\label{tab:radii}

\end{table}

\subsection{Usefulness of Mass Diagnostic}

In this subsection, we dive a bit deeper into the usefulness of the \NIIdiag{} diagnostic for determining SN progenitor masses. As discussed in Section \ref{sec:mass_determination}, using our diagnostic finds progenitor masses that for the majority of our sample agree well with prior estimates from the literature. This fact increases simultaneously the trust in both the literature estimates, as well as our own estimates. 

The key advantage of our diagnostic with respect to the often used [\ion{Ca}{II}]/[\ion{O}{I}] ratio, is that our diagnostic is normalised with respect to a large part of the optical spectrum, as opposed to the emission of a single line. This makes the diagnostic more robust with respect to object-to-object variations in the amounts of calcium synthesised in the explosion. Additionally, the fact that the diagnostic is mostly independent from the [\ion{Ca}{II}]/[\ion{O}{I}] ratio\footnote{It is not fully independent, as the amounts of [\ion{Ca}{II}] and [\ion{O}{I}] emission form a significant fraction of the emitted flux in the 5000 -- 8000 Å region.} means that both diagnostics can be used simultaneously to reduce the expected uncertainty on the resulting mass estimates. Considering the large scatter in Figure 2 of \citet{Fang_2019_NIIusage} in the relation between mass and [\ion{Ca}{II}]/[\ion{O}{I}], we expect these potential gains to be large. A more detailed analysis of this simultaneous usage of the two diagnostics is left to future works. 

As was also discussed in Section \ref{sec:mass_determination}, the diagnostic also has limitations in its use. The most limiting factor is the fact that it only becomes truly useful after about 200 -- 250 days, when the overlap between model \NIIdiag{} values has disappeared, as well as when by far most spectra can be said to be truly nebular and the \NIIdoublet{} emission emerges strong when significant amounts of N is present. Due to the typical decay of SN flux with time, this means that the usefulness of our diagnostic for a particular SN will be brightness limited, making it less likely to be used for far away or low luminosity events, also possibly introducing an observational bias. Additionally, it means that the sample of SNe that this diagnostic can (for now) be used on will only consist of about $\sim$ 100 objects, due to the limited availability of nebular spectra. Note however that these arguments can equally well be made for most nebular spectroscopy methods. 

Finally, the applicability of the diagnostic is limited in mass range. This can be understood both from the increasingly flat and overlapping tracks in Figure \ref{fig:NII_per_type}, as well as from the increased scatter around the diagonal in Figure \ref{fig:mass_estimates}. As the relative amount of emission from \NIIdoublet{} decreases with mass, it becomes increasingly difficult to distinguish this feature from others, increasing the relative uncertainty on \NIIdiag{} and our mass estimates. Even if we would extend our model grid to higher masses to extend our available mass range, this increased uncertainty would mean that we do not gain any new information regarding the progenitor masses. Our diagnostic can thus be said to only be useful for progenitors which have a mass of \Mf{} $\lesssim$ 5 \Msun{}. For higher mass progenitors the diagnostic will give a lower limit of about 5 \Msun{} for \Mf{}.

\subsection{Causes of Uncertainty}
\label{sec:uncertainties}

When considering Figures \ref{fig:NII_per_type} and \ref{fig:mass_estimates}, one should be aware of the uncertainties present. The selection of models we study is for the "standard" mass-loss case (see the definition in \citet[][their section 2.1]{Woosley_2019_models}), and our resulting absolute mass estimates will depend on this assumption. Figure \ref{fig:mass_loss_effect} shows the relation between the mass fraction of nitrogen in the ejecta (which traces \NIIdiag{}) and the resulting \Mf{} for three different mass-loss rates for progenitors from \citet{Woosley_2019_models}. While the relation is the same for each mass-loss prescription (i.e. relatively less nitrogen in the ejecta, giving weaker \NIIdoublet{} emission, for higher \Mf{}), an increased mass-loss rate (which might be preferred (see \citet{Yoon_2017_WRMassLoss} and \citet{Woosley_2019_models}, but also \citet{McClelland_2016_WRMassloss} and \citet{Vink_2017_Massloss})) will result in a lower \Mf{} value for the same mass fraction of nitrogen in the ejecta. This means that the values on the y-axis in Figure \ref{fig:mass_estimates} might be lower/higher, but that in any case the relative positions of these data points with respect to each other will not change. Related to this uncertainty in mass loss, the timing of the removal of the hydrogen envelope may also be of importance. While in the models from \citet{Woosley_2019_models} this envelope is removed promptly at the onset of core He burning, a smoother or slower removal (e.g. as the star further expands towards the end of He core burning and may then reach Robe Lobe filling) may protect some of the He/N zone from being lost to winds (as well as being replenished by shell H burning).

A further source of uncertainty is the metallicity of the exploding star. As discussed in Section \ref{sec:introduction}, the initial C, N and O present from the star's formation are all channeled into \el{14}{N} by the CNO cycle due to the bottleneck reaction \el{14}{N}($p, \gamma$)\el{15}{O}. As this abundance of C, N and O scales linearly with metallicity, there is a linear dependence between the final nitrogen abundance in the He/N zone and the star's metallicity. Seeing how our input models from \citet{Woosley_2019_models} all use the same metallicity (solar, $Z_{\odot} = 0.015$), this introduces uncertainty in our link between the strength of \NIIdoublet{} emission and \Mf{}.

This sensitivity is limited by the relatively  moderate spread in metallicity for CCSNe in the local universe (which are the only ones we obtain nebular spectra for), which all have recently formed progenitors. \citet{Anderson_2010_MetallicityCCSNE} found for a sample of 27 Type Ibc SNe a mean [12+log(O/H)] metallicity of 8.64 (0.89 $Z_{\odot}$), with the most extreme values at 8.33 (0.44 $Z_{\odot}$) and 8.83 (1.38 $Z_{\odot}$). In other words, an upper limit to the error on the nitrogen abundance in the progenitor induced by metallicity is about 50\%, with a more typical error more likely being 20 -- 30\%. Furthermore, the uncertainties induced by metallicity and mass-loss are not fully independent. If a star is born with a higher metallicity (and thus eventually will have a higher nitrogen abundance in the He/N layer), it will also experience a higher mass-loss by winds. This increased mass-loss will at the end of the stars life have peeled of a larger part of the outer envelope, thus resulting in a smaller fraction of the He/N zone retained. While exact numbers would need full stellar evolution modelling, the example above shows that the uncertainties induced by metallicity and mass-loss counteract each other to some extent. 

The new \NIIdiag{} diagnostic may also find use in its reverse direction application. For example, if the
progenitor mass and metallicity of a given SN are well determined by other means - one could find the best-fitting mass-loss rate to reproduce its \NIIdiag{} value.

\section{Conclusions}

In this work, we have used spectral synthesis modelling to study the formation of the \NIIdoublet{} line in nebular-phase stripped envelope supernovae. Based on this, we introduce a new diagnostic for determining the progenitor mass of SESNe using the \NIIdoublet{} line strength relative to the quasi-optical luminosity (5000 -- 8000 \AA). This metric is powerful because the He/N zone responsible for the N emission decreases in size for higher mass He cores, whereas the O and Ca nucleosynthesis, dominating the quasi-optical flux, increases. This new \NIIdiag{} diagnostic was computed for model spectra for a grid of five different model SN-progenitors at six different epochs (spanning 150 -- 400d), computed with the \texttt{KEPLER} code. Then, we applied it to a compilation of $\sim$ 40 SESNe and were able to determine estimates for \Mf{} for 33 of these. We now list our main findings:

\vspace{-0.1cm}

\begin{enumerate}[I]

    \item The models show that \NIIdoublet{} is a useful diagnostic line from about 200 days post-explosion. The \NIIdiag{} metric varies by up to factor 7 between the highest-mass models (about \NIIdiag{} = 2) to the lowest-mass ones (up to \NIIdiag{} = 17). True \NIIdoublet{} detection is indicated by a curve rising in time, and can also be strengthened by identification of \NIIsinglet{}. Low-mass He cores are predicted to have very strong nebular \NIIdoublet{} lines.

    \item Out of all SNe for which we predict a low progenitor mass (\Mf{} $\lesssim$ 3 \Msun{}), all but one are of Type IIb. Type Ib SNe generally show no strong \NIIdoublet{} emission, with the exception of SN 2007Y. This implicates that the majority of Type Ib SNe either are free of or have a very small He/N zone, or arise from massive enough He cores that the \NIIdoublet{} emission is weak. As \Mf{} estimates from light curve modelling require lower \Mf{} than we would need for such massive He cores, the first scenario would be preferred. This in turn can be achieved with a stronger mass loss rate than adopted in this work, which would decrease \Mf{} for the same strength of \NIIdoublet{} found. 
  
    \item As for Type Ib SNe, we find no clear indication of \NIIdoublet{} emission in Type Ic SNe. While this is consistent with these being He-zone free (although we do not probe the presence of a He/C layer), the fact that neither Type Ib SNe show clear \NIIdoublet{} emission means the finding does not rule out the presence of He. At the same time, we can not exclude the presence of a small amount of this emission. It has been proposed that Type Ic SNe may contain a helium-layer, just like Type IIb/Ib SNe, but that it is not visibly emitting because the helium is not excited enough. Looking for signatures of nitrogen, the second most abundant species in these helium-layers, opens up a new avenue for investigating this question.

    \item When comparing the He core progenitor mass estimates from our new diagnostic to estimates in the literature from a variety of methods, we find generally good agreement. At the same time, it should be stated that the diagnostic is applicable for only a limited amount of SNe, due to the fact that late time ($\gtrsim$ 200 days post-explosion) nebular spectra are required. Additionally, the disappearance of \NIIdoublet{} emission with increasing progenitor mass, means we can only probe progenitor masses up to \Mf{} $\simeq$ 5 \Msun{}; more massive ones give \NIIdoublet{} emission indistinguishable from zero. Nonetheless, the diagnostic will be a valuable new tool for uncovering which progenitors lead to which types of SNe.
    
\end{enumerate}

Further study of nitrogen lines in stripped-envelope SNe may also yield new ways to probe the late shell burning stages in stellar evolution, in turn sensitive to processes like semi-convection and overshooting. Nitrogen emission may also become a metallicity probe, as the nitrogen abundance (left over from CNO cycling) is directly proportional to this. The indications found here that Type IIb, Ib and Ic SNe have different properties of this outermost Helium layer holds promise to better understand the mass stripping mechanisms of these stars - whether by winds or binary companions.

\label{sec:conclusion}

\section*{Acknowledgements}
This research was funded by the Swedish Research Council (Starting Grant 2018-03799, PI: Jerkstrand). Ken'ichi Nomoto has been supported by the World Premier International Research Center
Initiative (WPI), MEXT, Japan, and the Japan Society for the Promotion of Science (JSPS) KAKENHI grants JP20K04024, JP21H044pp, and JP23K03452. The computations in this work were enabled by resources provided by the Swedish National Infrastructure for Computing (SNIC), the National
Academic Infrastructure for Supercomputing in Sweden (NAISS),
and at the Parallelldatorcentrum (PDC) Center for High Performance
Computing, Royal Institute of Technology (KTH), partially funded
by the Swedish Research Council through grant agreements no. 2022-06725 and no. 2018-05973.

We thank Bart van Baal, Priscila Pessi, Anamaria Gkini and Tassilo Schweyer for useful discussions. We thank Thomas Janka for useful discussion and for helping to provide full access to the Garching CCSN model suite at \url{https://wwwmpa.mpa-garching.mpg.de/ccsnarchive/}. Furthermore, we thank Yize Dong for providing access to the nebular spectra for SN 2022crv, and Tassilo Schweyer for providing access to the nebular spectra for SN 2019odp.

\section*{Data Availability}

In this work, nebular phase SESNe spectra were used that were easily obtained with the help of the web-tools WISeREP (\url{https://www.wiserep.org}) and the OSC (\url{https://github.com/astrocatalogs/supernovae}. The stellar evolution and explosion models from \citet{Woosley_2019_models} and \citet{Ertl_2020_models} were obtained from \url{https://wwwmpa.mpa-garching.mpg.de/ccsnarchive/}. The code necessary to produce the figures in this manuscript is available at \url{https://github.com/StanBarmentloo/NII_nebular_phase})

\bibliographystyle{mnras}
\bibliography{references}

\begin{thebibliography}{}
\makeatletter
\relax
\def\mn@urlcharsother{\let\do\@makeother \do\$\do\&\do\#\do\^\do\_\do\%\do\~}
\def\mn@doi{\begingroup\mn@urlcharsother \@ifnextchar [ {\mn@doi@}
  {\mn@doi@[]}}
\def\mn@doi@[#1]#2{\def\@tempa{#1}\ifx\@tempa\@empty \href
  {http://dx.doi.org/#2} {doi:#2}\else \href {http://dx.doi.org/#2} {#1}\fi
  \endgroup}
\def\mn@eprint#1#2{\mn@eprint@#1:#2::\@nil}
\def\mn@eprint@arXiv#1{\href {http://arxiv.org/abs/#1} {{\tt arXiv:#1}}}
\def\mn@eprint@dblp#1{\href {http://dblp.uni-trier.de/rec/bibtex/#1.xml}
  {dblp:#1}}
\def\mn@eprint@#1:#2:#3:#4\@nil{\def\@tempa {#1}\def\@tempb {#2}\def\@tempc
  {#3}\ifx \@tempc \@empty \let \@tempc \@tempb \let \@tempb \@tempa \fi \ifx
  \@tempb \@empty \def\@tempb {arXiv}\fi \@ifundefined
  {mn@eprint@\@tempb}{\@tempb:\@tempc}{\expandafter \expandafter \csname
  mn@eprint@\@tempb\endcsname \expandafter{\@tempc}}}

\bibitem[\protect\citeauthoryear{{Aldering}, {Humphreys}  \&
  {Richmond}}{{Aldering} et~al.}{1994}]{1993J_BB}
{Aldering} G.,  {Humphreys} R.~M.,   {Richmond} M.,  1994, \mn@doi [\aj]
  {10.1086/116886}, \href
  {https://ui.adsabs.harvard.edu/abs/1994AJ....107..662A} {107, 662}

\bibitem[\protect\citeauthoryear{{Anderson}, {Covarrubias}, {James}, {Hamuy}
  \& {Habergham}}{{Anderson} et~al.}{2010}]{Anderson_2010_MetallicityCCSNE}
{Anderson} J.~P.,  {Covarrubias} R.~A.,  {James} P.~A.,  {Hamuy} M.,
  {Habergham} S.~M.,  2010, \mn@doi [\mnras]
  {10.1111/j.1365-2966.2010.17118.x}, \href
  {https://ui.adsabs.harvard.edu/abs/2010MNRAS.407.2660A} {407, 2660}

\bibitem[\protect\citeauthoryear{{Arcavi} et~al.,}{{Arcavi}
  et~al.}{2011}]{2011dh_BB}
{Arcavi} I.,  et~al., 2011, \mn@doi [\apjl] {10.1088/2041-8205/742/2/L18},
  \href {https://ui.adsabs.harvard.edu/abs/2011ApJ...742L..18A} {742, L18}

\bibitem[\protect\citeauthoryear{{Ben-Ami} et~al.,}{{Ben-Ami}
  et~al.}{2012}]{PTF12gzk_a}
{Ben-Ami} S.,  et~al., 2012, \mn@doi [\apjl] {10.1088/2041-8205/760/2/L33},
  \href {https://ui.adsabs.harvard.edu/abs/2012ApJ...760L..33B} {760, L33}

\bibitem[\protect\citeauthoryear{{Benvenuto}, {Bersten}  \&
  {Nomoto}}{{Benvenuto} et~al.}{2013}]{2011dh_C}
{Benvenuto} O.~G.,  {Bersten} M.~C.,   {Nomoto} K.,  2013, \mn@doi [\apj]
  {10.1088/0004-637X/762/2/74}, \href
  {https://ui.adsabs.harvard.edu/abs/2013ApJ...762...74B} {762, 74}

\bibitem[\protect\citeauthoryear{{Bersten} et~al.,}{{Bersten}
  et~al.}{2012a}]{Bersten_2012_2011dh}
{Bersten} M.~C.,  et~al., 2012a, \mn@doi [\apj] {10.1088/0004-637X/757/1/31},
  \href {https://ui.adsabs.harvard.edu/abs/2012ApJ...757...31B} {757, 31}

\bibitem[\protect\citeauthoryear{{Bersten} et~al.,}{{Bersten}
  et~al.}{2012b}]{2011dh_F}
{Bersten} M.~C.,  et~al., 2012b, \mn@doi [\apj] {10.1088/0004-637X/757/1/31},
  \href {https://ui.adsabs.harvard.edu/abs/2012ApJ...757...31B} {757, 31}

\bibitem[\protect\citeauthoryear{{Bersten} et~al.,}{{Bersten}
  et~al.}{2014}]{iPTF13bvn_D}
{Bersten} M.~C.,  et~al., 2014, \mn@doi [\aj] {10.1088/0004-6256/148/4/68},
  \href {https://ui.adsabs.harvard.edu/abs/2014AJ....148...68B} {148, 68}

\bibitem[\protect\citeauthoryear{{Bethe} \& {Wilson}}{{Bethe} \&
  {Wilson}}{1985}]{Bethe_1985_Neutrinoheating}
{Bethe} H.~A.,  {Wilson} J.~R.,  1985, \mn@doi [\apj] {10.1086/163343}, \href
  {https://ui.adsabs.harvard.edu/abs/1985ApJ...295...14B} {295, 14}

\bibitem[\protect\citeauthoryear{{Bianco} et~al.,}{{Bianco}
  et~al.}{2014}]{2007C_a_2008bo_a}
{Bianco} F.~B.,  et~al., 2014, \mn@doi [\apjs] {10.1088/0067-0049/213/2/19},
  \href {https://ui.adsabs.harvard.edu/abs/2014ApJS..213...19B} {213, 19}

\bibitem[\protect\citeauthoryear{{Blinnikov}, {Eastman}, {Bartunov},
  {Popolitov}  \& {Woosley}}{{Blinnikov} et~al.}{1998}]{1993J_D}
{Blinnikov} S.~I.,  {Eastman} R.,  {Bartunov} O.~S.,  {Popolitov} V.~A.,
  {Woosley} S.~E.,  1998, \mn@doi [\apj] {10.1086/305375}, \href
  {https://ui.adsabs.harvard.edu/abs/1998ApJ...496..454B} {496, 454}

\bibitem[\protect\citeauthoryear{{Boty{\'a}nszki}, {Kasen}  \&
  {Plewa}}{{Boty{\'a}nszki} et~al.}{2018}]{Botyanszki2018}
{Boty{\'a}nszki} J.,  {Kasen} D.,   {Plewa} T.,  2018, \mn@doi [\apjl]
  {10.3847/2041-8213/aaa07b}, \href
  {https://ui.adsabs.harvard.edu/abs/2018ApJ...852L...6B} {852, L6}

\bibitem[\protect\citeauthoryear{{Bufano} et~al.,}{{Bufano}
  et~al.}{2014}]{2011hs}
{Bufano} F.,  et~al., 2014, \mn@doi [\mnras] {10.1093/mnras/stu065}, \href
  {https://ui.adsabs.harvard.edu/abs/2014MNRAS.439.1807B} {439, 1807}

\bibitem[\protect\citeauthoryear{{Burrows}}{{Burrows}}{2013}]{Burrows_2013_Neutrino_Explosions}
{Burrows} A.,  2013, \mn@doi [Reviews of Modern Physics]
  {10.1103/RevModPhys.85.245}, \href
  {https://ui.adsabs.harvard.edu/abs/2013RvMP...85..245B} {85, 245}

\bibitem[\protect\citeauthoryear{{Cao} et~al.,}{{Cao}
  et~al.}{2013}]{iPTF13bvn_a}
{Cao} Y.,  et~al., 2013, \mn@doi [\apjl] {10.1088/2041-8205/775/1/L7}, \href
  {https://ui.adsabs.harvard.edu/abs/2013ApJ...775L...7C} {775, L7}

\bibitem[\protect\citeauthoryear{{Chandra}, {Nayana}, {Bj{\"o}rnsson},
  {Taddia}, {Lundqvist}, {Ray}  \& {Shappee}}{{Chandra} et~al.}{2019}]{J1204_a}
{Chandra} P.,  {Nayana} A.~J.,  {Bj{\"o}rnsson} C.~I.,  {Taddia} F.,
  {Lundqvist} P.,  {Ray} A.~K.,   {Shappee} B.~J.,  2019, \mn@doi [\apj]
  {10.3847/1538-4357/ab1900}, \href
  {https://ui.adsabs.harvard.edu/abs/2019ApJ...877...79C} {877, 79}

\bibitem[\protect\citeauthoryear{{Chevalier}}{{Chevalier}}{1976}]{Chevalier_1976__Binaries}
{Chevalier} R.~A.,  1976, \mn@doi [\apj] {10.1086/154669}, \href
  {https://ui.adsabs.harvard.edu/abs/1976ApJ...208..826C} {208, 826}

\bibitem[\protect\citeauthoryear{{Chevalier} \& {Klein}}{{Chevalier} \&
  {Klein}}{1978}]{Chevalier_1978_RT}
{Chevalier} R.~A.,  {Klein} R.~I.,  1978, \mn@doi [\apj] {10.1086/155864},
  \href {https://ui.adsabs.harvard.edu/abs/1978ApJ...219..994C} {219, 994}

\bibitem[\protect\citeauthoryear{{Colgate} \& {White}}{{Colgate} \&
  {White}}{1966}]{Colgate_1966_SNcanon}
{Colgate} S.~A.,  {White} R.~H.,  1966, \mn@doi [\apj] {10.1086/148549}, \href
  {https://ui.adsabs.harvard.edu/abs/1966ApJ...143..626C} {143, 626}

\bibitem[\protect\citeauthoryear{{Dessart} \& {Hillier}}{{Dessart} \&
  {Hillier}}{2010}]{Dessart_2010_CMFGENb}
{Dessart} L.,  {Hillier} D.~J.,  2010, \mn@doi [\mnras]
  {10.1111/j.1365-2966.2010.16611.x}, \href
  {https://ui.adsabs.harvard.edu/abs/2010MNRAS.405.2141D} {405, 2141}

\bibitem[\protect\citeauthoryear{{Dessart}, {Hillier}, {Livne}, {Yoon},
  {Woosley}, {Waldman}  \& {Langer}}{{Dessart} et~al.}{2011}]{Dessart2011}
{Dessart} L.,  {Hillier} D.~J.,  {Livne} E.,  {Yoon} S.-C.,  {Woosley} S.,
  {Waldman} R.,   {Langer} N.,  2011, \mn@doi [\mnras]
  {10.1111/j.1365-2966.2011.18598.x}, \href
  {https://ui.adsabs.harvard.edu/abs/2011MNRAS.414.2985D} {414, 2985}

\bibitem[\protect\citeauthoryear{{Dessart}, {Hillier}, {Li}  \&
  {Woosley}}{{Dessart} et~al.}{2012}]{Dessart_2012_Heliumhidden}
{Dessart} L.,  {Hillier} D.~J.,  {Li} C.,   {Woosley} S.,  2012, \mn@doi
  [\mnras] {10.1111/j.1365-2966.2012.21374.x}, \href
  {https://ui.adsabs.harvard.edu/abs/2012MNRAS.424.2139D} {424, 2139}

\bibitem[\protect\citeauthoryear{{Dessart}, {Yoon}, {Aguilera-Dena}  \&
  {Langer}}{{Dessart} et~al.}{2020}]{Dessart_2020_Hiddenhe}
{Dessart} L.,  {Yoon} S.-C.,  {Aguilera-Dena} D.~R.,   {Langer} N.,  2020,
  \mn@doi [\aap] {10.1051/0004-6361/202038763}, \href
  {https://ui.adsabs.harvard.edu/abs/2020A&A...642A.106D} {642, A106}

\bibitem[\protect\citeauthoryear{{Dessart}, {Hillier}, {Sukhbold}, {Woosley}
  \& {Janka}}{{Dessart} et~al.}{2021a}]{Dessart2021}
{Dessart} L.,  {Hillier} D.~J.,  {Sukhbold} T.,  {Woosley} S.~E.,   {Janka}
  H.~T.,  2021a, \mn@doi [\aap] {10.1051/0004-6361/202140839}, \href
  {https://ui.adsabs.harvard.edu/abs/2021A&A...652A..64D} {652, A64}

\bibitem[\protect\citeauthoryear{{Dessart}, {Hillier}, {Sukhbold}, {Woosley}
  \& {Janka}}{{Dessart} et~al.}{2021b}]{Dessart_2021_Hestarexpl}
{Dessart} L.,  {Hillier} D.~J.,  {Sukhbold} T.,  {Woosley} S.~E.,   {Janka}
  H.~T.,  2021b, \mn@doi [\aap] {10.1051/0004-6361/202141927}, \href
  {https://ui.adsabs.harvard.edu/abs/2021A&A...656A..61D} {656, A61}

\bibitem[\protect\citeauthoryear{{Dessart}, {Hillier}, {Woosley}  \&
  {Kuncarayakti}}{{Dessart} et~al.}{2023}]{Dessart_2023_TimeEvolution}
{Dessart} L.,  {Hillier} D.~J.,  {Woosley} S.~E.,   {Kuncarayakti} H.,  2023,
  \mn@doi [\aap] {10.1051/0004-6361/202346626}, \href
  {https://ui.adsabs.harvard.edu/abs/2023A&A...677A...7D} {677, A7}

\bibitem[\protect\citeauthoryear{{Dong} et~al.,}{{Dong} et~al.}{2023}]{2022crv}
{Dong} Y.,  et~al., 2023, preprint, \href
  {https://ui.adsabs.harvard.edu/abs/2023arXiv230909433D} {p. arXiv:2309.09433}
  (\mn@eprint {arXiv} {2309.09433})

\bibitem[\protect\citeauthoryear{{Drout} et~al.,}{{Drout}
  et~al.}{2011}]{2004dk_a_2004gq_a}
{Drout} M.~R.,  et~al., 2011, \mn@doi [\apj] {10.1088/0004-637X/741/2/97},
  \href {https://ui.adsabs.harvard.edu/abs/2011ApJ...741...97D} {741, 97}

\bibitem[\protect\citeauthoryear{{Drout} et~al.,}{{Drout}
  et~al.}{2016}]{2013ge}
{Drout} M.~R.,  et~al., 2016, \mn@doi [\apj] {10.3847/0004-637X/821/1/57},
  \href {https://ui.adsabs.harvard.edu/abs/2016ApJ...821...57D} {821, 57}

\bibitem[\protect\citeauthoryear{{Ebisuzaki}, {Shigeyama}  \&
  {Nomoto}}{{Ebisuzaki} et~al.}{1989}]{Ebisuzaki_1989_asymmetry}
{Ebisuzaki} T.,  {Shigeyama} T.,   {Nomoto} K.,  1989, \mn@doi [\apjl]
  {10.1086/185532}, \href
  {https://ui.adsabs.harvard.edu/abs/1989ApJ...344L..65E} {344, L65}

\bibitem[\protect\citeauthoryear{{Eldridge}, {Fraser}, {Smartt}, {Maund}  \&
  {Crockett}}{{Eldridge} et~al.}{2013}]{Eldridge_2013_BinarymoreMass}
{Eldridge} J.~J.,  {Fraser} M.,  {Smartt} S.~J.,  {Maund} J.~R.,   {Crockett}
  R.~M.,  2013, \mn@doi [\mnras] {10.1093/mnras/stt1612}, \href
  {https://ui.adsabs.harvard.edu/abs/2013MNRAS.436..774E} {436, 774}

\bibitem[\protect\citeauthoryear{{Eldridge}, {Fraser}, {Maund}  \&
  {Smartt}}{{Eldridge} et~al.}{2015}]{iPTF13bvn_E}
{Eldridge} J.~J.,  {Fraser} M.,  {Maund} J.~R.,   {Smartt} S.~J.,  2015,
  \mn@doi [\mnras] {10.1093/mnras/stu2197}, \href
  {https://ui.adsabs.harvard.edu/abs/2015MNRAS.446.2689E} {446, 2689}

\bibitem[\protect\citeauthoryear{{Elias}, {Matthews}, {Neugebauer}  \&
  {Persson}}{{Elias} et~al.}{1985}]{Elias_1985_Ib}
{Elias} J.~H.,  {Matthews} K.,  {Neugebauer} G.,   {Persson} S.~E.,  1985,
  \mn@doi [\apj] {10.1086/163456}, \href
  {https://ui.adsabs.harvard.edu/abs/1985ApJ...296..379E} {296, 379}

\bibitem[\protect\citeauthoryear{{Elmhamdi}, {Tsvetkov}, {Danziger}  \&
  {Kordi}}{{Elmhamdi} et~al.}{2011}]{2004ao_a}
{Elmhamdi} A.,  {Tsvetkov} D.,  {Danziger} I.~J.,   {Kordi} A.,  2011, \mn@doi
  [\apj] {10.1088/0004-637X/731/2/129}, \href
  {https://ui.adsabs.harvard.edu/abs/2011ApJ...731..129E} {731, 129}

\bibitem[\protect\citeauthoryear{{Ensman} \& {Woosley}}{{Ensman} \&
  {Woosley}}{1988}]{Ensman1988}
{Ensman} L.~M.,  {Woosley} S.~E.,  1988, \mn@doi [\apj] {10.1086/166785}, \href
  {https://ui.adsabs.harvard.edu/abs/1988ApJ...333..754E} {333, 754}

\bibitem[\protect\citeauthoryear{{Ergon} \& {Fransson}}{{Ergon} \&
  {Fransson}}{2022}]{Ergon_2022_JEKYLLb}
{Ergon} M.,  {Fransson} C.,  2022, \mn@doi [\aap]
  {10.1051/0004-6361/202243448}, \href
  {https://ui.adsabs.harvard.edu/abs/2022A&A...666A.104E} {666, A104}

\bibitem[\protect\citeauthoryear{{Ergon} et~al.,}{{Ergon}
  et~al.}{2014}]{2011dh_a}
{Ergon} M.,  et~al., 2014, \mn@doi [\aap] {10.1051/0004-6361/201321850}, \href
  {https://ui.adsabs.harvard.edu/abs/2014A&A...562A..17E} {562, A17}

\bibitem[\protect\citeauthoryear{{Ergon} et~al.,}{{Ergon}
  et~al.}{2015a}]{2011dh_b}
{Ergon} M.,  et~al., 2015a, \mn@doi [\aap] {10.1051/0004-6361/201424592}, \href
  {https://ui.adsabs.harvard.edu/abs/2015A&A...580A.142E} {580, A142}

\bibitem[\protect\citeauthoryear{{Ergon} et~al.,}{{Ergon}
  et~al.}{2015b}]{2011dh_D}
{Ergon} M.,  et~al., 2015b, \mn@doi [\aap] {10.1051/0004-6361/201424592}, \href
  {https://ui.adsabs.harvard.edu/abs/2015A&A...580A.142E} {580, A142}

\bibitem[\protect\citeauthoryear{{Ergon}, {Fransson}, {Jerkstrand}, {Kozma},
  {Kromer}  \& {Spricer}}{{Ergon} et~al.}{2018}]{Ergon_2018_JEKYLLa}
{Ergon} M.,  {Fransson} C.,  {Jerkstrand} A.,  {Kozma} C.,  {Kromer} M.,
  {Spricer} K.,  2018, \mn@doi [\aap] {10.1051/0004-6361/201833043}, \href
  {https://ui.adsabs.harvard.edu/abs/2018A&A...620A.156E} {620, A156}

\bibitem[\protect\citeauthoryear{{Ergon} et~al.,}{{Ergon}
  et~al.}{2023}]{2020acat_b}
{Ergon} M.,  et~al., 2023, preprint, \href
  {https://ui.adsabs.harvard.edu/abs/2023arXiv230807158E} {p. arXiv:2308.07158}
  (\mn@eprint {arXiv} {2308.07158})

\bibitem[\protect\citeauthoryear{{Ertl}, {Woosley}, {Sukhbold}  \&
  {Janka}}{{Ertl} et~al.}{2020}]{Ertl_2020_models}
{Ertl} T.,  {Woosley} S.~E.,  {Sukhbold} T.,   {Janka} H.~T.,  2020, \mn@doi
  [\apj] {10.3847/1538-4357/ab6458}, \href
  {https://ui.adsabs.harvard.edu/abs/2020ApJ...890...51E} {890, 51}

\bibitem[\protect\citeauthoryear{{Falk} \& {Arnett}}{{Falk} \&
  {Arnett}}{1973}]{Falk_1973_RT}
{Falk} S.~W.,  {Arnett} W.~D.,  1973, \mn@doi [\apjl] {10.1086/181154}, \href
  {https://ui.adsabs.harvard.edu/abs/1973ApJ...180L..65F} {180, L65}

\bibitem[\protect\citeauthoryear{{Fang} \& {Maeda}}{{Fang} \&
  {Maeda}}{2018}]{Fang_2018_NIIdiscovery}
{Fang} Q.,  {Maeda} K.,  2018, \mn@doi [\apj] {10.3847/1538-4357/aad096}, \href
  {https://ui.adsabs.harvard.edu/abs/2018ApJ...864...47F} {864, 47}

\bibitem[\protect\citeauthoryear{{Fang}, {Maeda}, {Kuncarayakti}, {Sun}  \&
  {Gal-Yam}}{{Fang} et~al.}{2019}]{Fang_2019_NIIusage}
{Fang} Q.,  {Maeda} K.,  {Kuncarayakti} H.,  {Sun} F.,   {Gal-Yam} A.,  2019,
  \mn@doi [Nature Astronomy] {10.1038/s41550-019-0710-6}, \href
  {https://ui.adsabs.harvard.edu/abs/2019NatAs...3..434F} {3, 434}

\bibitem[\protect\citeauthoryear{{Filippenko}}{{Filippenko}}{1988}]{Filippenko_1988_IIb}
{Filippenko} A.~V.,  1988, \mn@doi [\aj] {10.1086/114940}, \href
  {https://ui.adsabs.harvard.edu/abs/1988AJ.....96.1941F} {96, 1941}

\bibitem[\protect\citeauthoryear{{Filippenko}}{{Filippenko}}{1997}]{Filippenko_1997_Taxonomy}
{Filippenko} A.~V.,  1997, \mn@doi [\araa] {10.1146/annurev.astro.35.1.309},
  \href {https://ui.adsabs.harvard.edu/abs/1997ARA&A..35..309F} {35, 309}

\bibitem[\protect\citeauthoryear{{Filippenko} \& {Sargent}}{{Filippenko} \&
  {Sargent}}{1986}]{1985F_b}
{Filippenko} A.~V.,  {Sargent} W.~L.~W.,  1986, \mn@doi [\aj] {10.1086/114051},
  \href {https://ui.adsabs.harvard.edu/abs/1986AJ.....91..691F} {91, 691}

\bibitem[\protect\citeauthoryear{{Filippenko}, {Porter}, {Sargent}  \&
  {Schneider}}{{Filippenko} et~al.}{1986}]{1985F_a}
{Filippenko} A.~V.,  {Porter} A.~C.,  {Sargent} W. L.~W.,   {Schneider} D.~P.,
  1986, \mn@doi [\aj] {10.1086/114266}, \href
  {https://ui.adsabs.harvard.edu/abs/1986AJ.....92.1341F} {92, 1341}

\bibitem[\protect\citeauthoryear{{Filippenko}, {Porter}  \&
  {Sargent}}{{Filippenko} et~al.}{1990}]{1987M}
{Filippenko} A.~V.,  {Porter} A.~C.,   {Sargent} W. L.~W.,  1990, \mn@doi [\aj]
  {10.1086/115618}, \href
  {https://ui.adsabs.harvard.edu/abs/1990AJ....100.1575F} {100, 1575}

\bibitem[\protect\citeauthoryear{{Filippenko}, {Matheson}  \&
  {Ho}}{{Filippenko} et~al.}{1993}]{Filippenko_1993_IIb}
{Filippenko} A.~V.,  {Matheson} T.,   {Ho} L.~C.,  1993, \mn@doi [\apjl]
  {10.1086/187043}, \href
  {https://ui.adsabs.harvard.edu/abs/1993ApJ...415L.103F} {415, L103}

\bibitem[\protect\citeauthoryear{{Filippenko}, {Matheson}  \&
  {Barth}}{{Filippenko} et~al.}{1994a}]{Filippenko_1994_1993J}
{Filippenko} A.~V.,  {Matheson} T.,   {Barth} A.~J.,  1994a, \mn@doi [\aj]
  {10.1086/117234}, \href
  {https://ui.adsabs.harvard.edu/abs/1994AJ....108.2220F} {108, 2220}

\bibitem[\protect\citeauthoryear{{Filippenko}, {Matheson}  \&
  {Barth}}{{Filippenko} et~al.}{1994b}]{1993J_b}
{Filippenko} A.~V.,  {Matheson} T.,   {Barth} A.~J.,  1994b, \mn@doi [\aj]
  {10.1086/117234}, \href
  {https://ui.adsabs.harvard.edu/abs/1994AJ....108.2220F} {108, 2220}

\bibitem[\protect\citeauthoryear{{Fitzpatrick}}{{Fitzpatrick}}{1999}]{Fitzpatrick_1999_Reddening}
{Fitzpatrick} E.~L.,  1999, \mn@doi [\pasp] {10.1086/316293}, \href
  {https://ui.adsabs.harvard.edu/abs/1999PASP..111...63F} {111, 63}

\bibitem[\protect\citeauthoryear{{Folatelli}, {Bersten}, {Kuncarayakti},
  {Benvenuto}, {Maeda}  \& {Nomoto}}{{Folatelli}
  et~al.}{2015}]{Folatelli_2015_2008ax}
{Folatelli} G.,  {Bersten} M.~C.,  {Kuncarayakti} H.,  {Benvenuto} O.~G.,
  {Maeda} K.,   {Nomoto} K.,  2015, \mn@doi [\apj]
  {10.1088/0004-637X/811/2/147}, \href
  {https://ui.adsabs.harvard.edu/abs/2015ApJ...811..147F} {811, 147}

\bibitem[\protect\citeauthoryear{{Fransson} \& {Chevalier}}{{Fransson} \&
  {Chevalier}}{1989}]{Fransson_1989_CaOIratio}
{Fransson} C.,  {Chevalier} R.~A.,  1989, \mn@doi [\apj] {10.1086/167707},
  \href {https://ui.adsabs.harvard.edu/abs/1989ApJ...343..323F} {343, 323}

\bibitem[\protect\citeauthoryear{{Fremling} et~al.,}{{Fremling}
  et~al.}{2014}]{iPTF13bvn_C}
{Fremling} C.,  et~al., 2014, \mn@doi [\aap] {10.1051/0004-6361/201423884},
  \href {https://ui.adsabs.harvard.edu/abs/2014A&A...565A.114F} {565, A114}

\bibitem[\protect\citeauthoryear{{Fremling} et~al.,}{{Fremling}
  et~al.}{2016}]{PTF12os_iPTF13bvn_b}
{Fremling} C.,  et~al., 2016, \mn@doi [\aap] {10.1051/0004-6361/201628275},
  \href {https://ui.adsabs.harvard.edu/abs/2016A&A...593A..68F} {593, A68}

\bibitem[\protect\citeauthoryear{{Fremling} et~al.,}{{Fremling}
  et~al.}{2018}]{Fremling_2018_pEW}
{Fremling} C.,  et~al., 2018, \mn@doi [\aap] {10.1051/0004-6361/201731701},
  \href {https://ui.adsabs.harvard.edu/abs/2018A&A...618A..37F} {618, A37}

\bibitem[\protect\citeauthoryear{{Galama} et~al.,}{{Galama}
  et~al.}{1998}]{1998bw_a}
{Galama} T.~J.,  et~al., 1998, \mn@doi [\nat] {10.1038/27150}, \href
  {https://ui.adsabs.harvard.edu/abs/1998Natur.395..670G} {395, 670}

\bibitem[\protect\citeauthoryear{{Gangopadhyay} et~al.,}{{Gangopadhyay}
  et~al.}{2023}]{funky}
{Gangopadhyay} A.,  et~al., 2023, \mn@doi [\apj] {10.3847/1538-4357/acfa94},
  \href {https://ui.adsabs.harvard.edu/abs/2023ApJ...957..100G} {957, 100}

\bibitem[\protect\citeauthoryear{{Groh}, {Georgy}  \& {Ekstr{\"o}m}}{{Groh}
  et~al.}{2013}]{iPTF13bvn_F}
{Groh} J.~H.,  {Georgy} C.,   {Ekstr{\"o}m} S.,  2013, \mn@doi [\aap]
  {10.1051/0004-6361/201322369}, \href
  {https://ui.adsabs.harvard.edu/abs/2013A&A...558L...1G} {558, L1}

\bibitem[\protect\citeauthoryear{{Hachinger}, {Mazzali}, {Taubenberger},
  {Hillebrandt}, {Nomoto}  \& {Sauer}}{{Hachinger}
  et~al.}{2012}]{Hachinger_2012_HiddenHe}
{Hachinger} S.,  {Mazzali} P.~A.,  {Taubenberger} S.,  {Hillebrandt} W.,
  {Nomoto} K.,   {Sauer} D.~N.,  2012, \mn@doi [\mnras]
  {10.1111/j.1365-2966.2012.20464.x}, \href
  {https://ui.adsabs.harvard.edu/abs/2012MNRAS.422...70H} {422, 70}

\bibitem[\protect\citeauthoryear{{Hachisu}, {Matsuda}, {Nomoto}  \&
  {Shigeyama}}{{Hachisu} et~al.}{1991}]{Hachisu_1991_Ib_Mixing_Models}
{Hachisu} I.,  {Matsuda} T.,  {Nomoto} K.,   {Shigeyama} T.,  1991, \mn@doi
  [\apjl] {10.1086/185940}, \href
  {https://ui.adsabs.harvard.edu/abs/1991ApJ...368L..27H} {368, L27}

\bibitem[\protect\citeauthoryear{{Hachisu}, {Matsuda}, {Nomoto}  \&
  {Shigeyama}}{{Hachisu} et~al.}{1992}]{Hachisu_1992_asymmetry}
{Hachisu} I.,  {Matsuda} T.,  {Nomoto} K.,   {Shigeyama} T.,  1992, \mn@doi
  [\apj] {10.1086/171274}, \href
  {https://ui.adsabs.harvard.edu/abs/1992ApJ...390..230H} {390, 230}

\bibitem[\protect\citeauthoryear{{Hamuy} et~al.,}{{Hamuy}
  et~al.}{2009}]{2003bg}
{Hamuy} M.,  et~al., 2009, \mn@doi [\apj] {10.1088/0004-637X/703/2/1612}, \href
  {https://ui.adsabs.harvard.edu/abs/2009ApJ...703.1612H} {703, 1612}

\bibitem[\protect\citeauthoryear{{Heger}, {Fryer}, {Woosley}, {Langer}  \&
  {Hartmann}}{{Heger} et~al.}{2003}]{Heger_2003_Massloss}
{Heger} A.,  {Fryer} C.~L.,  {Woosley} S.~E.,  {Langer} N.,   {Hartmann} D.~H.,
   2003, \mn@doi [\apj] {10.1086/375341}, \href
  {https://ui.adsabs.harvard.edu/abs/2003ApJ...591..288H} {591, 288}

\bibitem[\protect\citeauthoryear{{Hillier} \& {Dessart}}{{Hillier} \&
  {Dessart}}{2012}]{Hillier_2012_CMFGENc}
{Hillier} D.~J.,  {Dessart} L.,  2012, \mn@doi [\mnras]
  {10.1111/j.1365-2966.2012.21192.x}, \href
  {https://ui.adsabs.harvard.edu/abs/2012MNRAS.424..252H} {424, 252}

\bibitem[\protect\citeauthoryear{{Hillier} \& {Miller}}{{Hillier} \&
  {Miller}}{1998}]{Hillier_1998_CMFGENa}
{Hillier} D.~J.,  {Miller} D.~L.,  1998, \mn@doi [\apj] {10.1086/305350}, \href
  {https://ui.adsabs.harvard.edu/abs/1998ApJ...496..407H} {496, 407}

\bibitem[\protect\citeauthoryear{{Iwamoto}, {Young}, {Nakasato}, {Shigeyama},
  {Nomoto}, {Hachisu}  \& {Saio}}{{Iwamoto}
  et~al.}{1997}]{Iwamoto_1997_RTinstab}
{Iwamoto} K.,  {Young} T.~R.,  {Nakasato} N.,  {Shigeyama} T.,  {Nomoto} K.,
  {Hachisu} I.,   {Saio} H.,  1997, \mn@doi [\apj] {10.1086/303729}, \href
  {https://ui.adsabs.harvard.edu/abs/1997ApJ...477..865I} {477, 865}

\bibitem[\protect\citeauthoryear{{Iwamoto} et~al.,}{{Iwamoto}
  et~al.}{1998}]{1998bw_AA}
{Iwamoto} K.,  et~al., 1998, \mn@doi [\nat] {10.1038/27155}, \href
  {https://ui.adsabs.harvard.edu/abs/1998Natur.395..672I} {395, 672}

\bibitem[\protect\citeauthoryear{{Janka}}{{Janka}}{2012}]{Janka_2012_CCSNE_Neutrinos}
{Janka} H.-T.,  2012, \mn@doi [Annual Review of Nuclear and Particle Science]
  {10.1146/annurev-nucl-102711-094901}, \href
  {https://ui.adsabs.harvard.edu/abs/2012ARNPS..62..407J} {62, 407}

\bibitem[\protect\citeauthoryear{{Janka} \& {Mueller}}{{Janka} \&
  {Mueller}}{1996}]{Janka_1996_PHOTBa}
{Janka} H.~T.,  {Mueller} E.,  1996, \aap, \href
  {https://ui.adsabs.harvard.edu/abs/1996A&A...306..167J} {306, 167}

\bibitem[\protect\citeauthoryear{{Janka}, {Langanke}, {Marek},
  {Mart{\'\i}nez-Pinedo}  \& {M{\"u}ller}}{{Janka}
  et~al.}{2007}]{Janka_2007_CCSNE}
{Janka} H.~T.,  {Langanke} K.,  {Marek} A.,  {Mart{\'\i}nez-Pinedo} G.,
  {M{\"u}ller} B.,  2007, \mn@doi [\physrep] {10.1016/j.physrep.2007.02.002},
  \href {https://ui.adsabs.harvard.edu/abs/2007PhR...442...38J} {442, 38}

\bibitem[\protect\citeauthoryear{{Jerkstrand}}{{Jerkstrand}}{2011}]{Jerkstrand_2011_SUMOa}
{Jerkstrand} A.,  2011, PhD thesis, Stockholm University

\bibitem[\protect\citeauthoryear{{Jerkstrand}}{{Jerkstrand}}{2017}]{Jerkstrand_2017_book}
{Jerkstrand} A.,  2017, in {Alsabti} A.~W.,  {Murdin} P.,  eds, , Handbook of
  Supernovae.
p.~795, \mn@doi{10.1007/978-3-319-21846-5_29}

\bibitem[\protect\citeauthoryear{{Jerkstrand}, {Fransson}, {Maguire}, {Smartt},
  {Ergon}  \& {Spyromilio}}{{Jerkstrand} et~al.}{2012}]{Jerkstrand_2012_SUMOb}
{Jerkstrand} A.,  {Fransson} C.,  {Maguire} K.,  {Smartt} S.,  {Ergon} M.,
  {Spyromilio} J.,  2012, \mn@doi [\aap] {10.1051/0004-6361/201219528}, \href
  {https://ui.adsabs.harvard.edu/abs/2012A&A...546A..28J} {546, A28}

\bibitem[\protect\citeauthoryear{{Jerkstrand}, {Smartt}, {Fraser}, {Fransson},
  {Sollerman}, {Taddia}  \& {Kotak}}{{Jerkstrand}
  et~al.}{2014}]{Jerkstrand_2014_SUMOc}
{Jerkstrand} A.,  {Smartt} S.~J.,  {Fraser} M.,  {Fransson} C.,  {Sollerman}
  J.,  {Taddia} F.,   {Kotak} R.,  2014, \mn@doi [\mnras]
  {10.1093/mnras/stu221}, \href
  {https://ui.adsabs.harvard.edu/abs/2014MNRAS.439.3694J} {439, 3694}

\bibitem[\protect\citeauthoryear{{Jerkstrand}, {Ergon}, {Smartt}, {Fransson},
  {Sollerman}, {Taubenberger}, {Bersten}  \& {Spyromilio}}{{Jerkstrand}
  et~al.}{2015}]{Jerkstrand_2015_NII_discovery}
{Jerkstrand} A.,  {Ergon} M.,  {Smartt} S.~J.,  {Fransson} C.,  {Sollerman} J.,
   {Taubenberger} S.,  {Bersten} M.,   {Spyromilio} J.,  2015, \mn@doi [\aap]
  {10.1051/0004-6361/201423983}, \href
  {https://ui.adsabs.harvard.edu/abs/2015A&A...573A..12J} {573, A12}

\bibitem[\protect\citeauthoryear{{Jiang} \& {Shu}}{{Jiang} \&
  {Shu}}{1996}]{Jiang_1996_WENO}
{Jiang} G.-S.,  {Shu} C.-W.,  1996, \mn@doi [Journal of Computational Physics]
  {10.1006/jcph.1996.0130}, \href
  {https://ui.adsabs.harvard.edu/abs/1996JCoPh.126..202J} {126, 202}

\bibitem[\protect\citeauthoryear{{Kasen}, {Thomas}  \& {Nugent}}{{Kasen}
  et~al.}{2006}]{Kasen2006}
{Kasen} D.,  {Thomas} R.~C.,   {Nugent} P.,  2006, \mn@doi [\apj]
  {10.1086/506190}, \href
  {https://ui.adsabs.harvard.edu/abs/2006ApJ...651..366K} {651, 366}

\bibitem[\protect\citeauthoryear{{Kifonidis}, {Plewa}, {Janka}  \&
  {M{\"u}ller}}{{Kifonidis} et~al.}{2003}]{Kifonidis_2003_PHOTBb}
{Kifonidis} K.,  {Plewa} T.,  {Janka} H.~T.,   {M{\"u}ller} E.,  2003, \mn@doi
  [\aap] {10.1051/0004-6361:20030863}, \href
  {https://ui.adsabs.harvard.edu/abs/2003A&A...408..621K} {408, 621}

\bibitem[\protect\citeauthoryear{{Kifonidis}, {Plewa}, {Scheck}, {Janka}  \&
  {M{\"u}ller}}{{Kifonidis} et~al.}{2006}]{Kifonidis_2006_mixing}
{Kifonidis} K.,  {Plewa} T.,  {Scheck} L.,  {Janka} H.~T.,   {M{\"u}ller} E.,
  2006, \mn@doi [\aap] {10.1051/0004-6361:20054512}, \href
  {https://ui.adsabs.harvard.edu/abs/2006A&A...453..661K} {453, 661}

\bibitem[\protect\citeauthoryear{{Kozma} \& {Fransson}}{{Kozma} \&
  {Fransson}}{1998}]{Kozma1998}
{Kozma} C.,  {Fransson} C.,  1998, \mn@doi [\apj] {10.1086/305409}, \href
  {https://ui.adsabs.harvard.edu/abs/1998ApJ...496..946K} {496, 946}

\bibitem[\protect\citeauthoryear{{Kromer} \& {Sim}}{{Kromer} \&
  {Sim}}{2009}]{Kromer2009}
{Kromer} M.,  {Sim} S.~A.,  2009, \mn@doi [\mnras]
  {10.1111/j.1365-2966.2009.15256.x}, \href
  {https://ui.adsabs.harvard.edu/abs/2009MNRAS.398.1809K} {398, 1809}

\bibitem[\protect\citeauthoryear{{Kuncarayakti} et~al.,}{{Kuncarayakti}
  et~al.}{2015}]{iPTF13bvn_AA}
{Kuncarayakti} H.,  et~al., 2015, \mn@doi [\aap] {10.1051/0004-6361/201425604},
  \href {https://ui.adsabs.harvard.edu/abs/2015A&A...579A..95K} {579, A95}

\bibitem[\protect\citeauthoryear{{Leonard} \& {Filippenko}}{{Leonard} \&
  {Filippenko}}{2005}]{2003gf}
{Leonard} D.~C.,  {Filippenko} A.~V.,  2005, in {Turatto} M.,  {Benetti} S.,
  {Zampieri} L.,   {Shea} W.,  eds,  Astronomical Society of the Pacific
  Conference Series Vol. 342, 1604-2004: Supernovae as Cosmological
  Lighthouses. p.~330 (\mn@eprint {arXiv} {astro-ph/0409518}),
  \mn@doi{10.48550/arXiv.astro-ph/0409518}

\bibitem[\protect\citeauthoryear{{Li} \& {McCray}}{{Li} \&
  {McCray}}{1996}]{Li1996}
{Li} H.,  {McCray} R.,  1996, \mn@doi [\apj] {10.1086/176659}, \href
  {https://ui.adsabs.harvard.edu/abs/1996ApJ...456..370L} {456, 370}

\bibitem[\protect\citeauthoryear{{Limongi}, {Straniero}  \&
  {Chieffi}}{{Limongi} et~al.}{2000}]{Limongi2000}
{Limongi} M.,  {Straniero} O.,   {Chieffi} A.,  2000, \mn@doi [\apjs]
  {10.1086/313424}, \href
  {https://ui.adsabs.harvard.edu/abs/2000ApJS..129..625L} {129, 625}

\bibitem[\protect\citeauthoryear{{Liu}, {Modjaz}, {Bianco}  \& {Graur}}{{Liu}
  et~al.}{2016}]{Liu_2016_IbIIb}
{Liu} Y.-Q.,  {Modjaz} M.,  {Bianco} F.~B.,   {Graur} O.,  2016, \mn@doi [\apj]
  {10.3847/0004-637X/827/2/90}, \href
  {https://ui.adsabs.harvard.edu/abs/2016ApJ...827...90L} {827, 90}

\bibitem[\protect\citeauthoryear{{Maeda} et~al.,}{{Maeda}
  et~al.}{2008}]{2004dk_b}
{Maeda} K.,  et~al., 2008, \mn@doi [Science] {10.1126/science.1149437}, \href
  {https://ui.adsabs.harvard.edu/abs/2008Sci...319.1220M} {319, 1220}

\bibitem[\protect\citeauthoryear{{Maeda} et~al.,}{{Maeda}
  et~al.}{2015}]{Maeda_2015_2013df}
{Maeda} K.,  et~al., 2015, \mn@doi [\apj] {10.1088/0004-637X/807/1/35}, \href
  {https://ui.adsabs.harvard.edu/abs/2015ApJ...807...35M} {807, 35}

\bibitem[\protect\citeauthoryear{{Maguire} et~al.,}{{Maguire}
  et~al.}{2010}]{Maguire_2010_LCvariety}
{Maguire} K.,  et~al., 2010, \mn@doi [\mnras]
  {10.1111/j.1365-2966.2010.16332.x}, \href
  {https://ui.adsabs.harvard.edu/abs/2010MNRAS.404..981M} {404, 981}

\bibitem[\protect\citeauthoryear{{Matheson} et~al.,}{{Matheson}
  et~al.}{2000a}]{Matheson_2000_1993Jspectra}
{Matheson} T.,  et~al., 2000a, \mn@doi [\aj] {10.1086/301518}, \href
  {https://ui.adsabs.harvard.edu/abs/2000AJ....120.1487M} {120, 1487}

\bibitem[\protect\citeauthoryear{{Matheson}, {Filippenko}, {Ho}, {Barth}  \&
  {Leonard}}{{Matheson} et~al.}{2000b}]{Matheson_2000_1993Jboxy}
{Matheson} T.,  {Filippenko} A.~V.,  {Ho} L.~C.,  {Barth} A.~J.,   {Leonard}
  D.~C.,  2000b, \mn@doi [\aj] {10.1086/301519}, \href
  {https://ui.adsabs.harvard.edu/abs/2000AJ....120.1499M} {120, 1499}

\bibitem[\protect\citeauthoryear{{Matheson}, {Filippenko}, {Ho}, {Barth}  \&
  {Leonard}}{{Matheson} et~al.}{2000c}]{1993J_E}
{Matheson} T.,  {Filippenko} A.~V.,  {Ho} L.~C.,  {Barth} A.~J.,   {Leonard}
  D.~C.,  2000c, \mn@doi [\aj] {10.1086/301519}, \href
  {https://ui.adsabs.harvard.edu/abs/2000AJ....120.1499M} {120, 1499}

\bibitem[\protect\citeauthoryear{{Matheson}, {Filippenko}, {Li}, {Leonard}  \&
  {Shields}}{{Matheson} et~al.}{2001}]{1997dq_b}
{Matheson} T.,  {Filippenko} A.~V.,  {Li} W.,  {Leonard} D.~C.,   {Shields}
  J.~C.,  2001, \mn@doi [\aj] {10.1086/319390}, \href
  {https://ui.adsabs.harvard.edu/abs/2001AJ....121.1648M} {121, 1648}

\bibitem[\protect\citeauthoryear{{Maund}, {Smartt}, {Kudritzki},
  {Podsiadlowski}  \& {Gilmore}}{{Maund} et~al.}{2004a}]{Maund_2004_1993J}
{Maund} J.~R.,  {Smartt} S.~J.,  {Kudritzki} R.~P.,  {Podsiadlowski} P.,
  {Gilmore} G.~F.,  2004a, \mn@doi [\nat] {10.1038/nature02161}, \href
  {https://ui.adsabs.harvard.edu/abs/2004Natur.427..129M} {427, 129}

\bibitem[\protect\citeauthoryear{{Maund}, {Smartt}, {Kudritzki},
  {Podsiadlowski}  \& {Gilmore}}{{Maund} et~al.}{2004b}]{1993J_C}
{Maund} J.~R.,  {Smartt} S.~J.,  {Kudritzki} R.~P.,  {Podsiadlowski} P.,
  {Gilmore} G.~F.,  2004b, \mn@doi [\nat] {10.1038/nature02161}, \href
  {https://ui.adsabs.harvard.edu/abs/2004Natur.427..129M} {427, 129}

\bibitem[\protect\citeauthoryear{{Maund} et~al.,}{{Maund}
  et~al.}{2011}]{2011dh_AA}
{Maund} J.~R.,  et~al., 2011, \mn@doi [\apjl] {10.1088/2041-8205/739/2/L37},
  \href {https://ui.adsabs.harvard.edu/abs/2011ApJ...739L..37M} {739, L37}

\bibitem[\protect\citeauthoryear{{Mazzali}, {Deng}, {Maeda}, {Nomoto},
  {Filippenko}  \& {Matheson}}{{Mazzali} et~al.}{2004}]{1997dq_a}
{Mazzali} P.~A.,  {Deng} J.,  {Maeda} K.,  {Nomoto} K.,  {Filippenko} A.~V.,
  {Matheson} T.,  2004, \mn@doi [\apj] {10.1086/423888}, \href
  {https://ui.adsabs.harvard.edu/abs/2004ApJ...614..858M} {614, 858}

\bibitem[\protect\citeauthoryear{{Mazzali} et~al.,}{{Mazzali}
  et~al.}{2008}]{2008D_a}
{Mazzali} P.~A.,  et~al., 2008, \mn@doi [Science] {10.1126/science.1158088},
  \href {https://ui.adsabs.harvard.edu/abs/2008Sci...321.1185M} {321, 1185}

\bibitem[\protect\citeauthoryear{{Mazzali}, {Maurer}, {Valenti}, {Kotak}  \&
  {Hunter}}{{Mazzali} et~al.}{2010}]{2007gr_b}
{Mazzali} P.~A.,  {Maurer} I.,  {Valenti} S.,  {Kotak} R.,   {Hunter} D.,
  2010, \mn@doi [\mnras] {10.1111/j.1365-2966.2010.17133.x}, \href
  {https://ui.adsabs.harvard.edu/abs/2010MNRAS.408...87M} {408, 87}

\bibitem[\protect\citeauthoryear{{McClelland} \& {Eldridge}}{{McClelland} \&
  {Eldridge}}{2016}]{McClelland_2016_WRMassloss}
{McClelland} L.~A.~S.,  {Eldridge} J.~J.,  2016, \mn@doi [\mnras]
  {10.1093/mnras/stw618}, \href
  {https://ui.adsabs.harvard.edu/abs/2016MNRAS.459.1505M} {459, 1505}

\bibitem[\protect\citeauthoryear{{Medler} et~al.,}{{Medler}
  et~al.}{2022}]{2020acat_a}
{Medler} K.,  et~al., 2022, \mn@doi [\mnras] {10.1093/mnras/stac1192}, \href
  {https://ui.adsabs.harvard.edu/abs/2022MNRAS.513.5540M} {513, 5540}

\bibitem[\protect\citeauthoryear{{Milisavljevic}, {Fesen}, {Gerardy},
  {Kirshner}  \& {Challis}}{{Milisavljevic} et~al.}{2010}]{2008bo_b}
{Milisavljevic} D.,  {Fesen} R.~A.,  {Gerardy} C.~L.,  {Kirshner} R.~P.,
  {Challis} P.,  2010, \mn@doi [\apj] {10.1088/0004-637X/709/2/1343}, \href
  {https://ui.adsabs.harvard.edu/abs/2010ApJ...709.1343M} {709, 1343}

\bibitem[\protect\citeauthoryear{{Milisavljevic} et~al.,}{{Milisavljevic}
  et~al.}{2013a}]{2011ei}
{Milisavljevic} D.,  et~al., 2013a, \mn@doi [\apj]
  {10.1088/0004-637X/767/1/71}, \href
  {https://ui.adsabs.harvard.edu/abs/2013ApJ...767...71M} {767, 71}

\bibitem[\protect\citeauthoryear{{Milisavljevic} et~al.,}{{Milisavljevic}
  et~al.}{2013b}]{2012au}
{Milisavljevic} D.,  et~al., 2013b, \mn@doi [\apjl]
  {10.1088/2041-8205/770/2/L38}, \href
  {https://ui.adsabs.harvard.edu/abs/2013ApJ...770L..38M} {770, L38}

\bibitem[\protect\citeauthoryear{{Modjaz}, {Kirshner}, {Blondin}, {Challis}  \&
  {Matheson}}{{Modjaz} et~al.}{2008}]{2004ao_b_2004gq_b}
{Modjaz} M.,  {Kirshner} R.~P.,  {Blondin} S.,  {Challis} P.,   {Matheson} T.,
  2008, \mn@doi [\apjl] {10.1086/593135}, \href
  {https://ui.adsabs.harvard.edu/abs/2008ApJ...687L...9M} {687, L9}

\bibitem[\protect\citeauthoryear{{Modjaz} et~al.,}{{Modjaz}
  et~al.}{2009}]{2008D_b}
{Modjaz} M.,  et~al., 2009, \mn@doi [\apj] {10.1088/0004-637X/702/1/226}, \href
  {https://ui.adsabs.harvard.edu/abs/2009ApJ...702..226M} {702, 226}

\bibitem[\protect\citeauthoryear{{Modjaz} et~al.,}{{Modjaz}
  et~al.}{2014}]{2006ld_b_2007C_b_2007I_b_2008aq_b}
{Modjaz} M.,  et~al., 2014, \mn@doi [\aj] {10.1088/0004-6256/147/5/99}, \href
  {https://ui.adsabs.harvard.edu/abs/2014AJ....147...99M} {147, 99}

\bibitem[\protect\citeauthoryear{{Modjaz}, {Liu}, {Bianco}  \&
  {Graur}}{{Modjaz} et~al.}{2016}]{PTF12gzk_b}
{Modjaz} M.,  {Liu} Y.~Q.,  {Bianco} F.~B.,   {Graur} O.,  2016, \mn@doi [\apj]
  {10.3847/0004-637X/832/2/108}, \href
  {https://ui.adsabs.harvard.edu/abs/2016ApJ...832..108M} {832, 108}

\bibitem[\protect\citeauthoryear{{Moore}}{{Moore}}{1993}]{Moore_1993_NIIlevels}
{Moore} C.~E.,  1993, {Tables of Spectra of Hydrogen, Carbon, Nitrogen, and
  Oxygen Atoms and Ions}

\bibitem[\protect\citeauthoryear{{Morales-Garoffolo}
  et~al.,}{{Morales-Garoffolo} et~al.}{2014}]{2013df}
{Morales-Garoffolo} A.,  et~al., 2014, \mn@doi [\mnras]
  {10.1093/mnras/stu1837}, \href
  {https://ui.adsabs.harvard.edu/abs/2014MNRAS.445.1647M} {445, 1647}

\bibitem[\protect\citeauthoryear{{Murphy}, {Jennings}, {Williams}, {Dalcanton}
  \& {Dolphin}}{{Murphy} et~al.}{2011}]{2011dh_E}
{Murphy} J.~W.,  {Jennings} Z.~G.,  {Williams} B.,  {Dalcanton} J.~J.,
  {Dolphin} A.~E.,  2011, \mn@doi [\apjl] {10.1088/2041-8205/742/1/L4}, \href
  {https://ui.adsabs.harvard.edu/abs/2011ApJ...742L...4M} {742, L4}

\bibitem[\protect\citeauthoryear{{Nakamura}, {Mazzali}, {Nomoto}  \&
  {Iwamoto}}{{Nakamura} et~al.}{2001}]{1998bw_B}
{Nakamura} T.,  {Mazzali} P.~A.,  {Nomoto} K.,   {Iwamoto} K.,  2001, \mn@doi
  [\apj] {10.1086/319784}, \href
  {https://ui.adsabs.harvard.edu/abs/2001ApJ...550..991N} {550, 991}

\bibitem[\protect\citeauthoryear{{Nomoto} \& {Hashimoto}}{{Nomoto} \&
  {Hashimoto}}{1988}]{Nomoto_1988_Hestars}
{Nomoto} K.,  {Hashimoto} M.,  1988, \mn@doi [\physrep]
  {10.1016/0370-1573(88)90032-4}, \href
  {https://ui.adsabs.harvard.edu/abs/1988PhR...163...13N} {163, 13}

\bibitem[\protect\citeauthoryear{{Nomoto}, {Suzuki}, {Shigeyama}, {Kumagai},
  {Yamaoka}  \& {Saio}}{{Nomoto} et~al.}{1993}]{Nomoto_1993_IIb}
{Nomoto} K.,  {Suzuki} T.,  {Shigeyama} T.,  {Kumagai} S.,  {Yamaoka} H.,
  {Saio} H.,  1993, \mn@doi [\nat] {10.1038/364507a0}, \href
  {https://ui.adsabs.harvard.edu/abs/1993Natur.364..507N} {364, 507}

\bibitem[\protect\citeauthoryear{{Nomoto}, {Iwamoto}  \& {Suzuki}}{{Nomoto}
  et~al.}{1995}]{Nomoto_1995_Mixing}
{Nomoto} K.~I.,  {Iwamoto} K.,   {Suzuki} T.,  1995, \mn@doi [\physrep]
  {10.1016/0370-1573(94)00107-E}, \href
  {https://ui.adsabs.harvard.edu/abs/1995PhR...256..173N} {256, 173}

\bibitem[\protect\citeauthoryear{{Nomoto}, {Kobayashi}  \& {Tominaga}}{{Nomoto}
  et~al.}{2013}]{Nomoto_2013_O-amount}
{Nomoto} K.,  {Kobayashi} C.,   {Tominaga} N.,  2013, \mn@doi [\araa]
  {10.1146/annurev-astro-082812-140956}, \href
  {https://ui.adsabs.harvard.edu/abs/2013ARA&A..51..457N} {51, 457}

\bibitem[\protect\citeauthoryear{{Pastorello} et~al.,}{{Pastorello}
  et~al.}{2008}]{2008ax_a}
{Pastorello} A.,  et~al., 2008, \mn@doi [\mnras]
  {10.1111/j.1365-2966.2008.13618.x}, \href
  {https://ui.adsabs.harvard.edu/abs/2008MNRAS.389..955P} {389, 955}

\bibitem[\protect\citeauthoryear{{Patat} et~al.,}{{Patat}
  et~al.}{2001}]{1996aq_1998bw_b}
{Patat} F.,  et~al., 2001, \mn@doi [\apj] {10.1086/321526}, \href
  {https://ui.adsabs.harvard.edu/abs/2001ApJ...555..900P} {555, 900}

\bibitem[\protect\citeauthoryear{{Podsiadlowski}, {Joss}  \&
  {Hsu}}{{Podsiadlowski} et~al.}{1992}]{Podsiadlowski_1992_Binaries}
{Podsiadlowski} P.,  {Joss} P.~C.,   {Hsu} J.~J.~L.,  1992, \mn@doi [\apj]
  {10.1086/171341}, \href
  {https://ui.adsabs.harvard.edu/abs/1992ApJ...391..246P} {391, 246}

\bibitem[\protect\citeauthoryear{{Podsiadlowski}, {Hsu}, {Joss}  \&
  {Ross}}{{Podsiadlowski} et~al.}{1993}]{Podsiadlowski_1993_IIb}
{Podsiadlowski} P.,  {Hsu} J.~J.~L.,  {Joss} P.~C.,   {Ross} R.~R.,  1993,
  \mn@doi [\nat] {10.1038/364509a0}, \href
  {https://ui.adsabs.harvard.edu/abs/1993Natur.364..509P} {364, 509}

\bibitem[\protect\citeauthoryear{{Prentice} et~al.,}{{Prentice}
  et~al.}{2016}]{Prentice_2016_biggersample}
{Prentice} S.~J.,  et~al., 2016, \mn@doi [\mnras] {10.1093/mnras/stw299}, \href
  {https://ui.adsabs.harvard.edu/abs/2016MNRAS.458.2973P} {458, 2973}

\bibitem[\protect\citeauthoryear{{Prentice} et~al.,}{{Prentice}
  et~al.}{2019}]{Prentice_2019_Bigsample_2015ah}
{Prentice} S.~J.,  et~al., 2019, \mn@doi [\mnras] {10.1093/mnras/sty3399},
  \href {https://ui.adsabs.harvard.edu/abs/2019MNRAS.485.1559P} {485, 1559}

\bibitem[\protect\citeauthoryear{{Prentice}, {Maguire}, {Siebenaler}  \&
  {Jerkstrand}}{{Prentice} et~al.}{2022}]{2019yz}
{Prentice} S.~J.,  {Maguire} K.,  {Siebenaler} L.,   {Jerkstrand} A.,  2022,
  \mn@doi [\mnras] {10.1093/mnras/stac1657}, \href
  {https://ui.adsabs.harvard.edu/abs/2022MNRAS.514.5686P} {514, 5686}

\bibitem[\protect\citeauthoryear{{Qiu}, {Li}, {Qiao}  \& {Hu}}{{Qiu}
  et~al.}{1999}]{1996cb}
{Qiu} Y.,  {Li} W.,  {Qiao} Q.,   {Hu} J.,  1999, \mn@doi [\aj]
  {10.1086/300731}, \href
  {https://ui.adsabs.harvard.edu/abs/1999AJ....117..736Q} {117, 736}

\bibitem[\protect\citeauthoryear{{Roming} et~al.,}{{Roming}
  et~al.}{2009}]{2008ax_AA}
{Roming} P.~W.~A.,  et~al., 2009, \mn@doi [\apjl]
  {10.1088/0004-637X/704/2/L118}, \href
  {https://ui.adsabs.harvard.edu/abs/2009ApJ...704L.118R} {704, L118}

\bibitem[\protect\citeauthoryear{{Ryder}, {Sadler}, {Subrahmanyan}, {Weiler},
  {Panagia}  \& {Stockdale}}{{Ryder} et~al.}{2004}]{2001ig_a}
{Ryder} S.~D.,  {Sadler} E.~M.,  {Subrahmanyan} R.,  {Weiler} K.~W.,  {Panagia}
  N.,   {Stockdale} C.,  2004, \mn@doi [\mnras]
  {10.1111/j.1365-2966.2004.07589.x}, \href
  {https://ui.adsabs.harvard.edu/abs/2004MNRAS.349.1093R} {349, 1093}

\bibitem[\protect\citeauthoryear{{Sana} et~al.,}{{Sana}
  et~al.}{2012}]{Sana_2012_Binaries}
{Sana} H.,  et~al., 2012, \mn@doi [Science] {10.1126/science.1223344}, \href
  {https://ui.adsabs.harvard.edu/abs/2012Sci...337..444S} {337, 444}

\bibitem[\protect\citeauthoryear{{Sauer}, {Mazzali}, {Deng}, {Valenti},
  {Nomoto}  \& {Filippenko}}{{Sauer} et~al.}{2006}]{Sauer_2006_HiddenHelium}
{Sauer} D.~N.,  {Mazzali} P.~A.,  {Deng} J.,  {Valenti} S.,  {Nomoto} K.,
  {Filippenko} A.~V.,  2006, \mn@doi [\mnras]
  {10.1111/j.1365-2966.2006.10438.x}, \href
  {https://ui.adsabs.harvard.edu/abs/2006MNRAS.369.1939S} {369, 1939}

\bibitem[\protect\citeauthoryear{{Schmidt} et~al.,}{{Schmidt}
  et~al.}{1993}]{1993J_a}
{Schmidt} B.~P.,  et~al., 1993, \mn@doi [\nat] {10.1038/364600a0}, \href
  {https://ui.adsabs.harvard.edu/abs/1993Natur.364..600S} {364, 600}

\bibitem[\protect\citeauthoryear{{Schweyer} et~al.,}{{Schweyer}
  et~al.}{2023}]{2019odp}
{Schweyer} T.,  et~al., 2023, preprint, \href
  {https://ui.adsabs.harvard.edu/abs/2023arXiv230314146S} {p. arXiv:2303.14146}
  (\mn@eprint {arXiv} {2303.14146})

\bibitem[\protect\citeauthoryear{{Shahbandeh} et~al.,}{{Shahbandeh}
  et~al.}{2022}]{14az_a}
{Shahbandeh} M.,  et~al., 2022, \mn@doi [\apj] {10.3847/1538-4357/ac4030},
  \href {https://ui.adsabs.harvard.edu/abs/2022ApJ...925..175S} {925, 175}

\bibitem[\protect\citeauthoryear{{Shigeyama}, {Nomoto}, {Tsujimoto}  \&
  {Hashimoto}}{{Shigeyama} et~al.}{1990}]{Shigeyama_1990_Ib_Mixing}
{Shigeyama} T.,  {Nomoto} K.,  {Tsujimoto} T.,   {Hashimoto} M.-A.,  1990,
  \mn@doi [\apjl] {10.1086/185818}, \href
  {https://ui.adsabs.harvard.edu/abs/1990ApJ...361L..23S} {361, L23}

\bibitem[\protect\citeauthoryear{{Shigeyama}, {Nomoto}, {Yamaoka}  \&
  {Thielemann}}{{Shigeyama} et~al.}{1992}]{Shigeyama_1992_Lagrangian}
{Shigeyama} T.,  {Nomoto} K.,  {Yamaoka} H.,   {Thielemann} F.-K.,  1992,
  \mn@doi [\apjl] {10.1086/186281}, \href
  {https://ui.adsabs.harvard.edu/abs/1992ApJ...386L..13S} {386, L13}

\bibitem[\protect\citeauthoryear{{Shigeyama}, {Suzuki}, {Kumagai}, {Nomoto},
  {Saio}  \& {Yamaoka}}{{Shigeyama} et~al.}{1994}]{1993J_AA}
{Shigeyama} T.,  {Suzuki} T.,  {Kumagai} S.,  {Nomoto} K.,  {Saio} H.,
  {Yamaoka} H.,  1994, \mn@doi [\apj] {10.1086/173564}, \href
  {https://ui.adsabs.harvard.edu/abs/1994ApJ...420..341S} {420, 341}

\bibitem[\protect\citeauthoryear{{Shingles} et~al.,}{{Shingles}
  et~al.}{2020}]{Shingles2020}
{Shingles} L.~J.,  et~al., 2020, \mn@doi [\mnras] {10.1093/mnras/stz3412},
  \href {https://ui.adsabs.harvard.edu/abs/2020MNRAS.492.2029S} {492, 2029}

\bibitem[\protect\citeauthoryear{{Shivvers} et~al.,}{{Shivvers}
  et~al.}{2019}]{14az_b}
{Shivvers} I.,  et~al., 2019, \mn@doi [\mnras] {10.1093/mnras/sty2719}, \href
  {https://ui.adsabs.harvard.edu/abs/2019MNRAS.482.1545S} {482, 1545}

\bibitem[\protect\citeauthoryear{{Silverman}, {Mazzali}, {Chornock},
  {Filippenko}, {Clocchiatti}, {Phillips}, {Ganeshalingam}  \&
  {Foley}}{{Silverman} et~al.}{2009}]{2001ig_b}
{Silverman} J.~M.,  {Mazzali} P.,  {Chornock} R.,  {Filippenko} A.~V.,
  {Clocchiatti} A.,  {Phillips} M.~M.,  {Ganeshalingam} M.,   {Foley} R.~J.,
  2009, \mn@doi [\pasp] {10.1086/603653}, \href
  {https://ui.adsabs.harvard.edu/abs/2009PASP..121..689S} {121, 689}

\bibitem[\protect\citeauthoryear{{Singh} et~al.,}{{Singh}
  et~al.}{2019}]{J1204_b}
{Singh} M.,  et~al., 2019, \mn@doi [\mnras] {10.1093/mnras/stz752}, \href
  {https://ui.adsabs.harvard.edu/abs/2019MNRAS.485.5438S} {485, 5438}

\bibitem[\protect\citeauthoryear{{Smartt}}{{Smartt}}{2009}]{Smartt_2009_Progenitors}
{Smartt} S.~J.,  2009, \mn@doi [\araa] {10.1146/annurev-astro-082708-101737},
  \href {https://ui.adsabs.harvard.edu/abs/2009ARA&A..47...63S} {47, 63}

\bibitem[\protect\citeauthoryear{{Smartt}}{{Smartt}}{2015}]{Smartt2015}
{Smartt} S.~J.,  2015, \mn@doi [\pasa] {10.1017/pasa.2015.17}, \href
  {https://ui.adsabs.harvard.edu/abs/2015PASA...32...16S} {32, e016}

\bibitem[\protect\citeauthoryear{{Smith}}{{Smith}}{2014}]{Smith_2014_MassLoss}
{Smith} N.,  2014, \mn@doi [\araa] {10.1146/annurev-astro-081913-040025}, \href
  {https://ui.adsabs.harvard.edu/abs/2014ARA&A..52..487S} {52, 487}

\bibitem[\protect\citeauthoryear{{Smith}, {Li}, {Filippenko}  \&
  {Chornock}}{{Smith} et~al.}{2011}]{Smith_2011_BinarymoreMasstransfer}
{Smith} N.,  {Li} W.,  {Filippenko} A.~V.,   {Chornock} R.,  2011, \mn@doi
  [\mnras] {10.1111/j.1365-2966.2011.17229.x}, \href
  {https://ui.adsabs.harvard.edu/abs/2011MNRAS.412.1522S} {412, 1522}

\bibitem[\protect\citeauthoryear{{Soderberg}, {Chevalier}, {Kulkarni}  \&
  {Frail}}{{Soderberg} et~al.}{2006}]{Söderberg_2006_2003bg}
{Soderberg} A.~M.,  {Chevalier} R.~A.,  {Kulkarni} S.~R.,   {Frail} D.~A.,
  2006, \mn@doi [\apj] {10.1086/507571}, \href
  {https://ui.adsabs.harvard.edu/abs/2006ApJ...651.1005S} {651, 1005}

\bibitem[\protect\citeauthoryear{{Srivastav}, {Anupama}  \& {Sahu}}{{Srivastav}
  et~al.}{2014}]{iPTF13bvn_B}
{Srivastav} S.,  {Anupama} G.~C.,   {Sahu} D.~K.,  2014, \mn@doi [\mnras]
  {10.1093/mnras/stu1878}, \href
  {https://ui.adsabs.harvard.edu/abs/2014MNRAS.445.1932S} {445, 1932}

\bibitem[\protect\citeauthoryear{{Stevance} et~al.,}{{Stevance}
  et~al.}{2016}]{2008aq_a}
{Stevance} H.~F.,  et~al., 2016, \mn@doi [\mnras] {10.1093/mnras/stw1479},
  \href {https://ui.adsabs.harvard.edu/abs/2016MNRAS.461.2019S} {461, 2019}

\bibitem[\protect\citeauthoryear{{Stritzinger} et~al.,}{{Stritzinger}
  et~al.}{2009}]{2007Y}
{Stritzinger} M.,  et~al., 2009, \mn@doi [\apj] {10.1088/0004-637X/696/1/713},
  \href {https://ui.adsabs.harvard.edu/abs/2009ApJ...696..713S} {696, 713}

\bibitem[\protect\citeauthoryear{{Stritzinger} et~al.,}{{Stritzinger}
  et~al.}{2023}]{2009K_b}
{Stritzinger} M.~D.,  et~al., 2023, \mn@doi [\aap]
  {10.1051/0004-6361/202243376}, \href
  {https://ui.adsabs.harvard.edu/abs/2023A&A...675A..82S} {675, A82}

\bibitem[\protect\citeauthoryear{{Sukhbold}, {Ertl}, {Woosley}, {Brown}  \&
  {Janka}}{{Sukhbold} et~al.}{2016}]{Sukhbold_2016_explosions}
{Sukhbold} T.,  {Ertl} T.,  {Woosley} S.~E.,  {Brown} J.~M.,   {Janka} H.~T.,
  2016, \mn@doi [\apj] {10.3847/0004-637X/821/1/38}, \href
  {https://ui.adsabs.harvard.edu/abs/2016ApJ...821...38S} {821, 38}

\bibitem[\protect\citeauthoryear{{Tachiev} \& {Froese Fischer}}{{Tachiev} \&
  {Froese Fischer}}{2001}]{Tachiev_2001_tp}
{Tachiev} G.,  {Froese Fischer} C.,  2001, \mn@doi [Canadian Journal of
  Physics] {10.1139/p01-059}, \href
  {https://ui.adsabs.harvard.edu/abs/2001CaJPh..79..955T} {79, 955}

\bibitem[\protect\citeauthoryear{{Taddia} et~al.,}{{Taddia}
  et~al.}{2015}]{Taddia_2015_Risetime}
{Taddia} F.,  et~al., 2015, \mn@doi [\aap] {10.1051/0004-6361/201423915}, \href
  {https://ui.adsabs.harvard.edu/abs/2015A&A...574A..60T} {574, A60}

\bibitem[\protect\citeauthoryear{{Taddia} et~al.,}{{Taddia}
  et~al.}{2016}]{2015fn_a}
{Taddia} F.,  et~al., 2016, \mn@doi [\aap] {10.1051/0004-6361/201628703}, \href
  {https://ui.adsabs.harvard.edu/abs/2016A&A...592A..89T} {592, A89}

\bibitem[\protect\citeauthoryear{{Taddia} et~al.,}{{Taddia}
  et~al.}{2018}]{Taddia_2018_Nimasses_2009K_a}
{Taddia} F.,  et~al., 2018, \mn@doi [\aap] {10.1051/0004-6361/201730844}, \href
  {https://ui.adsabs.harvard.edu/abs/2018A&A...609A.136T} {609, A136}

\bibitem[\protect\citeauthoryear{{Taddia}, {Sollerman}, {Fremling},
  {Karamehmetoglu}, {Barbarino}, {Lunnan}, {West}  \& {Gal-Yam}}{{Taddia}
  et~al.}{2019}]{2015fn_b}
{Taddia} F.,  {Sollerman} J.,  {Fremling} C.,  {Karamehmetoglu} E.,
  {Barbarino} C.,  {Lunnan} R.,  {West} S.,   {Gal-Yam} A.,  2019, \mn@doi
  [\aap] {10.1051/0004-6361/201833688}, \href
  {https://ui.adsabs.harvard.edu/abs/2019A&A...621A..64T} {621, A64}

\bibitem[\protect\citeauthoryear{{Takaki} et~al.,}{{Takaki}
  et~al.}{2013}]{Takaki_2013_2012au}
{Takaki} K.,  et~al., 2013, \mn@doi [\apjl] {10.1088/2041-8205/772/2/L17},
  \href {https://ui.adsabs.harvard.edu/abs/2013ApJ...772L..17T} {772, L17}

\bibitem[\protect\citeauthoryear{{Taubenberger} et~al.,}{{Taubenberger}
  et~al.}{2006}]{2004aw}
{Taubenberger} S.,  et~al., 2006, \mn@doi [\mnras]
  {10.1111/j.1365-2966.2006.10776.x}, \href
  {https://ui.adsabs.harvard.edu/abs/2006MNRAS.371.1459T} {371, 1459}

\bibitem[\protect\citeauthoryear{{Taubenberger} et~al.,}{{Taubenberger}
  et~al.}{2009}]{2006ld_a_2007I_a}
{Taubenberger} S.,  et~al., 2009, \mn@doi [\mnras]
  {10.1111/j.1365-2966.2009.15003.x}, \href
  {https://ui.adsabs.harvard.edu/abs/2009MNRAS.397..677T} {397, 677}

\bibitem[\protect\citeauthoryear{{Taubenberger} et~al.,}{{Taubenberger}
  et~al.}{2011}]{2008ax_b}
{Taubenberger} S.,  et~al., 2011, \mn@doi [\mnras]
  {10.1111/j.1365-2966.2011.18287.x}, \href
  {https://ui.adsabs.harvard.edu/abs/2011MNRAS.413.2140T} {413, 2140}

\bibitem[\protect\citeauthoryear{{Thielemann}, {Nomoto}  \&
  {Hashimoto}}{{Thielemann} et~al.}{1996}]{Thielemann_1996_oxygen}
{Thielemann} F.-K.,  {Nomoto} K.,   {Hashimoto} M.-A.,  1996, \mn@doi [\apj]
  {10.1086/176980}, \href
  {https://ui.adsabs.harvard.edu/abs/1996ApJ...460..408T} {460, 408}

\bibitem[\protect\citeauthoryear{{Toro}, {Spruce}  \& {Speares}}{{Toro}
  et~al.}{1994}]{Toro_1994_HLLC}
{Toro} E.~F.,  {Spruce} M.,   {Speares} W.,  1994, \mn@doi [Shock Waves]
  {10.1007/BF01414629}, \href
  {https://ui.adsabs.harvard.edu/abs/1994ShWav...4...25T} {4, 25}

\bibitem[\protect\citeauthoryear{{Uomoto} \& {Kirshner}}{{Uomoto} \&
  {Kirshner}}{1985}]{Uomoto_1985_Ib}
{Uomoto} A.,  {Kirshner} R.~P.,  1985, \aap, \href
  {https://ui.adsabs.harvard.edu/abs/1985A&A...149L...7U} {149, L7}

\bibitem[\protect\citeauthoryear{{Valenti} et~al.,}{{Valenti}
  et~al.}{2008}]{2007gr_a}
{Valenti} S.,  et~al., 2008, \mn@doi [\apjl] {10.1086/527672}, \href
  {https://ui.adsabs.harvard.edu/abs/2008ApJ...673L.155V} {673, L155}

\bibitem[\protect\citeauthoryear{{Valenti} et~al.,}{{Valenti}
  et~al.}{2011}]{Valenti_2011_risetime_2009jf}
{Valenti} S.,  et~al., 2011, \mn@doi [\mnras]
  {10.1111/j.1365-2966.2011.19262.x}, \href
  {https://ui.adsabs.harvard.edu/abs/2011MNRAS.416.3138V} {416, 3138}

\bibitem[\protect\citeauthoryear{{Valenti} et~al.,}{{Valenti}
  et~al.}{2012}]{2011bm}
{Valenti} S.,  et~al., 2012, \mn@doi [\apjl] {10.1088/2041-8205/749/2/L28},
  \href {https://ui.adsabs.harvard.edu/abs/2012ApJ...749L..28V} {749, L28}

\bibitem[\protect\citeauthoryear{{Van Dyk} et~al.,}{{Van Dyk}
  et~al.}{2014}]{vanDyk_2014_2013df}
{Van Dyk} S.~D.,  et~al., 2014, \mn@doi [\aj] {10.1088/0004-6256/147/2/37},
  \href {https://ui.adsabs.harvard.edu/abs/2014AJ....147...37V} {147, 37}

\bibitem[\protect\citeauthoryear{{Vink}}{{Vink}}{2017}]{Vink_2017_Massloss}
{Vink} J.~S.,  2017, \mn@doi [\aap] {10.1051/0004-6361/201731902}, \href
  {https://ui.adsabs.harvard.edu/abs/2017A&A...607L...8V} {607, L8}

\bibitem[\protect\citeauthoryear{{Weaver}, {Zimmerman}  \& {Woosley}}{{Weaver}
  et~al.}{1978}]{Weaver_1978_KEPLERa}
{Weaver} T.~A.,  {Zimmerman} G.~B.,   {Woosley} S.~E.,  1978, \mn@doi [\apj]
  {10.1086/156569}, \href
  {https://ui.adsabs.harvard.edu/abs/1978ApJ...225.1021W} {225, 1021}

\bibitem[\protect\citeauthoryear{{Wheeler} \& {Harkness}}{{Wheeler} \&
  {Harkness}}{1990}]{Wheeler_1990_LCvariety}
{Wheeler} J.~C.,  {Harkness} R.~P.,  1990, \mn@doi [Reports on Progress in
  Physics] {10.1088/0034-4885/53/12/001}, \href
  {https://ui.adsabs.harvard.edu/abs/1990RPPh...53.1467W} {53, 1467}

\bibitem[\protect\citeauthoryear{{Wheeler} \& {Levreault}}{{Wheeler} \&
  {Levreault}}{1985}]{Wheeler_1985_Ib}
{Wheeler} J.~C.,  {Levreault} R.,  1985, \mn@doi [\apjl] {10.1086/184500},
  \href {https://ui.adsabs.harvard.edu/abs/1985ApJ...294L..17W} {294, L17}

\bibitem[\protect\citeauthoryear{{Wheeler}, {Harkness}, {Barker}, {Cochran}  \&
  {Wills}}{{Wheeler} et~al.}{1987}]{Wheeler_1987_Ic}
{Wheeler} J.~C.,  {Harkness} R.~P.,  {Barker} E.~S.,  {Cochran} A.~L.,
  {Wills} D.,  1987, \mn@doi [\apjl] {10.1086/184833}, \href
  {https://ui.adsabs.harvard.edu/abs/1987ApJ...313L..69W} {313, L69}

\bibitem[\protect\citeauthoryear{{Williamson}, {Kerzendorf}  \&
  {Modjaz}}{{Williamson} et~al.}{2021}]{Williamson_2021_HiddenHe}
{Williamson} M.,  {Kerzendorf} W.,   {Modjaz} M.,  2021, \mn@doi [\apj]
  {10.3847/1538-4357/abd244}, \href
  {https://ui.adsabs.harvard.edu/abs/2021ApJ...908..150W} {908, 150}

\bibitem[\protect\citeauthoryear{{Wongwathanarat}, {Janka}  \&
  {M{\"u}ller}}{{Wongwathanarat}
  et~al.}{2013}]{Wongwathanarat_2013_multidhydro_a}
{Wongwathanarat} A.,  {Janka} H.~T.,   {M{\"u}ller} E.,  2013, \mn@doi [\aap]
  {10.1051/0004-6361/201220636}, \href
  {https://ui.adsabs.harvard.edu/abs/2013A&A...552A.126W} {552, A126}

\bibitem[\protect\citeauthoryear{{Wongwathanarat}, {M{\"u}ller}  \&
  {Janka}}{{Wongwathanarat} et~al.}{2015}]{Wongwathanarat_2015_multidhydrob}
{Wongwathanarat} A.,  {M{\"u}ller} E.,   {Janka} H.~T.,  2015, \mn@doi [\aap]
  {10.1051/0004-6361/201425025}, \href
  {https://ui.adsabs.harvard.edu/abs/2015A&A...577A..48W} {577, A48}

\bibitem[\protect\citeauthoryear{{Woosley}}{{Woosley}}{2019}]{Woosley_2019_models}
{Woosley} S.~E.,  2019, \mn@doi [\apj] {10.3847/1538-4357/ab1b41}, \href
  {https://ui.adsabs.harvard.edu/abs/2019ApJ...878...49W} {878, 49}

\bibitem[\protect\citeauthoryear{{Woosley} \& {Heger}}{{Woosley} \&
  {Heger}}{2015}]{Woosley_2015_Initcomp}
{Woosley} S.~E.,  {Heger} A.,  2015, \mn@doi [\apj]
  {10.1088/0004-637X/810/1/34}, \href
  {https://ui.adsabs.harvard.edu/abs/2015ApJ...810...34W} {810, 34}

\bibitem[\protect\citeauthoryear{{Woosley} \& {Weaver}}{{Woosley} \&
  {Weaver}}{1995}]{Woosley1995}
{Woosley} S.~E.,  {Weaver} T.~A.,  1995, \mn@doi [\apjs] {10.1086/192237},
  \href {https://ui.adsabs.harvard.edu/abs/1995ApJS..101..181W} {101, 181}

\bibitem[\protect\citeauthoryear{{Woosley}, {Pinto}, {Martin}  \&
  {Weaver}}{{Woosley} et~al.}{1987}]{Woosley_1987_IIb}
{Woosley} S.~E.,  {Pinto} P.~A.,  {Martin} P.~G.,   {Weaver} T.~A.,  1987,
  \mn@doi [\apj] {10.1086/165402}, \href
  {https://ui.adsabs.harvard.edu/abs/1987ApJ...318..664W} {318, 664}

\bibitem[\protect\citeauthoryear{{Woosley}, {Pinto}  \& {Ensman}}{{Woosley}
  et~al.}{1988}]{Woosley_1988_IIb}
{Woosley} S.~E.,  {Pinto} P.~A.,   {Ensman} L.,  1988, \mn@doi [\apj]
  {10.1086/165908}, \href
  {https://ui.adsabs.harvard.edu/abs/1988ApJ...324..466W} {324, 466}

\bibitem[\protect\citeauthoryear{{Woosley}, {Langer}  \& {Weaver}}{{Woosley}
  et~al.}{1993}]{Woosley_1993_Massloss}
{Woosley} S.~E.,  {Langer} N.,   {Weaver} T.~A.,  1993, \mn@doi [\apj]
  {10.1086/172886}, \href
  {https://ui.adsabs.harvard.edu/abs/1993ApJ...411..823W} {411, 823}

\bibitem[\protect\citeauthoryear{{Woosley}, {Eastman}, {Weaver}  \&
  {Pinto}}{{Woosley} et~al.}{1994}]{Woosley_1994_1993J}
{Woosley} S.~E.,  {Eastman} R.~G.,  {Weaver} T.~A.,   {Pinto} P.~A.,  1994,
  \mn@doi [\apj] {10.1086/174319}, \href
  {https://ui.adsabs.harvard.edu/abs/1994ApJ...429..300W} {429, 300}

\bibitem[\protect\citeauthoryear{{Woosley}, {Langer}  \& {Weaver}}{{Woosley}
  et~al.}{1995}]{Woosley_1995_Binaries}
{Woosley} S.~E.,  {Langer} N.,   {Weaver} T.~A.,  1995, \mn@doi [\apj]
  {10.1086/175963}, \href
  {https://ui.adsabs.harvard.edu/abs/1995ApJ...448..315W} {448, 315}

\bibitem[\protect\citeauthoryear{{Woosley}, {Eastman}  \& {Schmidt}}{{Woosley}
  et~al.}{1999}]{Woosley_1999_1998bw}
{Woosley} S.~E.,  {Eastman} R.~G.,   {Schmidt} B.~P.,  1999, \mn@doi [\apj]
  {10.1086/307131}, \href
  {https://ui.adsabs.harvard.edu/abs/1999ApJ...516..788W} {516, 788}

\bibitem[\protect\citeauthoryear{{Woosley}, {Heger}  \& {Weaver}}{{Woosley}
  et~al.}{2002}]{Woosley_2002_SNcanon}
{Woosley} S.~E.,  {Heger} A.,   {Weaver} T.~A.,  2002, \mn@doi [Reviews of
  Modern Physics] {10.1103/RevModPhys.74.1015}, \href
  {https://ui.adsabs.harvard.edu/abs/2002RvMP...74.1015W} {74, 1015}

\bibitem[\protect\citeauthoryear{{Woosley}, {Sukhbold}  \& {Kasen}}{{Woosley}
  et~al.}{2021}]{Woosley_2021_LC}
{Woosley} S.~E.,  {Sukhbold} T.,   {Kasen} D.~N.,  2021, \mn@doi [\apj]
  {10.3847/1538-4357/abf3be}, \href
  {https://ui.adsabs.harvard.edu/abs/2021ApJ...913..145W} {913, 145}

\bibitem[\protect\citeauthoryear{{Yaron} \& {Gal-Yam}}{{Yaron} \&
  {Gal-Yam}}{2012}]{Yaron_2012_WISEREP}
{Yaron} O.,  {Gal-Yam} A.,  2012, \mn@doi [\pasp] {10.1086/666656}, \href
  {https://ui.adsabs.harvard.edu/abs/2012PASP..124..668Y} {124, 668}

\bibitem[\protect\citeauthoryear{{Yoon}}{{Yoon}}{2017}]{Yoon_2017_WRMassLoss}
{Yoon} S.-C.,  2017, \mn@doi [\mnras] {10.1093/mnras/stx1496}, \href
  {https://ui.adsabs.harvard.edu/abs/2017MNRAS.470.3970Y} {470, 3970}

\bibitem[\protect\citeauthoryear{{Yoon}, {Dessart}  \& {Clocchiatti}}{{Yoon}
  et~al.}{2017}]{Yoon_2017_BinaryRadii}
{Yoon} S.-C.,  {Dessart} L.,   {Clocchiatti} A.,  2017, \mn@doi [\apj]
  {10.3847/1538-4357/aa6afe}, \href
  {https://ui.adsabs.harvard.edu/abs/2017ApJ...840...10Y} {840, 10}

\bibitem[\protect\citeauthoryear{{van Baal}, {Jerkstrand}, {Wongwathanarat}  \&
  {Janka}}{{van Baal} et~al.}{2023}]{vanBaal_2023_EXTRASS}
{van Baal} B. F.~A.,  {Jerkstrand} A.,  {Wongwathanarat} A.,   {Janka} H.-T.,
  2023, \mn@doi [\mnras] {10.1093/mnras/stad1488}, \href
  {https://ui.adsabs.harvard.edu/abs/2023MNRAS.523..954V} {523, 954}

\makeatother
\end{thebibliography}

\appendix

\section{Fitting Procedure}
\label{appendix:fitting_procedure}

As was mentioned in Section \ref{sec:fitting_procedure}, we will here describe the constraints and rules used to determine the \NIIdiag{} diagnostic for SESN spectra in this work. 

\subsection{Allowed shapes}
\label{A:allowed_shapes}

Depending on the morphology of the SN ejecta, the shape of the observed emission line will be different. Examples of some of the possible shapes are shown in \citet[][their figure 2]{Jerkstrand_2017_book}. In the optically thin limit (a condition mostly met in SESNe $\gtrsim$ 150d post explosion), the emission strength ratio of the two components of \OIdoublet{} is 3:1. However, due to the high velocities of the lines and the close proximity of the components in wavelength-space, the resulting line profile is well represented by a single Gaussian. For the fitting of \OIdoublet{}, we thus have three free parameters. 

In some of the early nebular phase observations, typically for higher mass SN progenitors, the \OIdoublet{} is observed to have an extended red tail. This effect can be understood as due to electron scattering, which although coherent in the comoving frame introduces redshifts in the stellar frame. As described in e.g. \citet{Jerkstrand_2017_book}, this effect causes a red tail, a damped peak amplitude, and a blueshift of the peak. For a typical optical depth of $\tau_{e} \sim 1$, this blueshift will be roughly 13\% of the width of the line (e.g. 650 \kms{} for a 5000 \kms{} wide emission line), for a uniform sphere emission and absorption model. As we see this effect in a non-negligible amount of observations, as well as in some of our models, we opted to add  $\tau_{e}$ as an additional free parameter during the fitting. 

When it comes to \NIIdoublet{}, the story is somewhat different. As the emission mainly originates from the He/N envelope, the lower density and larger distance to the core makes the effective $\tau_{e}$ consistent with zero. In fact, for increasing progenitor mass, our models show an increasingly flat topped emission for \NIIdoublet{} (see Figure \ref{fig:emitting_ions}), representing more closely the thick and thin shell shapes. We opted to allow for \NIIdoublet{} the thin shell and Gaussian shapes (and not a thick shell, saving a free parameter), both of which add another three free parameters.

To summarise, we fit for \OIdoublet{} either a Gaussian or an electron-scattered profile, and for \NIIdoublet{} either a Gaussian or a thin shell (i.e. tophat) profile. In total, we have seven free parameters to fit for. These are the centroids, amplitudes and linewidths for \NIIdoublet{} and \OIdoublet{}, and finally the electron-scattering optical depth. 

\subsection{Fitting Constraints}
\label{A:constraints}

As the possible shapes and the different amounts of blending still leave us with a lot of degeneracies for the fitting, we constrain our fitting procedure further by using knowledge of the physics behind the SN spectrum. Figure \ref{fig:Multi_epoch_fit} exemplifies most of the constraints used (five in total) and the resulting best fits for a sequence of spectra of SN~2011dh. Below we describe each of the constraints in detail.

\bigskip

\begin{enumerate}

\item{\textbf{Time consistent line widths.}} 
The first and potentially most important constraint, is that observations at different epochs for a given supernova are not independent; the expansion velocity of a SN always evolves in a similar way. During the nebular phase, this evolution is mostly constant to slowly decreasing, as the ejecta will expand very close to homologously. This knowledge is implemented in our fitting procedure by fitting all epochs of a given SN simultaneously with a single, time varying range for the width (i.e. velocity) $\sigma_{[\ion{N}{II}]}(t)$ of \NIIdoublet{}. This procedure can be summarised for a single SN as follows:

\begin{enumerate}
    \item Give an initial guess for $\sigma_{[\ion{N}{II}]}(100d)$ (line width in Å).
    \item Using the resulting allowed range for $\sigma_{[\ion{N}{II}]}(t)$ for each epoch $t$ of the SN, perform an OLS fit to each individual spectrum. Call the resulting lowest sum of squared residuals for a single spectrum SSR$_{t}$.
    \item Add the SSR$_{t}$ for each epoch together, and call this value SSR$_{total}$.
    \item Repeat this procedure for all $\sigma_{[\ion{N}{II}]}(100d)$ between 80 to 150 Å. The $\sigma_{[\ion{N}{II}]}(100d)$ that resulted in the lowest SSR$_{total}$ is deemed the best fit, and the best fit parameters for each epoch of the SN are taken from when that $\sigma_{[\ion{N}{II}]}(100d)$ was used for the fitting.
\end{enumerate}

The exact range for $\sigma_{[\ion{N}{II}]}(t)$ mentioned above is given by:
\begin{equation}
\begin{split}
    \sigma_{[\ion{N}{II}]}(t) > \sigma_{[\ion{N}{II}]}(100d) - 0.08 \times (t-100d)  \\ 
    \sigma_{[\ion{N}{II}]}(t) < \sigma_{[\ion{N}{II}]}(100d) - 0.04 \times (t-100d)
\end{split}
\label{eq:velocity_range}
\end{equation}

This equation is better understood with a concrete example: say we have two observations for a given SN at epochs 200d and 250d post explosion. If we then test for $\sigma_{[\ion{N}{II}]}(100d)$ = 100 Å, then the range that $\sigma_{[\ion{N}{II}]}(200d)$ may take is between 92 Å and 96 Å, and between 88 Å and 94 Å for $\sigma_{[\ion{N}{II}]}(250d)$ (see Figure \ref{fig:Multi_epoch_fit} as an example). Similarly, when we test for $\sigma_{[\ion{N}{II}]}(100d)$ = 90 Å, $\sigma_{[\ion{N}{II}]}(200d)$ is limited to 82 Å and 86 Å, and for $\sigma_{[\ion{N}{II}]}(250d)$ the limits are 78 Å and 84 Å. Testing of the fitting procedure with and without this constraint convinced us that it is a necessity to obtain fits that are realistic when comparing to the underlying physics. As mentioned above, the total range of tested values for $\sigma_{[\ion{N}{II}]}(100d)$ is between 80 Å and 150 Å, in intervals of 2 Å. The same velocity constraint is not applied for the fitting of the \OIdoublet{}, as the fitting algorithm experienced no non-physically large variations when fitting for different epochs of the same SN.

\item{\textbf{Maximum velocities.}} To this constraint on the evolution of \NIIdoublet{}, we add another constraint: a maximum velocity (i.e. line width). Due to the blend on the blue side with \OIdoublet{}, an algorithm without a maximum velocity can fit just about any velocity for \NIIdoublet{} (and thus integrated luminosity) by simply slightly changing the amplitude of \OIdoublet{}. However, we know that there is a rough physical limit to the velocities at which SN-material is ejected. In our models, the he5p00 model has by far the highest velocities, with the fastest 1\% of the ejecta traveling at $>$ 19 700 \kms{} ($\sigma_{[\ion{N}{II}]}$ = 163 Å). However, this model seems to be somewhat of an outlier, with the other four models staying around $\sim$ 15000 \kms{}. As our models typically have broader lines than the SNe in our sample, we therefore decided to put the maximum velocity somewhat lower than our fastest model, at $\sigma_{[\ion{N}{II}]}$ = 150 Å.
When the algorithm decides that this maximum velocity gives the best fit, the fitting procedure instead starts over but now tries to fit a tophat line profile to the [\ion{N}{II}] doublet. Here we again give an allowed range for the width of the tophat, between 150 Å and 200 Å, with the difference that if the fitting procedure now hits the bounds, we accept those bounds as best fit values. The reasoning behind this is that this typically only occurs when the amount of [\ion{N}{II}] emission is so low that $F_{\lambda}$ is flat in the 6500 -- 6800 Å region. As we will see, when the amount of [\ion{N}{II}] is that low, obtaining its exact value is not very important (as opposed to when the amount of [\ion{N}{II}] is significant). 
Admittedly, the choice for 150 Å as the transition value is somewhat arbitrary. Furthermore, the integrated area of a 150 Å width Gaussian is not the same as that of a 150 Å width tophat (i.e. it will often give a discontinuity in the integrated area). The exact choice of this transition velocity should not be understood as an attempt to perfectly describe the underlying physics, but rather as a pragmatic choice that makes the transition from Gaussian fitting to tophat fitting as smooth as possible. To help smoothing this transition, an additional requirement is set for moving to tophat: if the ratio of the amplitudes between the fitted peaks of \OIdoublet{} and \NIIdoublet{} is over 25, we stick with a Gaussian, with a velocity equal to the transition velocity. This is done as with such a high ratio, the chance that any of the emission is truly [\ion{N}{II}] becomes very low (see e.g. the he8p0, 400d model in Figure \ref{fig:emitting_ions}), and the risk of overestimating the \NIIdoublet{} emission with a tophat turned out to be large in these situations.

\item{\textbf{Time dependency of electron scattering.}} The third constraint is built on the same idea as the first: time dependency between observations. The electron optical depth in the ejecta scales linearly with both the density of the material, as well as the radius/extent of the ejecta. Assuming homologous expansion once more, it is easy to show that the electron optical depth $\tau_{e}$ scales with time as $t^{-2}$, for a fixed degree of ionization. Similar to the procedure for determining the line widths above, we here test a range for $\tau_{e}$ at the first epoch of the SN of $0.0 < \tau_{e, init} < 3.0$, with intervals of 0.1. For each tested $\tau_{e, init}$, the remaining $\tau_{e}(t)$ are then automatically determined using this scaling relation of $t^{-2}$. We again add all the individual SSR$_{t}$ to come to a final SSR$_{total}$, and the $\tau_{e, init}$ that lead to the lowest SSR$_{total}$ is deemed the best fit.

\vspace{0.2 cm}

\item{\textbf{Allowed centroid values.}} The fourth applied constraint, is that of a fixed range for the centroids of the fitted shapes. Typically, an emission line will have a width of a few thousand \kms{} (a few hundred Å), as this is the velocity at which the outermost layers of the emitting element are expanding. However, as the ejecta to first order expand spherically, the centroid should be approximately at the rest wavelength value nonetheless. There are two mechanisms that may shift the centroid of an emission line away from its theoretical value. The first was mentioned before: due to radiative transfer effects. In the case of free electron scattering, we may see a red tail as well as a blueshift in the centroid. As shown in \cite{Jerkstrand_2017_book}, this can lead to a relative blue shift of about 15 -- 20\%, or in other words $\sim$ 750 -- 1000 \kms{} / 15 -- 20 Å for a typical 5000 \kms{} broad \OIdoublet{} emission line. The second effect that may shift the centroid is due to the viewing angle under which we observe the SN. Using their 3D SN spectral synthesis code EXTRASS, \cite{vanBaal_2023_EXTRASS} have shown that for a SN of a 3.3 \Msun{}\footnote{In private correspondence the author mentioned that similar shifts are found for a 6.0 \Msun{} helium star explosion.} helium star the centroid of \OIdoublet{} may shift by $\sim \pm$300 \kms{} depending on the viewing angle (the shift being caused by non-zero bulk motion of the ejecta, due to the asymmetrical explosion). Combining these two effects, a centroid range for \OIdoublet{} of 6290 to 6340 Å is chosen, or 25 Å from the $\sim$ 6315 Å centroid expected from the theoretical 3:1 line ratio for \OIdoublet{}. For \NIIdoublet{}, a 1:3 ratio is expected from theory. However, the even stronger Doppler broadening for this line means that the resulting line profile is once more symmetric, centred around 6565 Å. We therefore allow a centroid range of 6540 to 6590 Å so that the centroid may lie anywhere between the two individual emission lines at 6548 Å and 6583 Å.

\vspace{0.2 cm}

\item{\textbf{Quasi-continuum subtraction.}} Finally, to be able to correctly measure the amount of \NIIdoublet{} emission, we want as little contamination by other lines as possible. To this end, a "continuum" is removed in the fitting region. In nebular-phase SNe, there is no or little true continuum emission, but rather the sum of thousands of weak lines make up a so called quasi-continuum \citep{Li1996,Jerkstrand_2011_SUMOa}. We will not make the distinction but use the term continuum throughout. This continuum is defined by a straight line, running between the average measured flux between 6100 -- 6150 Å on the left side, and 6850 -- 6900 Å on the right side. This results in these two regions (which typically have flat $F_{\lambda}$) being put to zero flux, allowing for a good fit of our \OIdoublet{} and \NIIdoublet{} components. Examples of this continuum removal are visible in Figures \ref{fig:halpha_removal} and \ref{fig:Multi_epoch_fit}.
\end{enumerate}

\subsection{Defining the Diagnostic}
\label{a:diagnostic}
For comparison between our model grid and observations, with focus on \NIIdoublet{}, we define a diagnostic that is the \NIIdoublet{} luminosity relative to the total spectral luminosity over a selected wavelength range in the optical window. This definition is similar to the one of \citet{Dessart_2021_Hestarexpl}, who use the range 3500-9500 \AA\ for the total spectral luminosity. We choose a somewhat different wavelength window, and also apply a subtraction of the (quasi)-continuum, as described below.

Observed spectra vary quite significantly in the wavelength ranges they cover. As for the purposes of this paper it is important to compare the models to as large samples of Type IIb, Ib, and Ic SNe as possible, we prioritize having to exclude as few observed SNe as possible. By choosing a wavelength range 5000 to 8000 Å, we make sure to include the majority of the available archival SNe, while simultaneously including the most important emitting regions during the nebular phase. The lower limit choice has the additional effect of leaving out the emission from \ion{Mg}{I}] $\lambda4571$. The neutral fraction of Mg in SESNe is around $10^{-3}$, which is just at the boundary between when thermal collisional excitation and recombination dominates the emission \citep{Jerkstrand_2015_NII_discovery}. As such, the line is quite sensitive to model uncertainties. Because of this, leaving out this line removes a source of uncertainty from the diagnostic. We note here that for \NIIdoublet{}, this uncertainty is not present: all our models have $x_{\text{\ion{N}{II}}} > 0.98$ at all epochs, so that the resulting emission is robust to the ionisation balance.

While in Monte Carlo spectral models it is possible to see exactly how much of the total emission at a given wavelength comes from a certain element, this is, of course, impossible to do for observed spectra. Therefore, we will need to make some assumptions on how to best capture the amount of \NIIdoublet{} emission. In Figure \ref{fig:emitting_ions}, contributions to the flux in the \NIIdoublet{} area by the different emitting ions are shown for our five different models. It is clear that there is a correlation between the mass of the progenitor and the relative amount of true \NIIdoublet{} emission. It also shows that for the cases where there is a significant amount of emission in the 6500 -- 6800 Å range, most (but not all) of this emission does indeed come from \NIIdoublet{}. Therefore, during the fitting of the amount of \NIIdoublet{} emission in observed spectra, it will be assumed that the best fit emission profile for \NIIdoublet{} (as defined above in this subsection) is indeed fully caused by emission from \NIIdoublet{}. To make the comparisons between models and observations fair, we will apply the exact same fitting procedure to determine the diagnostic \NIIdoublet{} for the models, rather than using the available information on the exact amount of \NIIdoublet{} emission present in the model.

The [\ion{N}{II}] diagnostic that will be extracted and compared between modelled and observed spectra is then

\begin{equation}
    f_{[\ion{N}{II}]} = \frac{ \text{integrated best fit \NIIdoublet{} flux}}{\int^{8000\  \text{Å} }_{5000\ \text{Å}} (F_{\lambda} - F_{\text{pseudo}}) d\lambda} \times 100
    \label{eq:definition}        
\end{equation}

with the best fit \NIIdoublet{} emission as defined in Section \ref{sec:fitting_procedure}, $F_{\lambda}$ the observed spectral flux and $F_{\text{pseudo}}$ a pseudo-continuum flux. This pseudo continuum is defined as follows: for each individual spectrum, the average flux is determined in four spectral regions: 5740 -- 5790 Å, 6020 -- 6070 Å, 6850 -- 6900 Å and 7950 -- 8000 Å. Then, the average of the three lowest values is taken. This final average defines the (constant)  $F_{\text{pseudo}}$ value (to be clear, this $F_{\text{pseudo}}$ is different from the quasi-continuum defined before). The removal of this pseudo-continuum $F_{\text{pseudo}}$ was performed because especially at later epochs, a significant number of our spectra showed varying levels of background continuum contribution. Subtracting a pseudo-continuum estimate was found to be a relatively simple way to remove this contamination. The four specific spectral regions were selected because in our models, these regions typically show very little to zero emission; emission seen here in observations can be therefore be assumed to be mostly background continuum.

It should be noted here that a non-zero value for \NIIdiag{} should not be interpreted as an absolute claim of any significant presence of nitrogen in the ejecta. As shown in Figure \ref{fig:emitting_ions}, especially at earlier times and for higher mass progenitors, emission in the 6400 -- 6800 Å region will not always be dominated by \NIIdoublet{} emission.  

There is no trivial way of calculating the uncertainty on \NIIdiag{}. 
First, the constraints that we force on our simultaneous fitting algorithm mean that each individual spectrum does not have a guaranteed normalised-$\chi^{2}$ value close to 1. Second, the many free parameters makes bootstrap difficult for uncertainty determination. Therefore, to determine the uncertainty on \NIIdiag{} we divide our best fit amplitude of the \NIIdoublet{} profile by the standard deviation of the measured flux between 6850 and 6950 Å, and name this 'SNR'. The 1$\sigma$ uncertainty on \NIIdiag{} is then defined as (1/SNR) $\times$ \NIIdiag{}. It should be noted that this definition will underestimate the uncertainty on the model \NIIdiag{} tracks, as none of the model uncertainties are taken into account.

\section{Spectroscopic Sample}

A more detailed description for the collections of spectra per object in our sample is provided in Table \ref{tab:sample}. In Table \ref{tab:multi_estimates}, we mention additional works to those mentioned in Table \ref{tab:mass_predictions} that determined mass estimates for the same SN.

\begin{table*}
\centering
\begin{tabular}{lllll}
\hline
SN & Type & No. of & Phase range & Ref \\
 & & Spectra & [d] & \\
 \hline \hline  \\[-.25cm]
1985F & Ib & 1 & 295 & \citet{1985F_a, 1985F_b} \\
1987M & Ic & 1 & 168 & \citet{1987M} \\
1993J & IIb & 22 & 166 -- 347 & \citet{1993J_a, 1993J_b} \\
1996aq & Ib & 2 & 248 -- 320 & \citet{1996aq_1998bw_b} \\
1996cb & IIb & 3 & 196 -- 358 & \citet{1996cb} \\
1997dq & Ic & 2 & 241 -- 260 & \citet{1997dq_a, 1997dq_b} \\
1998bw & Ic & 3 & 352 -- 391 & \citet{1998bw_a, 1996aq_1998bw_b} \\
2001ig & IIb & 2 & 309 -- 340 & \citet{2001ig_a, 2001ig_b} \\
2003bg & IIb & 5 & 176 -- 301 & \citet{2003bg} \\
2003gf & Ic & 2 & 149 -- 181 & \citet{2003gf} \\
2004ao & Ib & 5 & 172 -- 298 & \citet{2004ao_a, 2004ao_b_2004gq_b} \\
2004aw & Ic & 1 & 249 & \citet{2004aw} \\
2004dk & Ib & 2 & 284 -- 343 & \citet{2004dk_a_2004gq_a, 2004dk_b} \\
2004gq & Ib & 1 & 366 & \citet{2004dk_a_2004gq_a, 2004ao_b_2004gq_b} \\
2006ld & Ib & 1 & 141 & \citet{2006ld_a_2007I_a, 2006ld_b_2007C_b_2007I_b_2008aq_b} \\
2007C & Ib & 2 & 141 -- 175 & \citet{2007C_a_2008bo_a, 2006ld_b_2007C_b_2007I_b_2008aq_b} \\
2007I & Ic & 2 & 180 -- 207 & \citet{2006ld_a_2007I_a, 2006ld_b_2007C_b_2007I_b_2008aq_b} \\
2007Y & Ib & 3 & 219 -- 288 & \citet{2007Y} \\
2007gr & Ic & 2 & 155 -- 183 & \citet{2007gr_a, 2007gr_b} \\
2008D & Ib & 1 & 149 & \citet{2008D_a, 2008D_b} \\
2008aq & IIb & 1 & 329 & \citet{2008aq_a, 2006ld_b_2007C_b_2007I_b_2008aq_b} \\
2008ax & IIb & 4 & 158 -- 359 & \citet{2008ax_a, 2008ax_b} \\
2008bo & IIb & 2 & 157 -- 217 & \citet{2007C_a_2008bo_a, 2008bo_b} \\
2009K & IIb & 1 & 307 & \citet{Taddia_2018_Nimasses_2009K_a, 2009K_b} \\
2009jf & Ib & 2 & 267 -- 381 & \citet{Valenti_2011_risetime_2009jf} \\
2011bm & Ic & 3 & 270 -- 303 & \citet{2011bm} \\
2011dh & IIb & 10 & 152 -- 415 & \citet{2011dh_a, 2011dh_b} \\
2011ei & IIb & 1 & 329 & \citet{2011ei} \\
2011hs & IIb & 4 & 174 -- 348 & \citet{2011hs} \\
2012au & Ib & 2 & 290 -- 338 & \citet{2012au} \\
PTF12gzk & Ic & 2 & 323 -- 414 & \citet{PTF12gzk_a, PTF12gzk_b} \\
PTF12os & IIb & 2 & 139 -- 215 & \citet{PTF12os_iPTF13bvn_b}\\
iPTF13bvn & IIb & 2 & 250 -- 346 & \citet{iPTF13bvn_a, PTF12os_iPTF13bvn_b} \\
2013df & IIb & 1 & 179 & \citet{2013df} \\
2013ge & Ic & 4 & 165 -- 433 & \citet{2013ge} \\
J1204 & Ib & 7 & 177 -- 271 & \citet{J1204_a, J1204_b} \\
ASASSN14az & IIb & 5 & 125 -- 190 & \citet{14az_a, 14az_b} \\
2015ah & Ib & 1 & 165 & \citet{Prentice_2019_Bigsample_2015ah} \\
2015fn & Ic & 1 & 359 & \citet{2015fn_a, 2015fn_b} \\
2019yz & Ic & 3 & 138 -- 256 & \citet{2019yz} \\
2019odp & Ib & 3 & 147 -- 368 & \citet{2019odp} \\
2020acat & IIb & 3 & 147 -- 194 & \citet{2020acat_a, 2020acat_b} \\
2022crv & IIb & 3 & 290 -- 373 & \citet{2022crv} \\
\hline \hline

\end{tabular}
    \caption{List of all SNe in our sample. The phases are given with respect to explosion epoch. The references are not a complete list of the works on a particular SN, but will include at least the source of our used explosion epoch (first ref) and the work presenting the spectra used (second ref). If a single reference is given, it contains both of these. } 
    \label{tab:sample}
\end{table*}

\begin{table}
\centering
\begin{tabular}{l|l}
\hline
SN & Ref \\
\hline \hline  \\[-.25cm]
1993J & \citet{1993J_AA, 1993J_BB} \\
 & \citet{1993J_C, 1993J_D} \\
  & \citet{1993J_E, Nomoto_1993_IIb} \\
1998bw & \citet{1998bw_AA, 1998bw_B} \\
2008ax & \citet{2008ax_AA} \\
2011dh & \citet{2011dh_AA, 2011dh_BB} \\
2011dh & \citet{2011dh_C, 2011dh_D} \\
 & \citet{2011dh_E, 2011dh_F} \\
iPTF13bvn & \citet{iPTF13bvn_AA, iPTF13bvn_B} \\
 & \citet{iPTF13bvn_C, iPTF13bvn_D} \\
 & \citet{iPTF13bvn_E, iPTF13bvn_F} \\
 2022crv & \citet{funky} \\
\hline \hline
\end{tabular}

    \caption{List of all SNe in our sample for which multiple progenitor mass estimates exist in literature, with the corresponding references. The list is most likely not extensive, so if the reader deems a reference to be missing here, please contact the corresponding authors.}
    \label{tab:multi_estimates}
\end{table}

\FloatBarrier

\section{Model Compositions}
\label{appendix:model_compositions}

In Table \ref{tab:rules}, we give the rules that were used to discretise the hundreds of radial cells for the initial ejecta models into seven graspable zones, as discussed in Section \ref{sec:inputmodels}. Tables \ref{tab:comp33}-\ref{tab:comp80} give the detailed compositions of the ejecta models, once they had been discretised.

\begin{table}
\centering
\begin{tabular}{l|l}
\hline
Zone name & When does zone end? \\ \hline \hline
Fe/He & $X_{\text{\el{4}{He}}} < 10^{-2}$  \\
Si/S & $X_{\text{\el{16}{O}}} > 10^{-2}$ \\
O/Si/S & $X_{\text{\el{20}{Ne}}} > X_{\text{\el{28}{Si}}}$ \\
O/Ne/Mg & $X_{\text{\el{12}{C}}} > X_{\text{\el{20}{Ne}}}$ \\
O/C & $X_{\text{\el{4}{He}}} > X_{\text{\el{16}{O}}}$ \\
He/C & $X_{\text{\el{14}{N}}} > X_{\text{\el{12}{C}}}$ \\
He/N & $\int^{R}_{0} M_{\text{ej}} \ dR > 0.99\int^{\infty}_{0} M_{\text{ej}} \ dR $ \\ \hline \hline

\end{tabular}

\caption{Rules used to determine when a new zone should start in a SUMO model. Following \citet{Dessart_2021_Hestarexpl}, the Fe/He zone starts as soon as the expansion velocity is above 50 \kms{} to prevent any fallback material.}
\label{tab:rules}
\end{table}

\begin{table*}
    \fontsize{7.2}{8}\selectfont
    \begin{tabular}{l|lllllllllllll}
    \hline
    & Fe/He & Si/S & O/Si/S & O/Ne/Mg & O/C & He/C$_{\text{core}}$ & He/C$_{\text{env}}$ & He/N$_{\text{env}}$ & He/N$_{\text{env}}$ & He/N$_{\text{env}}$ & He/N$_{\text{env}}$ & He/N$_{\text{env}}$ & He/N$_{\text{env}}$ \\ \hline \hline 
    
    $M_{\text{zone}} [M_{\odot}]$ & 0.073 & 0.019 & 0.071 & 0.11 & 0.079 & 0.061 & 0.092 & 0.16 & 0.17 & 0.15 & 0.11 & 0.065 & 0.049 \\ 
    $V_{\text{in}} [10^{3} \kms{}]$ & 0.050 & 2.4 & 2.6 & 3.0 & 3.4 & 3.9 & 4.3 & 4.7 & 5.7 & 6.8 & 8.2 & 9.8 & 12 \\
    $V_{\text{out}} [10^{3} \kms{}]$ & 2.4 & 2.6 & 3.0 & 3.4 & 3.9 & 4.3 & 4.7 & 5.7 & 6.8 & 8.2 & 9.8 & 12 & 17 \\ \hline
    $X_{^{56}\text{Ni}}$ & 0.66 & 0.32 & 4.6(-6) & 7.8(-7) & 7.2(-8) & 5.0(-8) & 5.0(-8) & 6.4(-9) & 6.4(-9) & 6.4(-9) & 6.4(-9) & 6.4(-9) & 6.4(-9) \\ 
    $X_{\text{He}}$ & 0.2 & 1.1(-5) & 8.4(-6) & 7.0(-6) & 0.053 & 0.83 & 0.83 & 0.99 & 0.99 & 0.99 & 0.99 & 0.99 & 0.99 \\ 
    $X_{\text{C}}$ & 7.0(-7) & 1.7(-6) & 2.4(-3) & 0.012 & 0.42 & 0.13 & 0.13 & 3.6(-4) & 3.6(-4) & 3.6(-4) & 3.6(-4) & 3.6(-4) & 3.6(-4) \\ 
    \rowcolor{Gray}
    $X_{\text{N}}$ & 1.1(-6) & 5.4(-8) & 7.2(-6) & 1.9(-5) & 2.7(-5) & 2.0(-4) & 2.0(-4) & 8.7(-3) & 8.7(-3) & 8.7(-3) & 8.7(-3) & 8.7(-3) & 8.7(-3) \\ 
    $X_{\text{O}}$ & 1.6(-5) & 1.5(-5) & 0.63 & 0.59 & 0.47 & 0.015 & 0.015 & 5.6(-4) & 5.6(-4) & 5.6(-4) & 5.6(-4) & 5.6(-4) & 5.6(-4) \\ 
    $X_{\text{Ne}}$ & 1.9(-5) & 1.4(-6) & 5.6(-3) & 0.19 & 0.046 & 0.013 & 0.013 & 1.2(-3) & 1.2(-3) & 1.2(-3) & 1.2(-3) & 1.2(-3) & 1.2(-3) \\ 
    $X_{\text{Na}}$ & 5.8(-7) & 7.9(-7) & 5.7(-5) & 2.0(-3) & 1.8(-4) & 1.5(-4) & 1.5(-4) & 1.5(-4) & 1.5(-4) & 1.5(-4) & 1.5(-4) & 1.5(-4) & 1.5(-4) \\ 
    $X_{\text{Mg}}$ & 2.4(-5) & 1.3(-4) & 0.045 & 0.13 & 0.015 & 1.2(-3) & 1.2(-3) & 7.3(-4) & 7.3(-4) & 7.3(-4) & 7.3(-4) & 7.3(-4) & 7.3(-4) \\ 
    $X_{\text{Si}}$ & 2.1(-4) & 0.34 & 0.25 & 0.045 & 1.0(-3) & 8.5(-4) & 8.5(-4) & 8.2(-4) & 8.2(-4) & 8.2(-4) & 8.2(-4) & 8.2(-4) & 8.2(-4) \\ 
    $X_{\text{S}}$ & 1.4(-4) & 0.19 & 0.049 & 1.4(-3) & 2.6(-4) & 4.0(-4) & 4.0(-4) & 4.2(-4) & 4.2(-4) & 4.2(-4) & 4.2(-4) & 4.2(-4) & 4.2(-4) \\ 
    $X_{\text{Ca}}$ & 1.8(-3) & 0.033 & 2.9(-3) & 4.5(-5) & 3.2(-5) & 6.8(-5) & 6.8(-5) & 7.3(-5) & 7.3(-5) & 7.3(-5) & 7.3(-5) & 7.3(-5) & 7.3(-5) \\ 
    $X_{\text{Fe}}$ & 1.8(-3) & 0.061 & 4.5(-3) & 1.1(-3) & 9.6(-4) & 1.4(-3) & 1.4(-3) & 1.4(-3) & 1.4(-3) & 1.4(-3) & 1.4(-3) & 1.4(-3) & 1.4(-3) \\
    \hline \hline

    \end{tabular}

    \caption{Table with the zone structure and composition for our \texttt{SUMO} input model for the he3p3 model from \citet{Ertl_2020_models}, as described in Section \ref{sec:inputmodels}. The elemental abundances are given in mass fractions for that specific zone. The He/N zone gets divided into multiple subzones of equal composition to avoid too large velocity differences within a single zone. Numbers ending on a number in brackets indicate powers of 10, e.g. 7.8(-7) = 7.8 $\times$ 10$^{-7}$.}
    \label{tab:comp33}

\end{table*}

\begin{table*}
    \fontsize{7.2}{8}\selectfont
    \begin{tabular}{l|llllllllllll}
    \hline
    & Fe/He & Si/S & O/Si/S & O/Ne/Mg & O/C & He/C$_{\text{core}}$ & He/C$_{\text{env}}$ & He/N$_{\text{env}}$ & He/N$_{\text{env}}$ & He/N$_{\text{env}}$ & He/N$_{\text{env}}$ & He/N$_{\text{env}}$ \\ \hline \hline 
    
    $M_{\text{zone}} [M_{\odot}]$ & 0.076 & 0.033 & 0.11 & 0.32 & 0.15 & 0.036 & 0.17 & 0.22 & 0.19 & 0.14 & 0.081 & 0.082 \\ 
    $V_{\text{in}} [10^{3} \kms{}]$ & 0.050 & 2.5 & 2.7 & 3.2 & 3.8 & 4.3 & 4.5 & 5.2 & 6.2 & 7.5 & 9.0 & 11 \\
    $V_{\text{out}} [10^{3} \kms{}]$ & 2.5 & 2.7 & 3.2 & 3.8 & 4.3 & 4.5 & 5.2 & 6.2 & 7.5 & 9.0 & 11 & 15 \\ \hline
    $X_{^{56}\text{Ni}}$ & 0.68 & 0.28 & 2.1(-6) & 7.0(-7) & 7.8(-8) & 3.4(-8) & 3.4(-8) & 1.9(-8) & 1.9(-8) & 1.9(-8) & 1.9(-8) & 1.9(-8) \\ 
    $X_{\text{He}}$ & 0.18 & 9.5(-6) & 6.1(-6) & 5.0(-6) & 0.048 & 0.81 & 0.81 & 0.99 & 0.99 & 0.99 & 0.99 & 0.99 \\ 
    $X_{\text{C}}$ & 7.4(-7) & 4.8(-6) & 2.5(-3) & 0.019 & 0.4 & 0.15 & 0.15 & 3.6(-4) & 3.6(-4) & 3.6(-4) & 3.6(-4) & 3.6(-4) \\ 
    \rowcolor{Gray}
    $X_{\text{N}}$ & 7.1(-7) & 1.1(-7) & 5.4(-6) & 3.1(-5) & 2.3(-5) & 1.3(-4) & 1.3(-4) & 8.6(-3) & 8.6(-3) & 8.6(-3) & 8.6(-3) & 8.6(-3) \\ 
    $X_{\text{O}}$ & 1.5(-5) & 1.3(-4) & 0.58 & 0.51 & 0.51 & 0.02 & 0.02 & 6.3(-4) & 6.3(-4) & 6.3(-4) & 6.3(-4) & 6.3(-4) \\ 
    $X_{\text{Ne}}$ & 1.8(-5) & 2.4(-6) & 5.7(-3) & 0.34 & 0.034 & 0.014 & 0.014 & 1.3(-3) & 1.3(-3) & 1.3(-3) & 1.3(-3) & 1.3(-3) \\ 
    $X_{\text{Na}}$ & 4.3(-7) & 7.9(-7) & 5.7(-5) & 7.9(-3) & 1.8(-4) & 1.5(-4) & 1.5(-4) & 1.5(-4) & 1.5(-4) & 1.5(-4) & 1.5(-4) & 1.5(-4) \\ 
    $X_{\text{Mg}}$ & 2.2(-5) & 1.5(-4) & 0.035 & 0.1 & 7.9(-3) & 9.7(-4) & 9.7(-4) & 7.3(-4) & 7.3(-4) & 7.3(-4) & 7.3(-4) & 7.3(-4) \\ 
    $X_{\text{Si}}$ & 2.1(-4) & 0.39 & 0.29 & 0.011 & 9.5(-4) & 8.4(-4) & 8.4(-4) & 8.2(-4) & 8.2(-4) & 8.2(-4) & 8.2(-4) & 8.2(-4) \\ 
    $X_{\text{S}}$ & 1.3(-4) & 0.19 & 0.061 & 4.9(-4) & 3.0(-4) & 4.1(-4) & 4.1(-4) & 4.2(-4) & 4.2(-4) & 4.2(-4) & 4.2(-4) & 4.2(-4) \\ 
    $X_{\text{Ca}}$ & 1.9(-3) & 0.027 & 2.0(-3) & 4.7(-5) & 4.1(-5) & 7.0(-5) & 7.0(-5) & 7.3(-5) & 7.3(-5) & 7.3(-5) & 7.3(-5) & 7.3(-5) \\ 
    $X_{\text{Fe}}$ & 1.5(-3) & 0.071 & 3.3(-3) & 1.1(-3) & 1.2(-3) & 1.4(-3) & 1.4(-3) & 1.4(-3) & 1.4(-3) & 1.4(-3) & 1.4(-3) & 1.4(-3) \\
    \hline \hline
    
    \end{tabular}

    \caption{Same as Table \ref{tab:comp33}, but now for the he4p0 model.}
    \label{tab:comp40}
\end{table*}

\begin{table*}
    \fontsize{7.2}{8}\selectfont
    \begin{tabular}{l|llllllllllll}
    \hline
    & Fe/He & Si/S & O/Si/S & O/Ne/Mg & O/C & He/C$_{\text{core}}$ & He/C$_{\text{env}}$ & He/N$_{\text{env}}$ & He/N$_{\text{env}}$ & He/N$_{\text{env}}$ & He/N$_{\text{env}}$ \\ \hline \hline 

    $M_{\text{zone}} (M_{\odot})$ & 0.15 & 0.034 & 0.17 & 0.64 & 0.22 & 0.026 & 0.23 & 0.28 & 0.2 & 0.14 & 0.11 \\ 
    $V_{\text{in}} [10^{3} \kms{}]$ & 0.050 & 3.4 & 3.6 & 4.4 & 6.1 & 6.9 & 7.0 & 8.0 & 9.6 & 12 & 14 \\
    $V_{\text{out}} [10^{3} \kms{}]$ & 3.4 & 3.6 & 4.4 & 6.1 & 6.9 & 7.0 & 8.0 & 9.6 & 12 & 14 & 20 \\ \hline
    $X_{^{56}	ext{Ni}}$ & 0.68 & 0.22 & 3.3(-6) & 3.1(-7) & 2.0(-8) & 1.1(-8) & 1.1(-8) & 4.1(-8) & 4.1(-8) & 4.1(-8) & 4.1(-8) \\ 
    $X_{\text{He}}$ & 0.21 & 6.3(-6) & 4.8(-6) & 2.6(-6) & 0.058 & 0.7 & 0.7 & 0.99 & 0.99 & 0.99 & 0.99 \\ 
    $X_{\text{C}}$ & 1.1(-6) & 1.8(-6) & 2.9(-3) & 0.02 & 0.41 & 0.2 & 0.2 & 3.3(-4) & 3.3(-4) & 3.3(-4) & 3.3(-4) \\ 
    \rowcolor{Gray}
    $X_{\text{N}}$ & 8.0(-7) & 3.8(-8) & 5.1(-6) & 1.9(-5) & 1.3(-5) & 1.5(-4) & 1.5(-4) & 8.6(-3) & 8.6(-3) & 8.6(-3) & 8.6(-3) \\ 
    $X_{\text{O}}$ & 1.6(-5) & 1.4(-5) & 0.65 & 0.55 & 0.5 & 0.084 & 0.084 & 6.2(-4) & 6.2(-4) & 6.2(-4) & 6.2(-4) \\ 
    $X_{\text{Ne}}$ & 1.8(-5) & 9.0(-7) & 4.7(-3) & 0.31 & 0.025 & 0.014 & 0.014 & 1.2(-3) & 1.2(-3) & 1.2(-3) & 1.2(-3) \\ 
    $X_{\text{Na}}$ & 5.4(-7) & 5.9(-7) & 4.9(-5) & 6.4(-3) & 1.8(-4) & 1.5(-4) & 1.5(-4) & 1.5(-4) & 1.5(-4) & 1.5(-4) & 1.5(-4) \\ 
    $X_{\text{Mg}}$ & 2.2(-5) & 9.6(-5) & 0.031 & 0.091 & 5.4(-3) & 9.5(-4) & 9.5(-4) & 7.3(-4) & 7.3(-4) & 7.3(-4) & 7.3(-4) \\ 
    $X_{\text{Si}}$ & 1.9(-4) & 0.4 & 0.23 & 9.6(-3) & 9.1(-4) & 8.3(-4) & 8.3(-4) & 8.2(-4) & 8.2(-4) & 8.2(-4) & 8.2(-4) \\ 
    $X_{\text{S}}$ & 1.4(-4) & 0.23 & 0.052 & 4.5(-4) & 3.2(-4) & 4.1(-4) & 4.1(-4) & 4.2(-4) & 4.2(-4) & 4.2(-4) & 4.2(-4) \\ 
    $X_{\text{Ca}}$ & 2.0(-3) & 0.041 & 3.2(-3) & 4.3(-5) & 4.8(-5) & 7.0(-5) & 7.0(-5) & 7.3(-5) & 7.3(-5) & 7.3(-5) & 7.3(-5) \\ 
    $X_{\text{Fe}}$ & 2.0(-3) & 0.055 & 3.6(-3) & 1.1(-3) & 1.2(-3) & 1.4(-3) & 1.4(-3) & 1.4(-3) & 1.4(-3) & 1.4(-3) & 1.4(-3) \\
    \hline \hline

    \end{tabular}

    \caption{Same as Table \ref{tab:comp33}, but now for the he5p0 model.}
    \label{tab:comp50}
\end{table*}

\begin{table*}
    \fontsize{7.2}{8}\selectfont
    \begin{tabular}{l|llllllllllll}
    \hline
    & Fe/He & Si/S & O/Si/S & O/Ne/Mg & O/Ne/Mg & O/C & He/C$_{\text{core}}$ & He/C$_{\text{env}}$ & He/N$_{\text{env}}$ & He/N$_{\text{env}}$ & He/N$_{\text{env}}$ & He/N$_{\text{env}}$ \\ \hline \hline 
    
    $M_{\text{zone}} [M_{\odot}]$ & 0.12 & 0.033 & 0.17 & 0.35 & 0.75 & 0.31 & 0.045 & 0.4 & 0.27 & 0.17 & 0.12 & 0.061 \\ 
    $V_{\text{in}} [10^{3} \kms{}]$ & 0.050 & 2.3 & 2.5 & 3.1 & 3.8 & 4.8 & 5.5 & 5.7 & 7.0 & 8.4 & 10 & 12 \\
    $V_{\text{out}} [10^{3} \kms{}]$ & 2.3 & 2.5 & 3.1 & 3.8 & 4.8 & 5.5 & 5.7 & 7.0 & 8.4 & 10 & 12 & 14 \\ \hline
    $X_{^{56}Ni}$ & 0.67 & 0.21 & 4.4(-6) & 2.9(-7) & 2.9(-7) & 8.0(-8) & 7.8(-8) & 7.8(-8) & 3.2(-8) & 3.2(-8) & 3.2(-8) & 3.2(-8) \\ 
    $X_{\text{He}}$ & 0.21 & 1.0(-5) & 4.7(-6) & 2.9(-6) & 2.9(-6) & 0.05 & 0.58 & 0.58 & 0.99 & 0.99 & 0.99 & 0.99 \\ 
    $X_{\text{C}}$ & 1.2(-6) & 2.2(-6) & 2.1(-3) & 9.8(-3) & 9.8(-3) & 0.39 & 0.27 & 0.27 & 2.8(-4) & 2.8(-4) & 2.8(-4) & 2.8(-4) \\ 
    \rowcolor{Gray}
    $X_{\text{N}}$ & 7.9(-7) & 4.5(-8) & 5.5(-6) & 1.4(-5) & 1.4(-5) & 1.8(-5) & 6.3(-4) & 6.3(-4) & 9.0(-3) & 9.0(-3) & 9.0(-3) & 9.0(-3) \\ 
    $X_{\text{O}}$ & 1.7(-5) & 1.9(-5) & 0.67 & 0.58 & 0.58 & 0.54 & 0.12 & 0.12 & 2.7(-4) & 2.7(-4) & 2.7(-4) & 2.7(-4) \\ 
    $X_{\text{Ne}}$ & 1.7(-5) & 1.3(-6) & 6.3(-3) & 0.28 & 0.28 & 0.016 & 0.014 & 0.014 & 1.2(-3) & 1.2(-3) & 1.2(-3) & 1.2(-3) \\ 
    $X_{\text{Na}}$ & 5.1(-7) & 6.6(-7) & 4.6(-5) & 3.9(-3) & 3.9(-3) & 1.7(-4) & 1.5(-4) & 1.5(-4) & 1.5(-4) & 1.5(-4) & 1.5(-4) & 1.5(-4) \\ 
    $X_{\text{Mg}}$ & 2.0(-5) & 1.1(-4) & 0.04 & 0.1 & 0.1 & 4.2(-3) & 7.9(-4) & 7.9(-4) & 7.3(-4) & 7.3(-4) & 7.3(-4) & 7.3(-4) \\ 
    $X_{\text{Si}}$ & 2.7(-4) & 0.41 & 0.22 & 0.013 & 0.013 & 8.9(-4) & 8.3(-4) & 8.3(-4) & 8.2(-4) & 8.2(-4) & 8.2(-4) & 8.2(-4) \\ 
    $X_{\text{S}}$ & 2.1(-4) & 0.24 & 0.044 & 4.1(-4) & 4.1(-4) & 3.3(-4) & 4.1(-4) & 4.1(-4) & 4.2(-4) & 4.2(-4) & 4.2(-4) & 4.2(-4) \\ 
    $X_{\text{Ca}}$ & 2.1(-3) & 0.04 & 2.7(-3) & 4.1(-5) & 4.1(-5) & 5.3(-5) & 7.2(-5) & 7.2(-5) & 7.4(-5) & 7.4(-5) & 7.4(-5) & 7.4(-5) \\ 
    $X_{\text{Fe}}$ & 1.7(-3) & 0.054 & 3.2(-3) & 1.0(-3) & 1.0(-3) & 1.2(-3) & 1.4(-3) & 1.4(-3) & 1.4(-3) & 1.4(-3) & 1.4(-3) & 1.4(-3) \\
    \hline \hline
    
    \end{tabular}
    
    \caption{Same as Table \ref{tab:comp33}, but now for the he6p0 model.}
    \label{tab:comp60}
\end{table*}

\begin{table*}
    \fontsize{7.1}{8}\selectfont
    \begin{tabular}{l|lllllllllllll}
    \hline
    & Fe/He & Si/S & O/Si/S & O/Ne/Mg & O/Ne/Mg & O/Ne/Mg & O/C & He/C$_{\text{core}}$ & He/C$_{\text{env}}$ & He/C$_{\text{env}}$ & He/N$_{\text{env}}$ & He/N$_{\text{env}}$ & He/N$_{\text{env}}$ \\ \hline \hline 
    
    $M_{\text{zone}} [M_{\odot}]$ & 0.083 & 0.034 & 0.14 & 0.38 & 0.55 & 1 & 0.36 & 0.087 & 0.38 & 0.4 & 0.23 & 0.15 & 0.083 \\ 
    $V_{\text{in}} [10^{3} \kms{}]$ & 0.050 & 1.3 & 1.4 & 1.7 & 2.1 & 2.5 & 3.3 & 3.9 & 4.0 & 4.8 & 5.9 & 7.1 & 8.5 \\
    $V_{\text{out}} [10^{3} \kms{}]$ & 1.3 & 1.4 & 1.7 & 2.1 & 2.5 & 3.3 & 3.9 & 4.0 & 4.8 & 5.9 & 7.1 & 8.5 & 11 \\ \hline
    $X_{^{56}Ni}$ & 0.65 & 0.21 & 1.5(-6) & 2.0(-7) & 2.0(-7) & 2.0(-7) & 6.8(-8) & 6.9(-8) & 6.9(-8) & 6.9(-8) & 2.2(-8) & 2.2(-8) & 2.2(-8) \\ 
    $X_{\text{He}}$ & 0.23 & 1.3(-5) & 4.5(-6) & 3.4(-6) & 3.4(-6) & 3.4(-6) & 0.017 & 0.38 & 0.38 & 0.38 & 0.99 & 0.99 & 0.99 \\ 
    $X_{\text{C}}$ & 1.5(-6) & 9.9(-6) & 3.0(-3) & 0.018 & 0.018 & 0.018 & 0.31 & 0.39 & 0.39 & 0.39 & 3.7(-4) & 3.7(-4) & 3.7(-4) \\ 
    \rowcolor{Gray}
    $X_{\text{N}}$ & 5.5(-7) & 1.4(-7) & 6.5(-6) & 2.4(-5) & 2.4(-5) & 2.4(-5) & 1.5(-5) & 8.5(-4) & 8.5(-4) & 8.5(-4) & 9.0(-3) & 9.0(-3) & 9.0(-3) \\ 
    $X_{\text{O}}$ & 2.0(-5) & 6.5(-4) & 0.7 & 0.6 & 0.6 & 0.6 & 0.65 & 0.21 & 0.21 & 0.21 & 2.8(-4) & 2.8(-4) & 2.8(-4) \\ 
    $X_{\text{Ne}}$ & 1.9(-5) & 5.8(-6) & 7.0(-3) & 0.29 & 0.29 & 0.29 & 0.016 & 0.014 & 0.014 & 0.014 & 1.2(-3) & 1.2(-3) & 1.2(-3) \\ 
    $X_{\text{Na}}$ & 6.0(-7) & 6.9(-7) & 6.0(-5) & 7.8(-3) & 7.8(-3) & 7.8(-3) & 1.8(-4) & 1.5(-4) & 1.5(-4) & 1.5(-4) & 1.5(-4) & 1.5(-4) & 1.5(-4) \\ 
    $X_{\text{Mg}}$ & 1.8(-5) & 1.3(-4) & 0.037 & 0.07 & 0.07 & 0.07 & 5.8(-3) & 9.8(-4) & 9.8(-4) & 9.8(-4) & 7.2(-4) & 7.2(-4) & 7.2(-4) \\ 
    $X_{\text{Si}}$ & 3.3(-4) & 0.4 & 0.2 & 5.4(-3) & 5.4(-3) & 5.4(-3) & 9.1(-4) & 8.3(-4) & 8.3(-4) & 8.3(-4) & 8.2(-4) & 8.2(-4) & 8.2(-4) \\ 
    $X_{\text{S}}$ & 2.7(-4) & 0.23 & 0.038 & 3.2(-4) & 3.2(-4) & 3.2(-4) & 2.9(-4) & 4.1(-4) & 4.1(-4) & 4.1(-4) & 4.2(-4) & 4.2(-4) & 4.2(-4) \\ 
    $X_{\text{Ca}}$ & 2.7(-3) & 0.04 & 2.0(-3) & 3.8(-5) & 3.8(-5) & 3.8(-5) & 4.5(-5) & 7.1(-5) & 7.1(-5) & 7.1(-5) & 7.4(-5) & 7.4(-5) & 7.4(-5) \\ 
    $X_{\text{Fe}}$ & 1.9(-3) & 0.054 & 2.3(-3) & 9.0(-4) & 9.0(-4) & 9.0(-4) & 1.1(-3) & 1.4(-3) & 1.4(-3) & 1.4(-3) & 1.4(-3) & 1.4(-3) & 1.4(-3) \\
    \hline \hline
    
    \end{tabular}
    
    \caption{Same as Table \ref{tab:comp33}, but now for the he8p0 model.}
    \label{tab:comp80}
\end{table*}

\FloatBarrier

\section{Model Spectra}

\begin{figure*}
        \centering
        \includegraphics[width=.98\linewidth,angle=0]{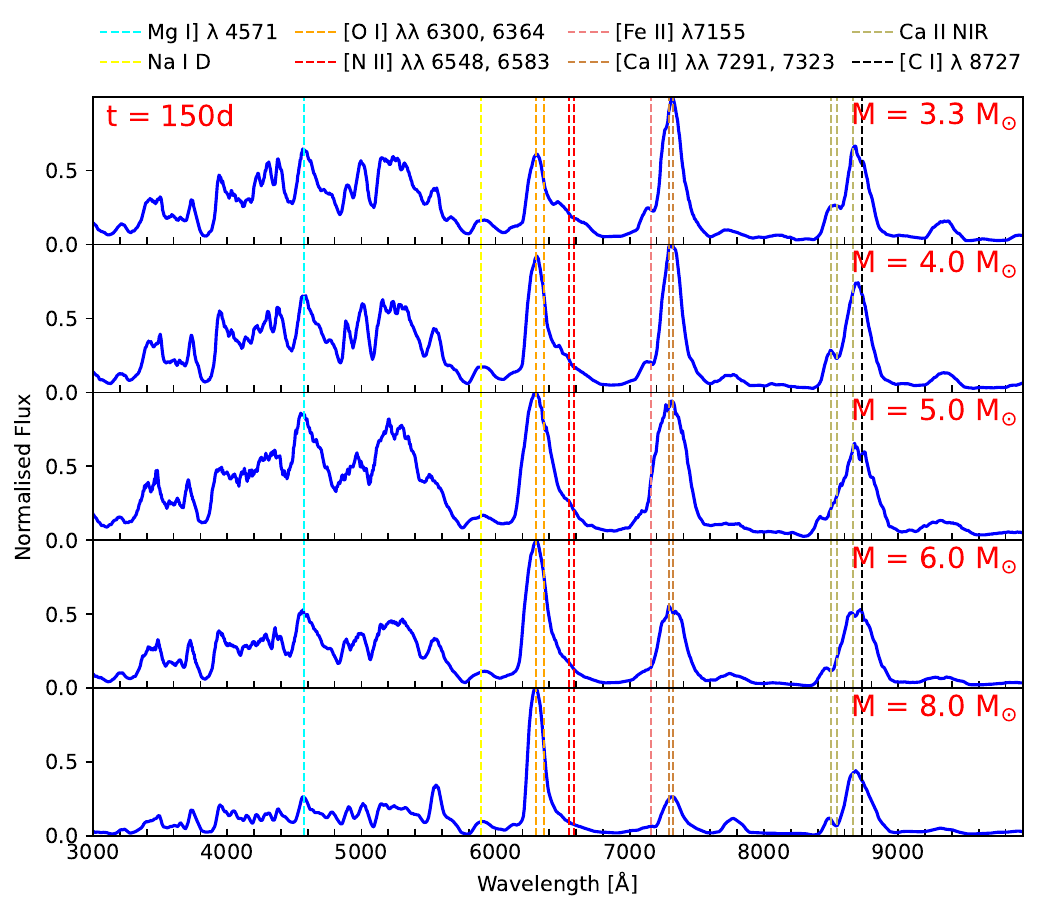}
        \vskip -.2cm
        \caption{Same as Figure \ref{fig:models300d}, but for the spectra at 150 days post-explosion.}
          \label{fig:models150d}
\end{figure*}

\begin{figure*}
        \centering
        \includegraphics[width=.98\linewidth,angle=0]{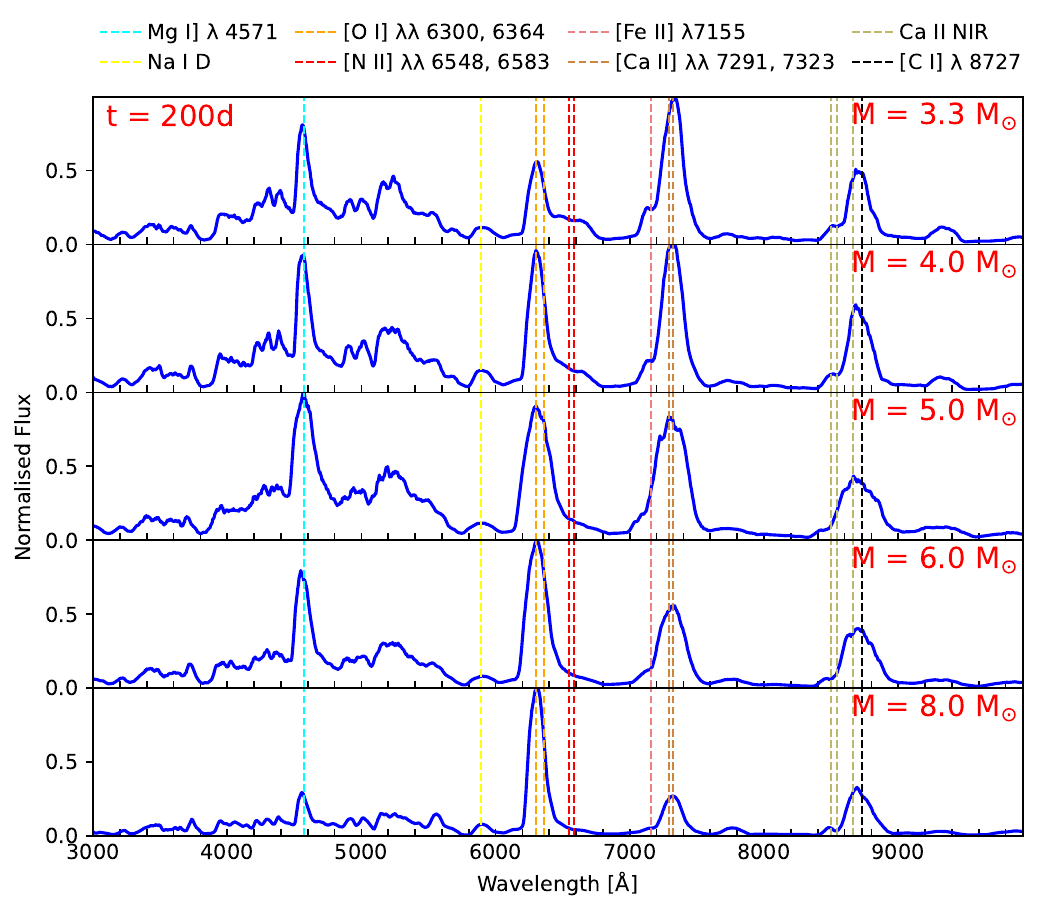}
        \vskip -.2cm
        \caption{Same as Figure \ref{fig:models300d}, but for the spectra at 200 days post-explosion.}
          \label{fig:models200d}
\end{figure*}

\begin{figure*}
        \centering
        \includegraphics[width=.98\linewidth,angle=0]{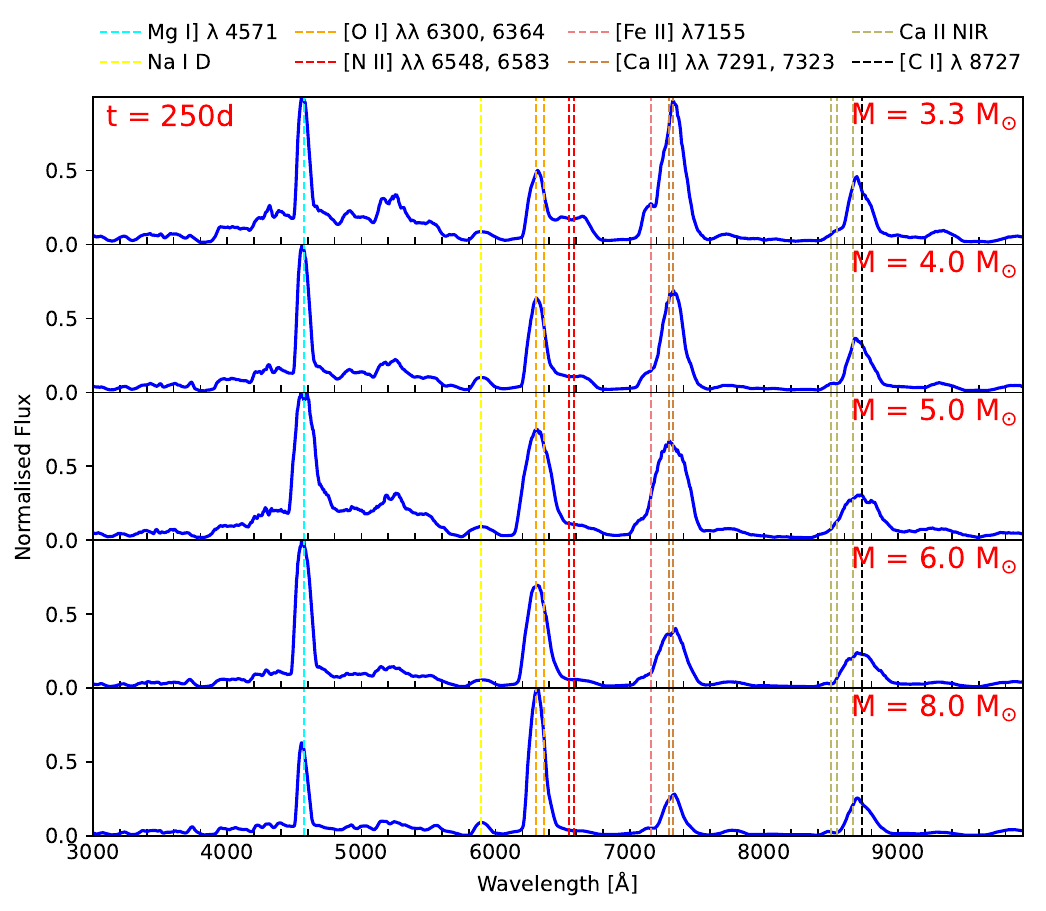}
        \vskip -.2cm
        \caption{Same as Figure \ref{fig:models300d}, but for the spectra at 250 days post-explosion.}
          \label{fig:models250d}
\end{figure*}

\begin{figure*}
        \centering
        \includegraphics[width=.98\linewidth,angle=0]{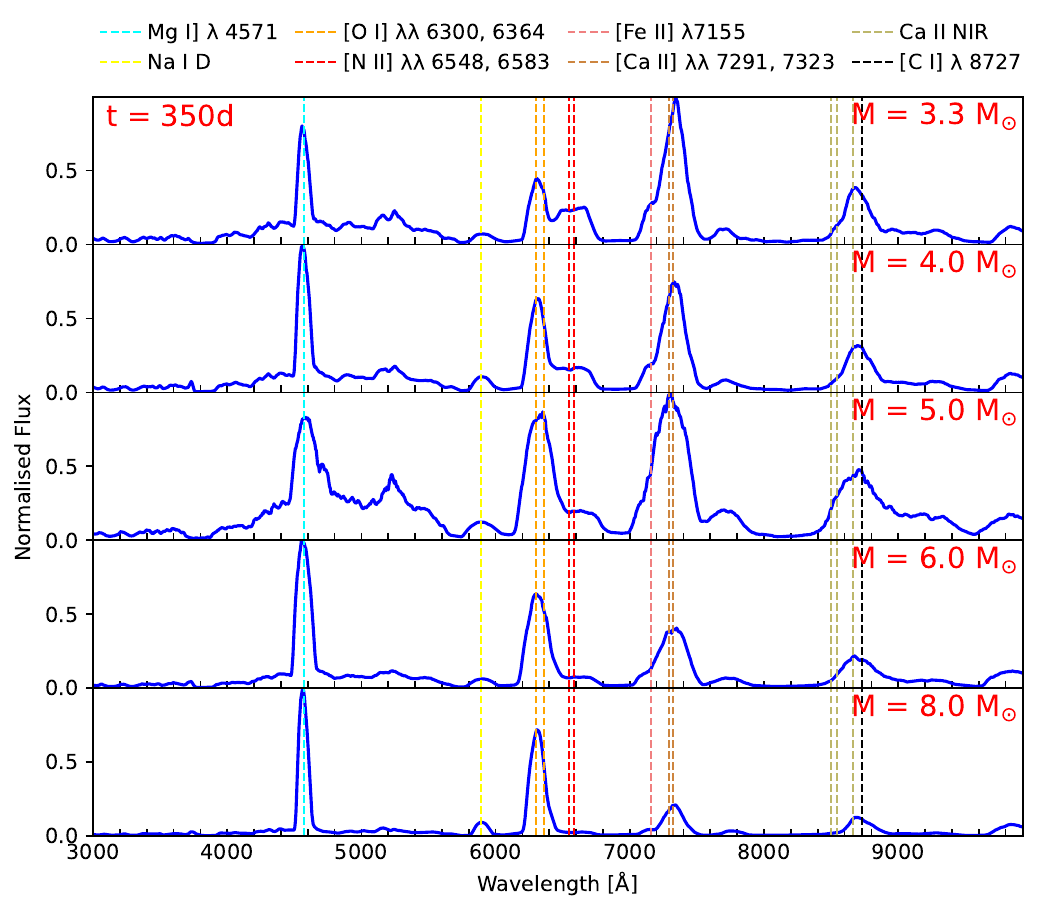}
        \vskip -.2cm
        \caption{Same as Figure \ref{fig:models300d}, but for the spectra at 350 days post-explosion}
          \label{fig:models350d}
\end{figure*}

\begin{figure*}
        \centering
        \includegraphics[width=.98\linewidth,angle=0]{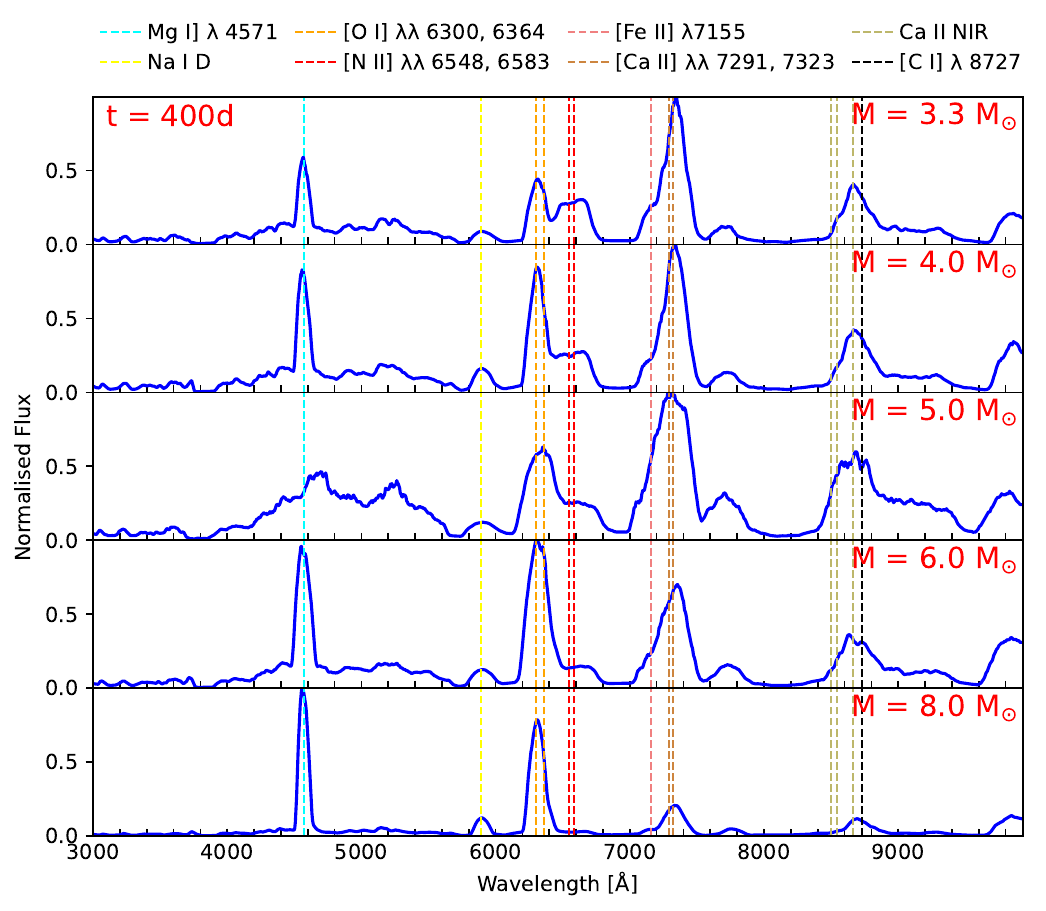}
        \vskip -.2cm
        \caption{Same as Figure \ref{fig:models300d}, but for the spectra at 400 days post-explosion}
          \label{fig:models400d}
\end{figure*}

Figures \ref{fig:models150d}-\ref{fig:models400d} show the emergent spectra as calculated by \texttt{SUMO} for our five progenitors, with each figure covering a single epoch. The figure covering the models at 300 days was presented as Figure \ref{fig:models300d}

\clearpage

\bsp	
\label{lastpage}

\end{document}